\documentstyle[epsfig,natbib2,natbibmnfix]{mn}

\newcommand{\be}{\begin{equation}}
\newcommand{\ee}{\end{equation}}

\def\ltsima{$\; \buildrel < \over \sim \;$}
\def\simlt{\lower.5ex\hbox{\ltsima}}
\def\gtsima{$\; \buildrel > \over \sim \;$}
\def\simgt{\lower.5ex\hbox{\gtsima}}

\renewcommand{\vec}[1]{ {\bmath #1} }

\title{The Aquarius Project: the subhalos of galactic halos}

\author[Springel et al.] {\parbox{18cm}{
V.~Springel$^{1}$, 
J.~Wang$^{1}$, 
M.~Vogelsberger$^{1}$,
A.~Ludlow$^{2},$ 
A.~Jenkins$^{3}$, 
A.~Helmi$^{4}$,
J.~F.~Navarro$^{2,5}$,
C.~S.~Frenk$^{3}$, %
and S.~D.~M.~White$^{1}$ %
}\vspace{0.3cm}\\
$^1$Max-Planck-Institut f\"{u}r Astrophysik,
Karl-Schwarzschild-Stra\ss{}e 1, 85740 Garching bei M\"{u}nchen,
Germany\\
$^{2}${Dep. of Physics \& Astron., University of
    Victoria, Victoria, BC, V8P 5C2, Canada}\\
$^{3}${Institute for Computational Cosmology, Dep. of Physics, Univ. of
  Durham, South Road, Durham  DH1 3LE, UK}\\
$^{4}${Kapteyn Astronomical Institute, Univ. of Groningen,
P.O. Box 800, 9700 AV Groningen, The Netherlands}\\
$^{5}${Department of Astronomy, University of Massachusetts, Amherst,
    MA 01003-9305, USA}\\
}

\begin{document}

\maketitle 
\begin{abstract}
  We have performed the largest ever particle simulation of a Milky Way-sized
  dark matter halo, and present the most comprehensive convergence study for
  an individual dark matter halo carried out thus far.  We have also simulated
  a {\em sample} of 6 ultra-highly resolved Milky-way sized halos, allowing us
  to estimate the halo-to-halo scatter in substructure statistics.  In our
  largest simulation, we resolve nearly 300,000 gravitationally bound subhalos
  within the virialized region of the halo.  Simulations of the same object
  differing in mass resolution by factors up to 1800 accurately reproduce the
  largest subhalos with the same mass, maximum circular velocity and position,
  and yield good convergence for the abundance and internal properties of dark
  matter substructures.  We detect up to four generations of subhalos within
  subhalos, but contrary to recent claims, we find less substructure in
  subhalos than in the main halo when regions of equal mean overdensity are
  compared. The overall substructure mass fraction is much lower in subhalos
  than in the main halo. Extrapolating the main halo's subhalo mass spectrum
  down to an Earth mass, we predict the mass fraction in substructure to be
  well below 3\% within 100 kpc, and to be below 0.1\% within the Solar
  Circle.  The inner density profiles of subhalos show no sign of converging
  to a fixed asymptotic slope and are well fit by gently curving profiles of
  Einasto form. The mean concentrations of isolated halos are accurately
  described by the fitting formula of Neto et al.~down to maximum circular
  velocities of $1.5\,{\rm km\,s^{-1}}$, an extrapolation over some 5 orders
  of magnitude in mass. However, at equal maximum circular velocity, subhalos
  are more concentrated than field halos, with a characteristic density that
  is typically $\sim 2.6$ times larger and increases with decreasing distance
  from halo centre.
\end{abstract}


\section{Introduction}
\label{intro}

\renewcommand{\thefootnote}{\fnsymbol{footnote}}
\footnotetext[1]{E-mail: volker@mpa-garching.mpg.de}

A major puzzle in Cosmology is that the main matter component in
today's Universe appears to be in the form of a yet-undiscovered
elementary particle whose contribution to the cosmic density is more
than 5 times that of ordinary baryonic matter
\citep[e.g.][]{Komatsu2008}.  This particle interacts extremely weakly
with atoms and photons, so that gravity alone has affected its
distribution since very early times. Recent observations have
established a standard paradigm in which dark matter emerged from the
early Universe with negligible thermal velocities and a gaussian and
scale-free distribution of density fluctuations. In this ``Cold Dark
Matter'' (CDM) hypothesis, quantum fluctuations during a very early
period of cosmic inflation determine the statistics of the dark matter
distribution at early epochs when the Universe was almost uniform
\citep{Guth1981,Starobinsky1982,Hawking1982,Bardeen1983}. Galaxies
form from these initial conditions through the condensation of gas at
the centres of a hierarchically aggregating population of
quasi-equilibrium dark matter halos \citep{White1978,White1991}.

When the effects of the baryons can be neglected, the nonlinear growth of dark
matter structure is a well-posed problem where both the initial conditions and
the evolution equations are known. This is an N-body problem {\it par
  excellence}. The faithfulness of late-time predictions (which must be
confronted directly with observation to test the paradigm) is limited purely
by numerical technique and by the available computing resources.

Over the past two decades, numerical simulations have played a pivotal role in
establishing the viability of the CDM paradigm \citep[e.g.][]{Davis1985,Frenk1988,
  Warren1992,Gelb1994,Cen1994,Hernquist1996,Jenkins2001,Wambsganss2004}. They
have led to the discovery of a universal internal structure for dark matter
halos \citep[][hereafter NFW]{Navarro1996,Navarro1997}, and they have supplied
precise predictions for the expected large-scale structure of the Universe.
The matter distribution on scales from $\sim 50$ kpc to the size of the
observable Universe and the galaxy population predicted by hierarchical CDM
scenarios have been compared directly with a wide array of observations.  A
recent example is the ``Millennium Run'' \citep{Springel2005a}, still one of
the largest cosmological simulations ever carried out, which followed the
formation and evolution of over 10 million galaxies by post-processing the
stored simulation outputs \citep{Croton2006,Bower2006}.  So far, the
$\Lambda$CDM paradigm has passed these tests successfully, particularly
those that consider the large-scale matter distribution.

Given this success in reproducing the large-scale structure of the Universe,
it is important to test CDM predictions also on smaller scales, not least
because these are sensitive to the nature of the dark matter.  Indeed, a
number of serious challenges to the paradigm have emerged on the scale of
individual galaxies and their central structure.  The realisation that CDM
halos have cuspy dark matter density profiles led to a fierce debate about
whether these are consistent with the rotation curves observed for low surface
brightness, apparently dark matter dominated galaxies
\citep{Flores1994,Moore1994,McGaugh1998,deBlok2001,Hayashi2004,Hayashi2006}.
The abundance of small dark matter subhalos predicted within CDM halos has
also drawn much attention \citep{Klypin1999,Moore1999b}. The total number
is much larger than the number of known satellite galaxies surrounding the
Milky Way, even accounting for the many recently discovered faint systems. It
is still unclear whether this reflects an absence of the predicted low mass
objects or merely the fact that no stars were able to form within them. The
issue of dark matter substructure within halos is made more urgent by the
prospect of observing dark matter particles in the near future, either by
their annihilation radiation \citep[e.g.][]{Bergstrom1998}, or through direct
detection in experiments here on Earth \citep[reviewed, e.g.,
  by][]{Gaitskell2004}. For both, a precise and quantitative understanding of
the small-scale dark matter distribution within our Galaxy is needed.

\begin{table*}
\begin{tabular}{lccrrccccr}
\hline
Name & $m_{\rm p}$ & $\epsilon$ & $N_{\rm hr}$ & $N_{\rm lr}$ &
$M_{\rm 200}$ & $r_{\rm 200}$ & $M_{\rm 50}$ & $r_{\rm 50}$ &   $N_{\rm 50}$\\
   & $[{\rm M}_\odot]$ & $[{\rm pc}]$ & & & $[{\rm M}_\odot]$ & $[{\rm kpc}]$ & 
 $[{\rm M}_\odot]$ &  $[{\rm kpc}]$    \\
\hline
Aq-A-1   &  $1.712\times 10^3$  & 20.5   & 4,252,607,000  &  144,979,154 & $1.839\times 10^{12}$ &  245.76 & $2.523\times 10^{12}$ &  433.48 & 1,473,568,512 \\
Aq-A-2   &  $1.370\times 10^4$  & 65.8   &   531,570,000  &   75,296,170 & $1.842\times 10^{12}$ &  245.88 & $2.524\times 10^{12}$ &  433.52 &   184,243,536 \\
Aq-A-3   &  $4.911\times 10^4$  & 120.5  &   148,285,000  &   20,035,279 & $1.836\times 10^{12}$ &  245.64 & $2.524\times 10^{12}$ &  433.50 &    51,391,468 \\
Aq-A-4   &  $3.929\times 10^5$  & 342.5  &    18,535,972  &      634,793 & $1.838\times 10^{12}$ &  245.70 & $2.524\times 10^{12}$ &  433.52 &     6,424,399 \\
Aq-A-5   &  $3.143\times 10^6$  & 684.9  &     2,316,893  &      634,793 & $1.853\times 10^{12}$ &  246.37 & $2.541\times 10^{12}$ &  434.50 &       808,479 \\
\hline
Aq-B-2   &  $6.447\times 10^3$  & 65.8   &   658,815,010  &   80,487,598 & $8.194\times 10^{11}$ &  187.70 & $1.045\times 10^{12}$ &  323.12 &   162,084,992 \\
Aq-B-4   &  $2.242\times 10^5$  & 342.5  &    18,949,101  &      648,874 & $8.345\times 10^{11}$ &  188.85 & $1.050\times 10^{12}$ &  323.60 &     4,683,037 \\
\hline
Aq-C-2   &  $1.399\times 10^4$  & 65.8   &   612,602,795  &   78,634,854 & $1.774\times 10^{12}$ &  242.82 & $2.248\times 10^{12}$ &  417.09 &   160,630,624 \\
Aq-C-4   &  $3.213\times 10^5$  & 342.5  &    26,679,146  &      613,141 & $1.793\times 10^{12}$ &  243.68 & $2.285\times 10^{12}$ &  419.36 &     7,110,775 \\
\hline
Aq-D-2   &  $1.397\times 10^4$  & 65.8   &   391,881,102  &   79,615,274 & $1.774\times 10^{12}$ &  242.85 & $2.519\times 10^{12}$ &  433.21 &   180,230,512 \\
Aq-D-4   &  $2.677\times 10^5$  & 342.4  &    20,455,156  &      625,272 & $1.791\times 10^{12}$ &  243.60 & $2.565\times 10^{12}$ &  435.85 &     9,579,672 \\
\hline
Aq-E-2   &  $9.593\times 10^3$  & 65.8   &   465,905,916  &   74,119,996 & $1.185\times 10^{12}$ &  212.28 & $1.548\times 10^{12}$ &  368.30 &   161,323,676 \\
Aq-E-4   &  $2.604\times 10^5$  &  342.5 &    17,159,996  &      633,106 & $1.208\times 10^{12}$ &  213.63 & $1.558\times 10^{12}$ &  369.14 &     5,982,797 \\ 
\hline
Aq-F-2   &  $6.776\times 10^3$  & 65.8   &   414,336,000  &      712,839 & $1.135\times 10^{12}$ &  209.21 & $1.517\times 10^{12}$ &  365.87 &   223,901,216 \\
Aq-F-3   &  $2.287\times 10^4$  & 120.5  &   122,766,400  &      712,839 & $1.101\times 10^{12}$ &  207.15 & $1.494\times 10^{12}$ &  363.98 &    65,320,572 \\
\hline
\end{tabular}
\caption{Basic parameters of the Aquarius simulations. We have simulated 6
  different halos, each at several different numerical resolutions. The
  leftmost column gives the simulation name, encoding the halo (A to F),
  and the resolution level (1 to 5).  $m_{\rm p}$ is the particle mass,
  $\epsilon$ is the  Plummer equivalent gravitational softening length, $N_{\rm
    hr}$ is the number of high resolution particles, and $N_{\rm lr}$ the
  number of low resolution particles filling the rest of the volume. $M_{\rm
    200}$ is the virial mass of the halo, defined as the mass enclosed in a
  sphere with mean density 200 times the critical value.  $r_{\rm 200}$ gives
  the corresponding virial radius. We also give the mass and radius for a
  sphere of overdensity 50 times the critical density, denoted as $M_{\rm 50}$
  and $r_{\rm 50}$. Note that this radius encloses a mean density $200$
  times the {\em background density}; in some studies
  \citep[e.g.][]{Diemand2007} $M_{\rm 50}$ and $r_{\rm 50}$ have been defined
  as virial mass and radius.  Finally, $N_{\rm 50}$ gives the number of
  simulation particles within $r_{\rm 50}$.
  \label{TabSims}}
\end{table*} 
  
Our Aquarius Project addresses these questions by studying the highly
nonlinear structure of Galaxy-sized CDM halos in unprecedented detail using
state-of-the art numerical simulations.  We are particularly interested in the
inner regions of these halos and of their substructures, where the density
contrast exceeds $10^6$ and the astrophysical consequences of the nature of
dark matter may be most clearly apparent. Quantifying such structure reliably
through simulation is an acute challenge to numerical technique. In order to
address this challenge, we use a newly developed version of our parallel
simulation code, {\small GADGET-3} \citep[based on][]{Springel2001a,
  Springel2005b}, which allows us to cover an unprecedented dynamic range at
high numerical accuracy. We carefully validate our simulation techniques and
establish their range of numerical convergence through systematic convergence
studies, thereby building confidence in the reliability of our results.

In order to evaluate the scatter in structural properties between halos, we
have simulated six different systems at high resolution, each having between
160 and 224 million particles within $r_{\rm 50}$, the radius with mean
enclosed overdensity 50 times the critical value.\footnote{We use this
  unconventional outer radius for our halos, rather than the standard $r_{\rm
    200}$, to facilitate comparison with \citet{Diemand2007, Diemand2008} who
  quote results for their Via Lactea simulations within $r_{\rm 50}$ although
  they refer to this radius as ``$r_{\rm 200}$''.} Each of these simulations
is better resolved than any previously published high-resolution halo
simulation except the very recent `Via Lactea II' run of \citet{Diemand2008}
which has 470 million particles within $r_{\rm 50}$. For one of our halos, we
have increased the resolution by a further factor of 8, pushing the particle
number within $r_{\rm 50}$ to 1.47 billion. The gravitational softening length
of this largest run is just $20.5\,{\rm pc}$. We collectively refer to this
suite of simulations as the Aquarius Project.

In the present paper, we describe our simulation techniques and analyze a
number of basic properties of our $z=0$ halos. We focus in particular on
the abundance of substructures, their radial distribution, and their internal
density profiles. We also give results for the concentration of substructures,
and for the fraction of the mass they contain at different radii. In a
companion paper \citep{Springel2008}, we study implications for the
detectability of dark matter annihilation within the Milky Way's dark matter
halo, and in \citet{Navarro2008} we study the structure of the central density
cusps of the main halos. Future papers will study the evolution of our halos
and their substructure.

This paper is organized as follows. In Section~\ref{SecSims}, we introduce our
simulation set and describe our numerical techniques. The abundance of dark
matter substructures and their radial distribution within our halos are
analyzed in Section~\ref{SecSubAbundance}. Then, in
Section~\ref{SecSubsInSubs}, we turn to an analysis of the abundance of
substructure within subhalos.  In Section~\ref{SecSubInternalStructure}, we
consider the density profiles of subhalos and their concentrations.  Finally,
we summarize our conclusions in Section~\ref{SecSummary}.

\section{Simulation set and numerical techniques} \label{SecSims}

All our simulations follow halo formation within a periodic cube of side
$100\,h^{-1}{\rm Mpc}\simeq 137\,{\rm Mpc}$ in a cosmology with parameters
$\Omega_m = 0.25$, $\Omega_\Lambda=0.75$, $\sigma_8=0.9$, $n_s=1$, and Hubble
constant $H_0 =100\,h\,{\rm km\,s^{-1}\,Mpc^{-1}} = 73\,{\rm
  km\,s^{-1}\,Mpc^{-1}}$. These cosmological parameters are the same as used
in the Millennium Simulation project, and are consistent with the current set
of cosmological constraints within their uncertainties, in particular those
from the WMAP 1- and 5-year data analyses.

\subsection{Setting the initial conditions}

The linear power spectrum used for making the initial conditions is based on a
transfer function made by {\small CMBFAST} \citep[v4.5.1,][]{Seljak1996} with
$\Omega_{\rm baryon}=0.045$.  The transfer function was evaluated at $z=0$
where the CDM and baryon transfer functions are virtually identical for
wavenumbers $k$ up to $\log_{10}(kh/{\rm Mpc})=2.5$. For higher wavenumbers
the effects of pressure are important for the baryons and are reflected in a
feature present also in the CDM transfer function.  Given that our simulations
model only a CDM component and cannot account for the separate evolution of
the baryons, we have chosen to ignore this baryon induced feature for our CDM
power spectrum and have instead created a smooth composite transfer function
(explained in detail below) which lacks this feature. In practice, for the
Aquarius simulations, the difference between using our composite transfer
function and the {\small CMBFAST} transfer function is relatively modest.  The
size of the difference can be gauged by considering the fractional change in
the {\it rms} linear density fluctuations for a spherical top-hat filter
enclosing a mass of mean density corresponding to 32 particles (our resolution
limit in the subsequent analyses). For our highest resolution simulation the
{\it rms} using our composite transfer function is less than 2\% above that using the
{\small CMBFAST} CDM transfer function. For all other simulations the
difference is even smaller.
 
Our composite transfer function was formed as follows.  It was set equal to
the {\small CMBFAST} transfer function for $\log_{10}(kh/{\rm Mpc}) < 1$, and
equal to an analytic form based on the CDM only transfer function of
\citet[][hereafter BBKS]{BBKS} for $\log_{10}(kh/{\rm Mpc}) > 2$.  Over the
intermediate range the composite transfer function is given by the linear
combination of $(1-w)$ times the {\small CMBFAST} and $w$ times the BBKS
transfer function, where the weighting function $w$ changes smoothly from 0 to
1.  By using a value of $\Gamma=0.16$ for the shape parameter in the BBKS
formula and scaling the overall amplitude it proved possible to match the
amplitudes of the {\small CMBFAST} and BBKS functions everywhere in the
transition region, $ 1 <\log_{10}(kh/{\rm Mpc}) < 2$, to better than 0.5\%,
with the result that the overall transfer function is very smooth.

\begin{figure}
\begin{center}
\resizebox{8.5cm}{!}{\includegraphics{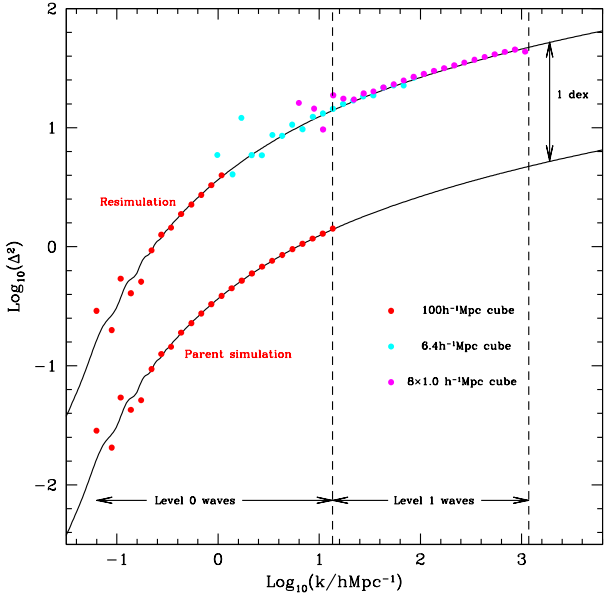}}
\end{center}
\caption{Measured power spectrum (dimensionless variance
  $\Delta^2(k)\sim k^3P(k)$ per natural log interval) in our highest
  resolution `zoom' initial conditions, Aq-A-1, linearly extrapolated
  to the values expected at $z=0$. The lower red points show the power
  spectrum measured in our homogeneously sampled parent simulation,
  shifted down by one dex for clarity. The upper red circles show our
  measurement for the zoom initial conditions for the whole box, while
  cyan and magenta circles show measurements from smaller boxes
  centred within the high-resolution region. The solid black lines
  show the linear theory input spectrum. The vertical dashed line on
  the right marks the Nyquist frequency of the high-resolution region,
  while the left vertical line is the joining point between the long
  wavelength modes from the parent simulation and the high frequency
  modes added to the high resolution cube. \label{FigPowerSpec}}
\end{figure}

We selected target halos for resimulation from a parent simulation carried out
at homogeneous resolution with $900^3$ particles in this same box. We targeted
halos of roughly Milky Way mass and without a massive close neighbour at
$z=0$. We also checked that semi-analytic modelling applied to our target
halos predicated them to host late-type galaxies. Otherwise our selection was
random. New initial conditions for the selected objects were then constructed
by identifying the Lagrangian region from which each halo formed.  The high
resolution region, which has an `amoeba'-like shape, was defined as the union
of a set of small identical cubes joined face to face covering the whole of
the Lagrangian region and forming a simply connected volume. Within this
region the mass distribution was represented by a much larger number of lower
mass particles.  On making the initial conditions, additional small-scale
power was added to the high resolution region as dictated by the higher local
particle Nyquist frequency. More distant regions were sampled with
progressively more massive particles, but retaining sufficient resolution to
ensure an accurate representation of the tidal field at all times. The initial
displacements were imprinted using the Zeldovich approximation, and a
`glass-like' uniform particle load \citep{White1996} was used within the
high-resolution regions of all our initial conditions.  We also invested
particular care to guarantee that all our final halos are unaffected by
contamination by heavier boundary particles. In fact, all our halos are free
of {\em any} boundary particles within the radius $r_{\rm 50}$, except for
simulations Aq-E-2, Aq-F-2, and Aq-F-3, where 71, 9, and 3 heavier particles
are found within this radius, respectively, corresponding to a fraction
$\simlt 10^{-5}$ of the total mass within this radius.  Typically, about 30\%
of the high-resolution particles in the new initial conditions end up in the
virialized region of the final halo.

In setting up these zoomed initial conditions, all power from the parent
simulation is deleted beyond some wavenumber which is smaller than the Nyquist
wavenumber of the parent simulation, but substantially larger than the
fundamental wavenumber of the cube enclosing the high resolution region in
which we generate the additional high frequency waves. The latter replace the
waves deleted from the fluctuation field of the parent simulation and extend
its power spectrum up to the Nyquist frequency corresponding to the mean
interparticle separation in the high-resolution region. When we create a
series of simulations of the same object at differing mass resolution
(i.e. with different particle masses in the high resolution region) we are
careful to ensure that all the waves used to create a lower resolution
simulation are present with identical amplitude and phase in all higher
resolution simulations. This means that every object which forms in the
lower resolution simulation should also be present with identical mass and
position in its higher resolution counterparts. This allows us to make
detailed convergence tests on the properties of every nonlinear object in our
simulations, not just on the main halo.

In Figure~\ref{FigPowerSpec}, we show the power spectrum as measured
from the initial conditions of our highest resolution resimulation,
Aq-A-1 from Table~\ref{TabSims}. All the measurements are made using a
$1000^3$ Fourier transform using a cloud-in-cell assignment scheme and
taking into account for each mode the expected smoothing effect of the
assignment scheme.  Measuring the power spectrum accurately however
requires a further measure.  Even in the absence of the imposed
perturbations, the unperturbed particle distribution has some
measurable power. This power in a discrete particle representation,
unlike a truely uniform mass distribution, does not grow by
gravitational instability.  Over most scales this extra power is
negligible compared to that introduced when the density perturbations
are added by displacing the particles which we wish to
measure. However at scales approaching the particle Nyquist frequency
the contribution from the unperturbed particle distribution starts to
become significant compared to the imposed power.  To allow for this
effect, which would otherwise lead to an overestimate of the input
power, the values for all the points plotted are computed by
differencing the power of the perturbed and unperturbed particle
positions.

 In Figure~\ref{FigPowerSpec} the lower sequence of red filled circles
shows the measured power spectrum (linearly extrapolated to $z=0$ and
offset by exactly one dex for clarity) from the parent $900^3$
particle simulation, plotted down to the shortest wavelength included
in the refined initial conditions. The two black curves show the
theoretical linear power spectrum, discussed above, with the upper
curve extrapolated to $z=0$ and the lower curve shifted down from
it by one dex. The power spectrum of the parent simulation is
evidently a very good match to the theoretical power spectrum except
at low $k$ where close agreement is not expected because only a few
modes contribute.  The upper red circles show the power spectrum
measured in the initial conditions of the zoom simulation (linearly
extrapolated to $z=0$). The most massive particles in the refined
initial conditions have a mass $1/41^3$ of the entire $100\,h^{-1}{\rm
Mpc}$ box. As the Nyquist frequency for particles of this mass
corresponds to $\log(k\,h\, {\rm Mpc}^{-1}) = 0.11$, only points to
the left of this limit are plotted.  The cyan circles show the power
spectrum measured from a box of side length $6.4\,h^{-1}{\rm Mpc}$
centred on the high resolution region. The masses of the particles in
this region vary, with the most massive particles having a mass of
$\sim 7.23\times 10^{-8}$ of the region within which high-frequency
power is added. (The latter is cubic with a side length of
$7.06\,h^{-1}{\rm Mpc}$.)  The Nyquist frequency corresponding to the
most massive particles is here $\log(k\,h\,{\rm Mpc}^{-1}) = 2.03$,
and the cyan points are shown only to the left of this limit.

 Finally, the magenta circles show the average power spectrum measured
from eight boxes of side-length $1\,h^{-1}{\rm Mpc}$ inside the
central high resolution region, where all the particles are of the
same mass.  The largest possible cube that can be extracted where all the
particles have the same mass is about $2.4\,h^{-1}{\rm Mpc}$ on a
side.  A $1000^3$ Fourier transform is not large enough to make an
accurate measurement of the power spectrum over this size of cube
because the particle and Fourier mesh Nyquist frequencies are very
close. Instead we placed 8 non-overlapping $1\,h^{-1}{\rm Mpc}$ cubes
inside this volume.  Averaging the eight regions reduces the expected
scatter at low wavenumber, but makes no significant difference
compared to a single measurement at high wave-numbers.  Again the
points are only plotted up to their Nyquist frequency, which is marked
by the rightmost vertical dashed line.  The leftmost vertical dashed
line marks the joining point in the high-resolution cube between the
long-wavelength waves from the parent simulation and the
high-frequency waves in the high-resolution cube. The joining point
corresponds to about 15 waves across the high-resolution cube. 

These measurements of the power spectrum of our initial conditions
show that we clearly achieve an excellent match to the desired linear
input spectrum over many decades in spatial scale. We stress that such
tests of the initial conditions are essential, as their quality is
obviously of paramount importance for the accuracy of the evolved
simulations.  Note that in the following, ``high-resolution region''
refers to the amoeba-shaped region where all the particles have the
same mass rather than to the region to which high frequency waves are
added (which is larger). Similarly, the low-resolution region is
everything outside the amoeba.

\subsection{The Aquarius simulation suite}

In Table~\ref{TabSims}, we provide an overview of the basic numerical
parameters of our simulations. This includes a symbolic simulation name, the
particle mass in the high-resolution region, the gravitational softening
length, the total particle numbers in the high- and low-resolution regions, as
well as various characteristic masses and radii for the final halos, and the
corresponding particle numbers.  Our naming convention is such that we use
the tags ``Aq-A'' to ``Aq-F'' to refer to simulations of the six Aquarius
halos. An additional suffix ``1'' to ``5'' denotes the resolution
level. ``Aq-A-1'' is our highest resolution calculation with $\sim 1.5$
billion halo particles. We have level 2 simulations of all 6 halos,
corresponding to 160 to 224 million particles per halo.  

We kept the gravitational softening length fixed in comoving coordinates
throughout the evolution of all our halos. The dynamics is then governed by a
Hamiltonian and the phase-space density of the discretized particle system
should be strictly conserved as a function of time \citep{Springel2005b}, modulo
the noise introduced by finite force and time integration errors.  Timestepping
was carried out with a kick-drift-kick leap-frog integrator where the timesteps
were based on the local gravitational acceleration, together with a
conservatively chosen maximum allowed timestep for all particles.

\begin{figure*}
\begin{center}
\begin{tabular}{cc}
\vspace*{-0.4cm}\hspace*{-0.2cm}\parbox[b]{3.2cm}{\resizebox{3.2cm}{!}{\includegraphics{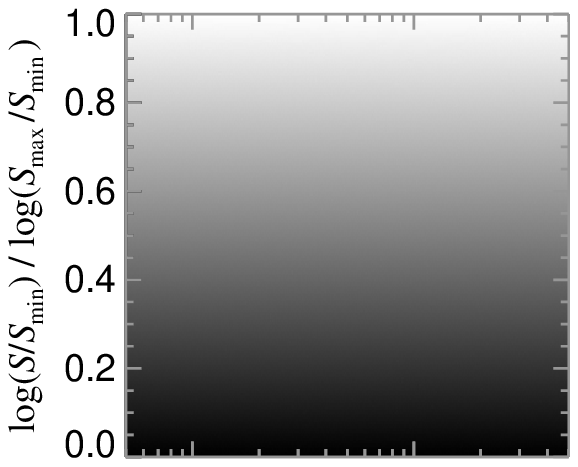}}\\%
\resizebox{3.2cm}{!}{\includegraphics{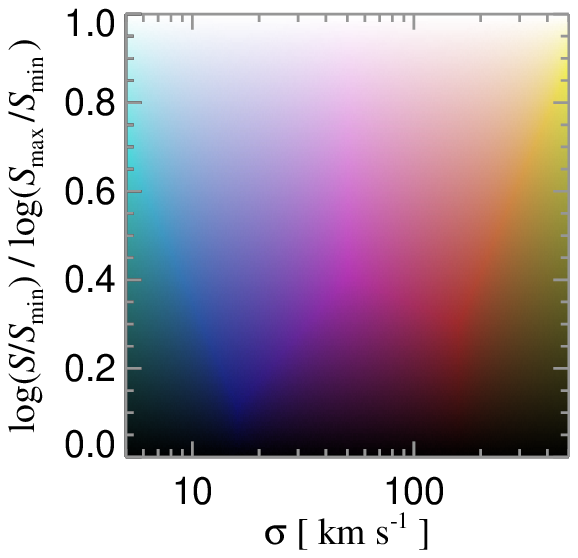}}}&
\resizebox{14.0cm}{!}{\includegraphics{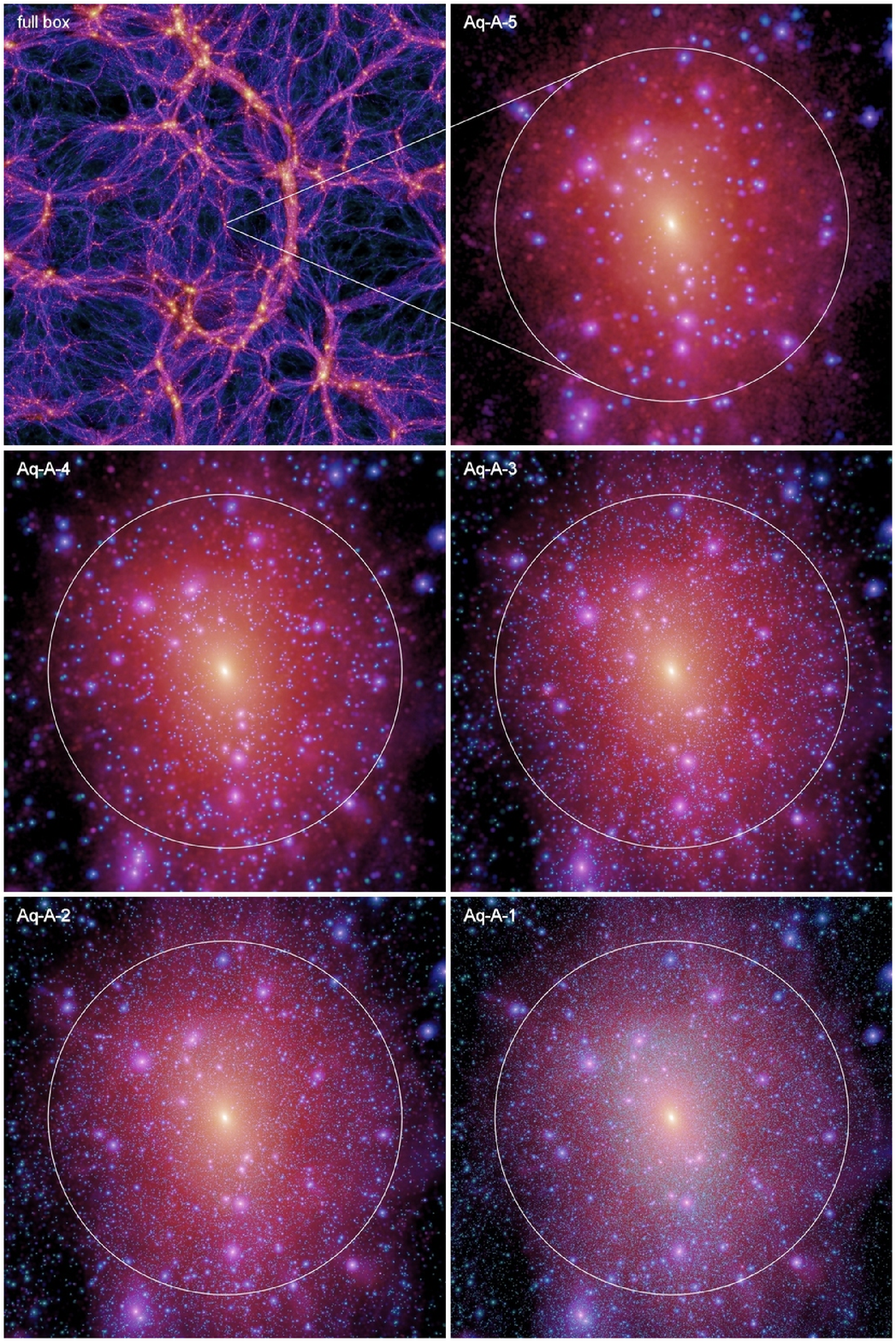}}
\end{tabular}
\end{center}
\caption{The top left panel shows the projected dark matter density at
$z=0$ in a slice of thickness $13.7\,{\rm Mpc}$ through the full box
($137\,{\rm Mpc}$ on a side) of our $900^3$ parent simulation,
centred on the `Aq-A' halo that was selected for resimulation. The
other five panels show this halo resimulated at different numerical
resolutions.  In these panels, all particles within a cubic box of
side-length $2.5\times r_{50}$ centred on the halo are shown. The
image brightness is proportional to the logarithm of the squared dark
matter density $S(x,y)$ projected along the line-of-sight, and the
colour hue encodes the local velocity dispersion weighted by the
squared density along the line-of-sight.  We use a two-dimensional
colour table (as shown on the left) to show both of these quantities
simultaneously. The colour hue information is orthogonal to the
brightness information; when converted to black and white, only the
density information remains, with a one-dimensional grey-scale colour
map as shown on the left.  The circles mark $r_{50}$.}
\label{FigDMDist}
\end{figure*}

\begin{figure*}
\begin{center}
\resizebox{17.6cm}{!}{\includegraphics{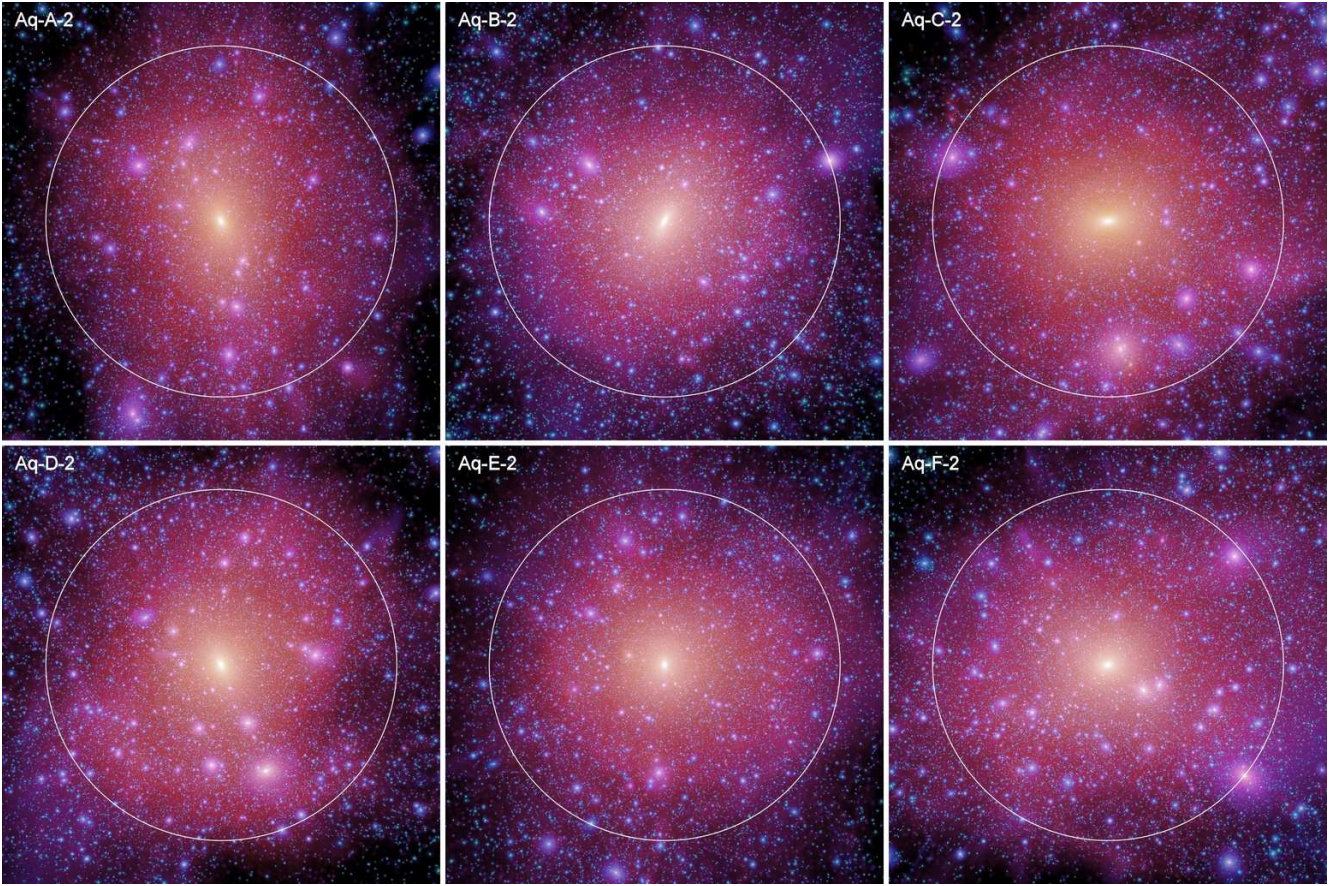}}
\end{center}
\caption{Projected dark matter density in our six different
high-resolution halos at $z=0$, at the `2' resolution level. In each
panel, all particles within a cubic box of side length $2.5\times
r_{50}$ centred on the halo are shown, and the circles mark the radius
$r_{50}$.  The image brightness is proportional to the logarithm of
the squared dark matter density, and the colour hue encodes the local
particle velocity dispersion, with the same colour map as in
Figure~\ref{FigDMDist}.  }
\label{FigDMDist2}
\end{figure*}

\begin{table*}
\begin{tabular}{lcccccrr}
\hline
Name &   $V_{\rm max}$ & $r_{\rm max}$  & $\delta_V$ & $c^*_{\rm NFW}$ & $z_{\rm
  form}$ & $N_{\rm sub}$ & $f_{\rm sub}^{\rm cumul}$ \\
     &   $[{\rm km\,s^{-1}}]$ & $[{\rm kpc}]$ &  & &  &   \\
\hline
Aq-A-1   &  208.75 &  28.35  & $2.035\times 10^4$  & 16.11 &  1.93 &   297791 &   13.20 \% \\
Aq-A-2   &  208.49 &  28.14  & $2.060\times 10^4$  & 16.19 &  1.93 &    45024 &   12.16 \% \\
Aq-A-3   &  209.22 &  27.88  & $2.114\times 10^4$  & 16.35 &  1.93 &    13854 &   11.34 \% \\
Aq-A-4   &  209.24 &  28.20  & $2.067\times 10^4$  & 16.21 &  1.93 &     1960 &    9.68 \% \\
Aq-A-5   &  209.17 &  28.55  & $2.015\times 10^4$  & 16.04 &  1.93 &      299 &    8.64 \% \\
\hline
Aq-B-2   &  157.68 &  40.15  & $5.788\times 10^3$  &  9.72 &  1.39 &    42537 &   10.54 \% \\
Aq-B-4   &  159.03 &  44.31  & $4.834\times 10^3$  &  9.02 &  1.39 &     1614 &    8.26 \% \\
\hline
Aq-C-2   &  222.40 &  32.47  & $1.761\times 10^4$  & 15.21 &  2.23 &    35022 &    7.17 \% \\
Aq-C-4   &  223.20 &  33.63  & $1.654\times 10^4$  & 14.84 &  2.23 &     1972 &    6.02 \% \\
\hline
Aq-D-2   &  203.20 &  54.08  & $5.299\times 10^3$  &  9.37 &  1.51 &    47014 &   13.06 \% \\
Aq-D-4   &  204.47 &  55.76  & $5.046\times 10^3$  &  9.18 &  1.51 &     3116 &   10.67 \% \\
\hline
Aq-E-2   &  179.00 &  55.50  & $3.904\times 10^3$  &  8.26 &  2.26 &    42725 &   10.75 \% \\
Aq-E-4   &  182.68 &  54.59  & $4.202\times 10^3$  &  8.52 &  2.26 &     2024 &    7.53 \% \\ 
\hline
Aq-F-2   &  169.08 &  42.67  & $5.892\times 10^3$  &  9.79 &  0.55 &    52503 &   13.39 \% \\
Aq-F-3   &  174.05 &  43.76  & $5.937\times 10^3$  &  9.82 &  0.55 &    12950 &    9.15 \% \\
\hline
\end{tabular}
\caption{Basic structural properties of the main halos in our various
  simulations.  The leftmost column gives the simulation name, $V_{\rm max}$
  is the maximum circular velocity, $r_{\rm max}$ is the radius where this
  maximum is reached, $\delta_V$ gives the characteristic density contrast 
  based on the peak of the
  circular velocity curve, while $c_{\rm NFW}^\star$ is the same value
  converted to an equivalent NFW concentration under the assumption that
  the halo is reasonably well fit by an NFW profile. $z_{\rm form}$ gives the
  formation redshift of the halo, defined as the earliest epoch at which the
  $M_{200}$ mass of the main halo progenitor exceeds half its final
  value. Finally, $N_{\rm sub}$ gives the total number of subhalos that we
  resolve inside $r_{50}$, and $f_{\rm sub}^{\rm cumul}$ is their total mass fraction
  relative to all the mass inside $r_{50}$.\label{TabProperties}}
\end{table*}

We define the virial mass $M_{\Delta}$ and viral radius $r_{\Delta}$ as the
mass and radius of a sphere that encloses a mean density $\Delta \times
\rho_{\rm crit}$, where $\rho_{\rm crit}$ is the critical density.  Different
choices for $\Delta$ are used in the literature. The most common ones are (1)
a fixed value of $\Delta = 200$ as in NFW's original work, (2) a value of
$\Delta\sim 178\, \Omega_m^{0.45}$ based on a generalization of the spherical
top-hat collapse model to low density cosmologies \citep{Eke1996,Bryan1998},
or (3) a value of $\Delta = 200\, \Omega_m(z)$, which corresponds to a fixed
overdensity relative to the background density.

We will frequently give results for the radius according to convention (3),
for which $\Delta = 200\, \Omega_m = 50$ at $z=0$, simply because this
yields the largest radius and hence the largest number of substructures, which
improves statistics.  The corresponding radius is designated as $r_{50}$,
while $r_{200}$ refers to the radius that encloses an overdensity of 200 with
respect to the critical density, as is customary in the literature
\citep[except for][and collaborators, who use $r_{200}$ to refer to a radius
that encloses 200 times the {\em mean density}, which is equivalent to
$r_{50}$ in our notation]{Diemand2007}.

\subsection{Integration technique}

All our simulations were started at redshift $z=127$, and were evolved with a
new parallel TreePM code {\small GADGET-3} written especially for the Aquarius
project.  This code offers much better scalability to large numbers of compute
cores as well as higher basic speed than its parent code \mbox{{\small
    GADGET-2}} \citep{Springel2005b}.  The gravitational field on large scales
is calculated with a particle-mesh (PM) algorithm, while the short-range
forces are delivered by a tree-based hierarchical multipole expansion, such
that a very accurate and fast gravitational solver results.  The scheme
combines the high spatial resolution and relative insensitivity to clustering
of tree algorithms with the unmatched speed and accuracy of the PM method to
calculate the long range gravitational field. We note that achieving our force
resolution with a single mesh in a standard PM approach would require a grid
with $(10^7)^3$ cells -- storing such a mesh would require several million
petabytes. This illustrates the enormous dynamic range we are aiming for with
our simulations.

\begin{figure}
\resizebox{8.5cm}{!}{\includegraphics{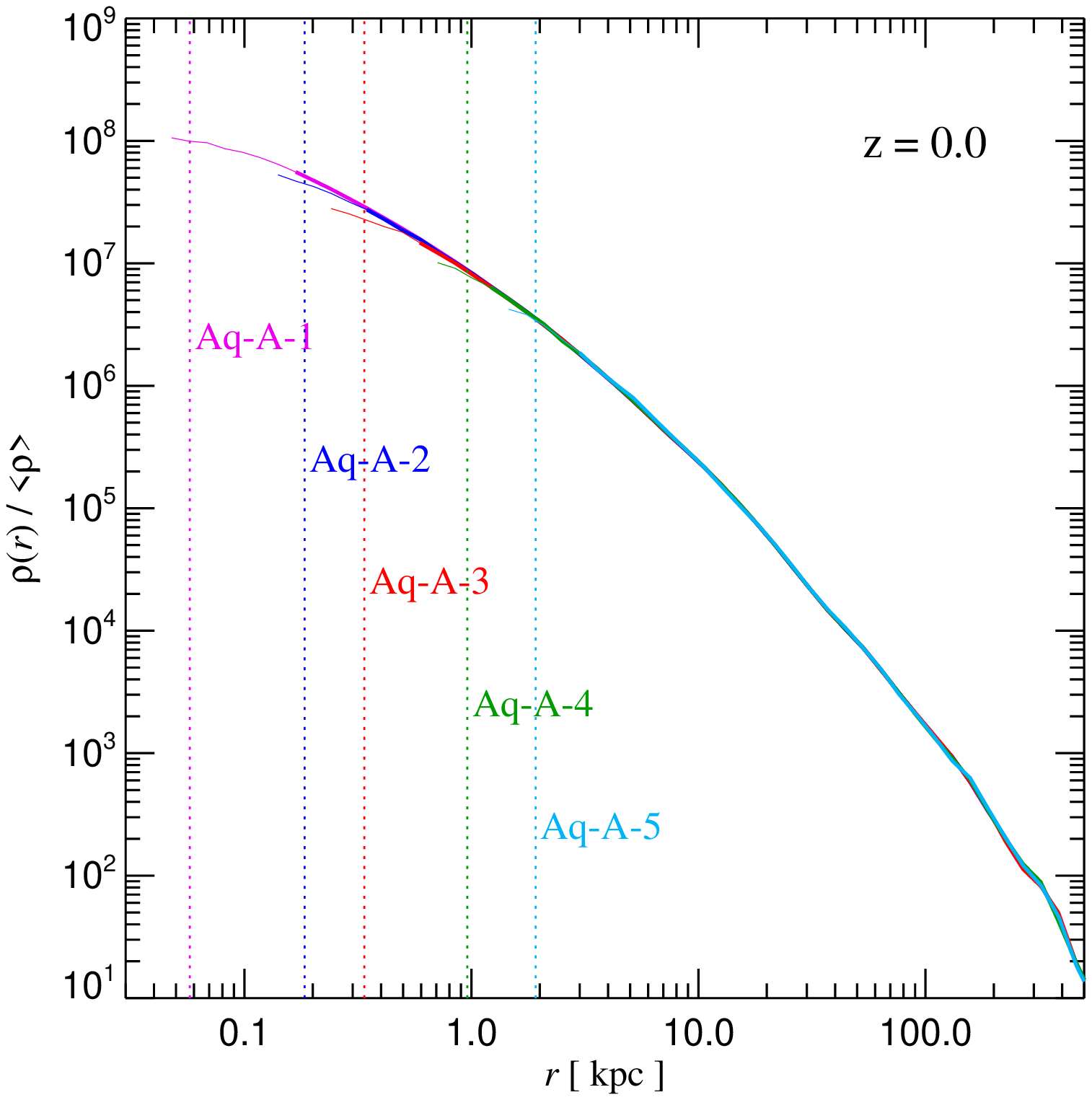}}
\caption{Spherically averaged density profile of the Aq-A halo at $z=0$,
  at different numerical resolutions. Each of the profiles is
  plotted as a thick line for radii that are expected to be converged
  according to the resolution criteria of \citet{Power2003}. These work very
  well for our simulation set. We continue the measurements as thin solid
  lines down to $2\,\epsilon$, where $\epsilon$ is the Plummer-equivalent
  gravitational softening length in the notation of \citet{Springel2001a}.
  The dotted vertical lines mark the scale $2.8\,\epsilon$, beyond which the
  gravitational force law is Newtonian. The mass resolution changes by a
  factor of $1835$ from the lowest to the highest resolution simulation in
  this series. Excellent convergence is achieved over the entire radial
  range where it is expected.
\label{DensProfilesC02}}
\end{figure}

\begin{figure}
\resizebox{8.5cm}{!}{\includegraphics{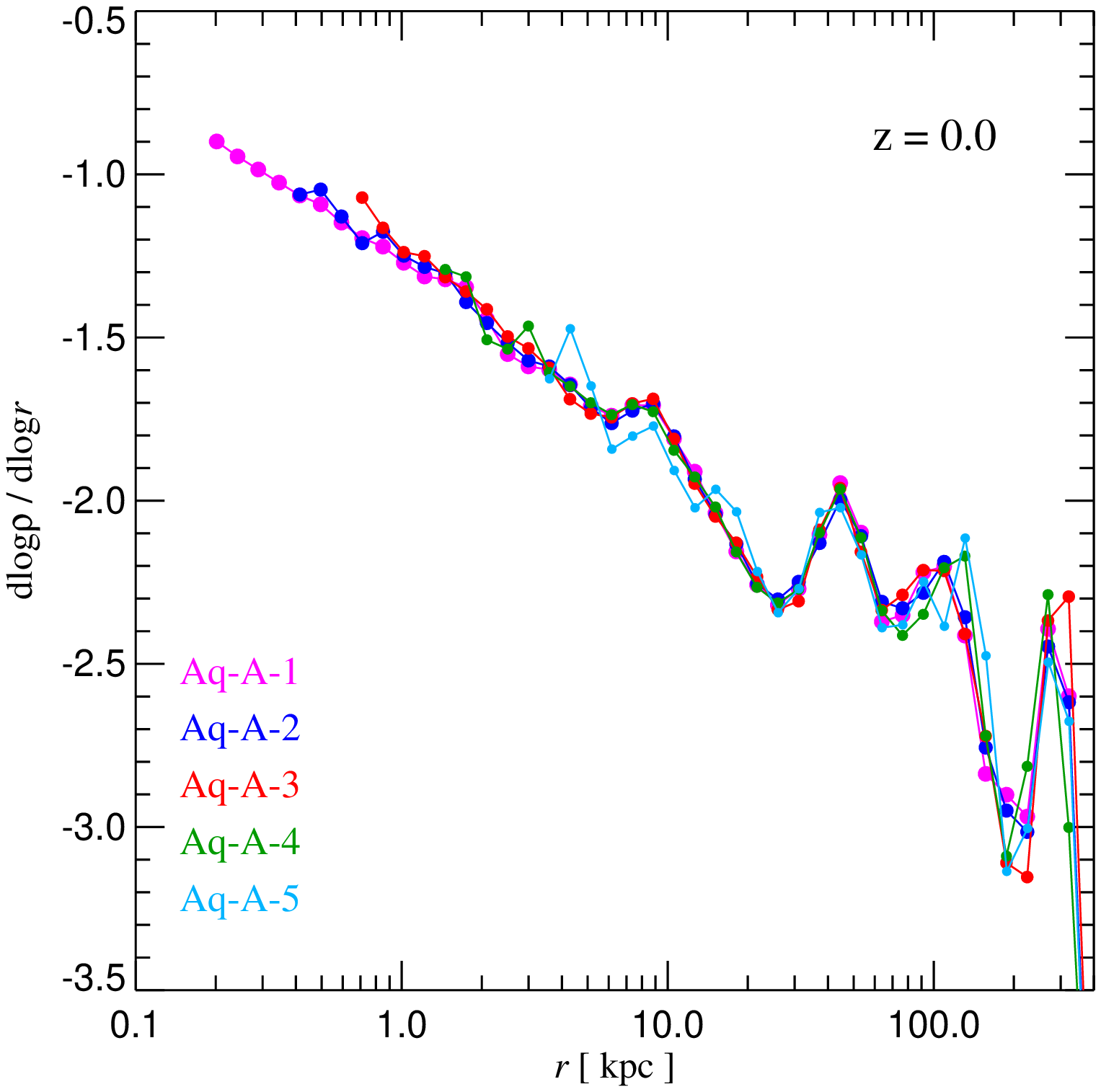}}
\caption{Local logarithmic slope of the density profiles as a function of
  radius for the Aq-A halo simulated at different numerical resolution. Only
  the radial region that should be converged according to the criteria of
  \citet{Power2003} is shown.
  Note that the large
  fluctuations in the outer parts are caused by substructures but
  nevertheless reproduce well between simulations. In this regime, we expect
  significant halo-to-halo scatter.
  \label{SlopesC02}}
\end{figure}

In fact, the numerical challenge of the calculations is substantial.  One
challenge simply reflects the large dynamic range involved: gravitational
timescales are inversely proportional to the square root of the density, so
simulating a CDM halo means dealing with a system where different regions
evolve on timescales which may differ by factors of thousands.  A code with
spatially-dependent, adaptive timestepping is mandatory; otherwise the most
rapidly evolving regions -- which usually include only a tiny fraction of the
mass -- force timesteps so short that the calculation grinds to a halt.  A
second challenge stems from the highly clustered spatial distribution of
matter and affects, in particular, the scalability of parallel algorithms. A
CDM halo is a nearly monolithic, highly concentrated structure with a
well-defined centre. There is no obvious geometrical decomposition which can
separate it into the large number of computationally equivalent domains
required for optimal exploitation of the many processors available in
high-performance parallel architectures. In our highest resolution
calculation, the clustering is so extreme that far more than a billion
particles (amounting to about one third of all particles in the simulation)
collect in a region that encompasses less than $10^{-8}$ of the simulated
volume. In addition, gravity couples the dynamics of matter throughout the
halo and beyond, requiring efficient communication between all parts of the
simulated region.

It is clear that parallelization of such calculations for distributed memory
machines is difficult, yet is mandatory to make them feasible on today's
supercomputers. We have carried out our most expensive calculation, the Aq-A-1
run, on the Altix 4700 supercomputer of the Leibniz Computing Center (LRZ) in
Garching/Germany, using 1024 CPUs and about 3 TB of main memory.  The
calculation took more than 3.5 million CPU hours to carry out about 101400
timesteps that involved $6.72\times 10^{13}$ force calculations in total.  We
have stored 128 simulation dumps for this calculation, amounting to a data
volume of about 45 TB.  The other simulations of the Aquarius Project were in
part calculated on the LRZ system, and in part on other supercomputers across
Europe.  These were the COSMA computer at Durham University/UK, the Bluegene/L
system STELLA of the LOFAR consortium in Groningen/Netherlands, and a
Bluegene/P system of the Max-Planck Computing Center in Garching. For all
these simulations we also stored at least 128 outputs, but for Aq-A-2 and
Aq-A-4 we kept 1024 dumps, and for Aq-A-3 half this number. This provides
exquisite time resolution for studies of the detailed formation history of
halos and the evolution of their substructure. In the present study, however,
we focus on an analysis of the objects at $z=0$.

\subsection{A first view of the simulations}

In Figures~\ref{FigDMDist} and \ref{FigDMDist2}, we show
  images\footnote{Further images and videos of the formation process
  of the halos are available at
  http://www.mpa-garching.mpg.de/aquarius} of the dark matter
  distribution in our 6 high resolution halos at redshift $z=0$. The
  brightness of each pixel is proportional to the logarithm of the
  squared dark matter density projected along the line-of-sight,
\begin{equation}
S(x,y) = \int \rho^2(\vec{r}) \,{\rm d}z ,
\end{equation}
while the colour hue encodes the mean dark matter velocity dispersion, weighted
as
\begin{equation}
\sigma(x,y) = \frac{1}{S(x,y)} \int \sigma_{\rm loc}(\vec{r})\,
\rho^2(\vec{r}) \,{\rm d}z .
\end{equation}
Here the local dark matter density $\rho(\vec{r})$ and the local velocity
dispersion $\sigma_{\rm loc}(\vec{r})$ of the particles are estimated with an
SPH kernel interpolation scheme based on 64 neighbours.  We use a
two-dimensional colour-table (see Fig.~\ref{FigDMDist}) in which the
information about the local dark matter `temperature' is orthogonal to the
density information; conversion of the images into grey-scale eliminates the
velocity information but leaves the density information intact, with the
latter being proportional to the dark matter annihilation luminosity.

Looking at these images it is clear that our halos are filled with a sea of
dark matter substructures of many different sizes. Figure~\ref{FigDMDist}
shows that these repeat closely, as they should, between simulations of the
same object at different resolution, albeit at slightly different
positions. As we will show later in more detail, there are even substructures
inside subhalos. In fact, up to four such generations are resolved by our
highest resolution calculation, the Aq-A-1 version of the `A' object.  It is
clear that an important task in analyzing this complex phase-space structure
lies in finding the gravitationally bound substructures that orbit within the
virialized regions of the halos.

We address this complex problem with our {\small SUBFIND} algorithm
\citep{Springel2001b} which finds substructures using a topological excursion
set method. Based on local dark matter density estimates calculated with the SPH
kernel interpolation approach for all high resolution particles, we first
identify a set of subhalo candidates, which are locally overdense structures
found within a given input group of particles identified with a {\small FOF}
(friends-of-friends) group finder \citep{Davis1985}. These are then subjected to
a gravitational unbinding procedure that iteratively eliminates all unbound
particles. Provided more than 20 bound particles remain, we record the particle
group as a genuine subhalo in our group catalogue. For each subhalo, we
calculate a number of physical properties such as the maximum circular velocity,
spin and velocity dispersion, and we store the particles in order of the
gravitational binding energy, which is useful for tracking subhalos between
simulation outputs at different times. We have fully parallelized the {\small
SUBFIND} and {\small FOF} algorithms for distributed memory systems and inlined
them in our simulation code {\small GADGET-3}. Thus group-finding can be done on
the fly during the simulation, if desired. This is often advantageous as these
calculations are computationally quite intense and require equally large memory
as the dynamical simulation code itself.

The density values used by {\small SUBFIND} were based on SPH density
estimates $\rho_i$ with smoothing lengths $h_i$ that satisfy the implicit
equation $(4\pi/3) h_i^3 \rho_i = 64 \,m_{\rm p}$ \citep{Springel2002}, where
$m_{\rm p}$ is the particle mass and an effective neighbour number of 64 was
adopted. In order to test the sensitivity of inferred dark matter
annihilation rates to the dark matter density estimator
\citep[see][]{Springel2008}, we have also calculated densities based on a
three-dimensional Voronoi tessellation of the simulation volume, where the
density estimate was defined as the mass of a particle divided by its Voronoi
volume. To construct the Voronoi tessellation, we have written a parallel code
that can rapidly calculate the Delaunay triangulation, and from it its
topological dual, the Voronoi tessellation.

\begin{figure}
\resizebox{8.5cm}{!}{\includegraphics{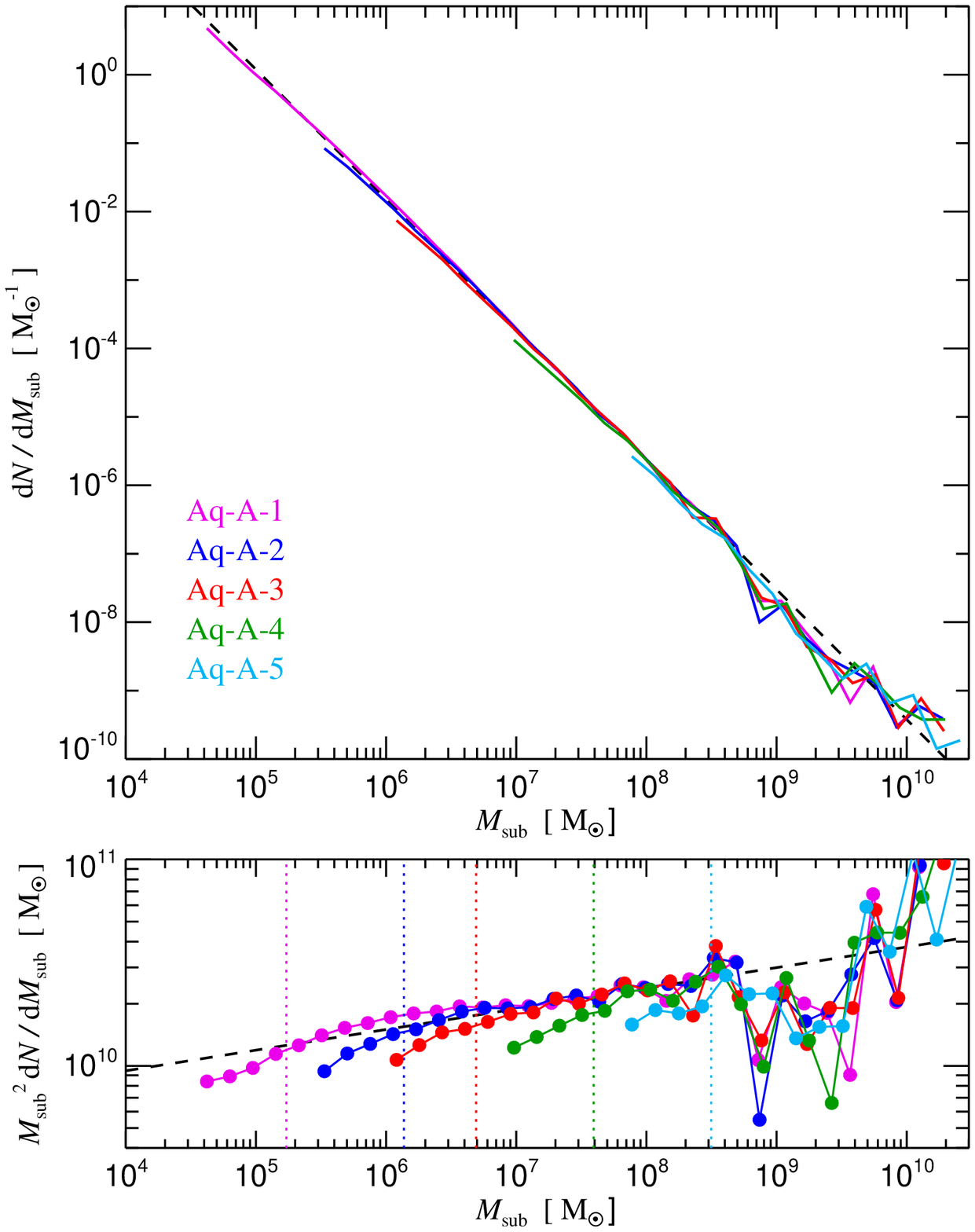}}
\caption{Differential subhalo abundance by mass in the `A' halo within the
  radius $r_{\rm 50}$. We show the count of subhalos per logarithmic mass
  interval for different resolution simulations of the same halo. The bottom
  panel shows the same data but multiplied by a factor $M_{\rm sub}^2$ to
  compress the vertical dynamic range. The dashed lines in both panels show a
  power-law ${\rm d}N/{\rm d}M\propto M^{-1.9}$. For each of the resolutions,
  the vertical dotted lines in the lower panel mark the masses of subhalos
  that contain 100 particles.  \label{FigSubMassSpectrum}}
\end{figure}

As an illustration of the extreme dynamic range of our simulation set and
the degree of numerical convergence between the different resolutions, we show
in Figure~\ref{DensProfilesC02} the spherically averaged density profiles for
the five simulations of the `Aq-A' halo. In each case, we draw
the measured density profile as a thick solid line down to the smallest radius
where convergence is expected based on the criteria of \citet{Power2003}, and
we continue the measurements as thin lines for scales where the force law is
unaffected by the gravitational softening.  \citet{Power2003} have shown that
convergence of the density profile at a given radius requires that the
two-body relaxation time at this radius be larger than the Hubble time. This
condition can be cast into the form
\begin{equation}
\frac{\sqrt{200}}{8}
\frac{N(r)}{\ln N(r)}
\left(\frac{\overline{\rho}(r)}{\rho_{\rm crit}}\right)^{-1/2} \ge 1, \label{EqnPower}
\end{equation}
where $N(r)$ is the number of particles inside $r$, and $\overline{\rho}(r)$
is the average enclosed density.  Note that this form of the convergence
criterion is in principle also applicable to dark matter subhalos (see below),
but in this regime it has not been empirically validated so far.

We find that there is very good agreement between the densities and enclosed
masses for all radii larger than the convergence radius estimated in this
way. The quality of this convergence is impressively demonstrated by
Figure~\ref{SlopesC02}, where we show the local logarithmic slope of the
density profile, for the radial range where convergence is expected according
to the Power criterion.  There are some large fluctuations of the local slope
in the outer parts of the halo, caused by substructures, which are remarkably
well reproduced at the different resolutions.  In the more relaxed inner
regions, the local logarithmic slope varies smoothly with radius. In
particular, it becomes gradually shallower towards the centre, as suggested by
\citet{Navarro2004}.  In fact, the local slope becomes clearly shallower than
$-1$ at the innermost converged radius. This has important implications for
the structure of the central cusp which will be analyzed in full detail in
Navarro et al.~(2008, in preparation). For the rest of this paper, we focus on
an analysis of the dark matter substructures.

\section{Substructure abundance and spatial distribution}  \label{SecSubAbundance}

In this section, we investigate the abundance of dark matter substructures as
measured by the {\small SUBFIND} algorithm. All our substructures consist of
particle groups that are gravitationally self-bound and are overdense with
respect to the local background. Every simulation particle can be part only of
one subhalo, but we are able to detect substructure within substructure (see
below). We count substructures down to a minimum of 20 bound particles.

\subsection{Subhalo counts and substructure mass fraction}

In Figure~\ref{FigSubMassSpectrum}, we show the differential abundance of
subhalos by mass (i.e.~the number of subhalos per unit mass interval) in our
`A' halo within $r_{\rm 50}$, and we compare results for simulations of the
same object at different mass resolution. For masses above $\sim 5\times
10^8\,{\rm M}_\odot$, the number of subhalos is small and large halo-to-halo
scatter may be expected (see below). However, for lower masses a smooth mass
spectrum is present that is well described by a power law over many orders of
magnitude.  Multiplication by $M_{\rm sub}^2$ compresses the vertical scale
drastically, so that the slope of this power-law and deviations from it can be
better studied. This is shown in the bottom panel of
Figure~\ref{FigSubMassSpectrum}. We see that resolution effects become
noticeable as a reduction in the number of objects at masses below a few
hundred particles, but for sufficiently well resolved subhalos, very good
convergence is reached. There is good evidence from the fully converged part
of the differential mass function that it exhibits a true power-law behaviour,
and that the slope of this power law is shallower than $-2$, though not by
much.  Our results are best fit by a power law ${\rm d}N/{\rm d}M \propto
M^{-1.9}$, the same slope found by \citet{Gao2004}, but significantly steeper
than \citet{Helmi2002} found for their rich cluster halo. The exact value
obtained for the slope in a formal fit varies slightly between $-1.87$ and
$-1.93$, depending on the mass range selected for the fit; the steepest value
of $-1.93$ is obtained when the fit is restricted to the mass range
$10^6\,{\rm M}_\odot$ to $10^7\,{\rm M}_\odot$ for the Aq-A-1 simulation.

The small tilt of the slope $n=-1.9$ away from $-2$ is quite important. For
$n=-2$, the total predicted mass in substructures smaller than a given limit
$m_0$ would be logarithmically divergent when extrapolated to arbitrarily
small masses. If realized, this might suggest that there is no smooth halo at
all, and that ultimately all the mass is contained in subhalos. However, even
for the logarithmically divergent case the total mass in substructures does
not become large enough for this to happen, because a sharp cut-off in the
subhalo mass spectrum is expected at the thermal free-streaming limit of the
dark matter. Depending on the specific particle physics model, this cut-off
lies around an Earth mass, at $\sim 10^{-6}\,{\rm M}_\odot$, but could be as
low as $10^{-12}\,{\rm M}_\odot$ in certain scenarios
\citep{Hofmann2001,Green2004}.

\begin{figure*}
\resizebox{15.0cm}{!}{\includegraphics{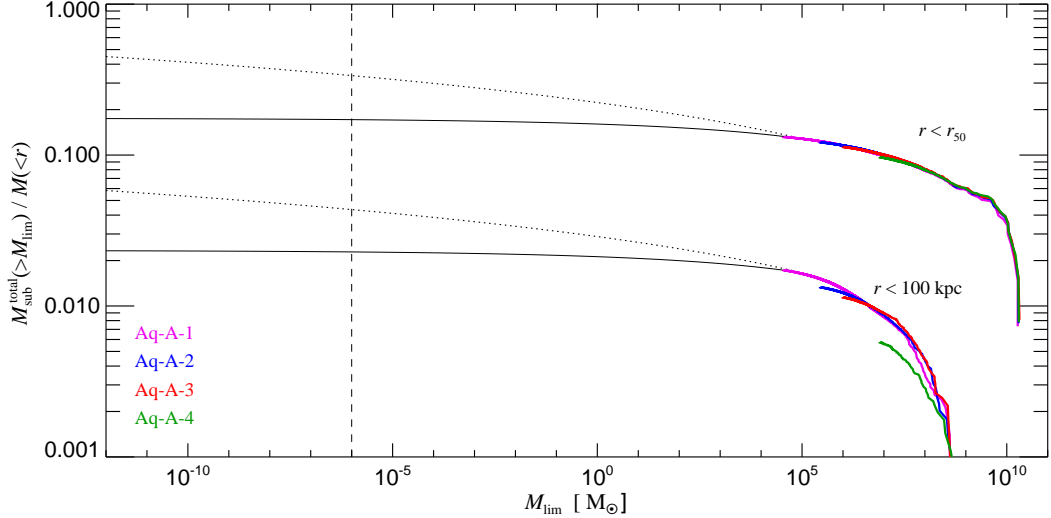}}
\caption{Expected mass fraction in subhalos as a function of the limiting mass
  $M_{\rm lim}$, inside $r_{50}$ (top curves) and inside $100\,{\rm kpc}$
  (bottom curves). The solid thin lines show an extrapolation of the direct
  simulation result with an $n=-1.9$ power-law for the differential subhalo
  mass function. In this case, the total substructure mass converges at the
  low-mass end. The dotted lines show the prediction for the logarithmically
  divergent case, $n=-2$. In this case, we would expect the mass in
  substructures down to an Earth mass (vertical dashed line) to be about twice
  what we can resolve directly. This mass is a reasonable estimate for the
  thermal free-streaming limit in many supersymmetric theories where the dark
  matter particle is a neutralino. However, the parameters of these theories
  are sufficiently uncertain that the thermal limit could lie as low as
  $10^{-12}\,{\rm M}_\odot$. Even in this case, the lumpy component of the
  halo would still be subdominant within $r_{\rm 50}$, and would be a small
  fraction of the total mass of $9.32\times 10^{11}\,{\rm M}_\odot$ within
  $100\,{\rm kpc}$.
\label{FigExtrapolatedMassFraction}
}
\end{figure*}

Our measured mass function for the `A' halo is well approximated by
\begin{equation}
\frac{{\rm d}N}{{\rm d}M} = a_0 \left(\frac{M}{m_0}\right)^n, \label{eqnfitA}
\end{equation}
with $n=-1.9$, and an amplitude of $a_0 = 8.21\times 10^7/M_{50}=3.26\times
10^{-5}\,{\rm M}_\odot^{-1}$ for a pivot point of $m_0 = 10^{-5}\,M_{50} =
2.52\times 10^{7}\, M_\odot$.  This means that the expected total mass in all
subhalos less massive than our resolution limit $m_{\rm res}$ is
\begin{equation}
M_{\rm tot}(<m_{\rm res}) = \int_{m_{\rm lim}}^{m_{\rm res}} M \frac{{\rm
    d}N}{{\rm d}M} {\rm d}M  = \frac{a_0}{n+2}\,\frac{m_{\rm res}^{n+2}- m_{\rm lim}^{n+2}}{m_0^n},
\end{equation}
where $m_{\rm lim}$ is the thermal dark matter limit. For $m_{\rm
lim}\to 0$ and our nominal subhalo resolution limit of $m_{\rm
res}=3.24\times 10^{4}\,{\rm M}_\odot$ in the Aq-A-1 simulation, this
gives $M_{\rm tot} = 1.1\times 10^{11}\,{\rm M}_\odot$, corresponding
to about 4.5\% of the mass of the halo within $r_{\rm 50}$.  While
non-negligible, this is considerably smaller than the total mass in
the substructures that are {\em already resolved} by the
simulation. The latter is 13.2\% of the mass within $r_{\rm 50}$ for
the Aq-A-1 simulation. We hence conclude that despite the very broad
mass spectrum assumed in this extrapolation, the total mass in
subhalos is still dominated by the most massive substructures, and an
upper limit for the total mass fraction in subhalos is $\sim 18\%$
within $r_{\rm 50}$ for the `Aq-A' halo.

We caution, however, that the extrapolation to the thermal limit extends over
10 orders of magnitude! This is illustrated explicitly in
Figure~\ref{FigExtrapolatedMassFraction}, where we show the mass fraction in
substructures above a given mass limit, combining the direct simulation
results with the extrapolation above. We also include an alternative
extrapolation in which a steeper slope of $-2$ is assumed. In this case, the
total mass fraction in substructures would approximately double if the thermal
limit lies around one Earth mass. If it is much smaller, say at $m_{\rm
  lim}\sim 10^{-12}\,{\rm M}_\odot$, the mass fraction in substructure could
grow to $\sim 50\%$ within $r_{\rm 50}$, still leaving room for a substantial
smooth halo component. Notice, however, that within $100\,{\rm kpc}$ even this
extreme extrapolation results in a substructure mass fraction of only about
5\%.  Most of the mass of the inner halo is smoothly distributed.

Within $r_{\rm 50}$ the mass fraction in resolved substructures varies around
11\% for our 6 simulations at resolution level 2, each of which has at least
160 million particles in this region.  Table~\ref{TabProperties} lists these
numbers, which are 12.2\% (Aq-A-2 simulation), 10.5\% (Aq-B-2), 7.2\%
(Aq-C-2), 13.1\% (Aq-D-2), 10.8\% (Aq-E-2), and 13.4\% (Aq-F-2). This gives an
average of 11.2\% within $r_{\rm 50}$ down to the relevant subhalo mass
resolution limit, $\sim 2\times 10^5\,{\rm M}_\odot$. This is similar to the
substructure mass fractions found by earlier work on galaxy cluster halos
\citep[e.g.][]{Ghigna1998,Springel2001b,DeLucia2004} and Galaxy-sized halos
\citep{Stoehr2003} once the different limiting radius ($r_{200}$ instead of
$r_{50}$) is corrected for. However, it is larger than the 5.3\% inside
$r_{50}$ reported by \citet{Diemand2007} for a Milky Way-sized halo.

\begin{figure}
\resizebox{8.5cm}{!}{\includegraphics{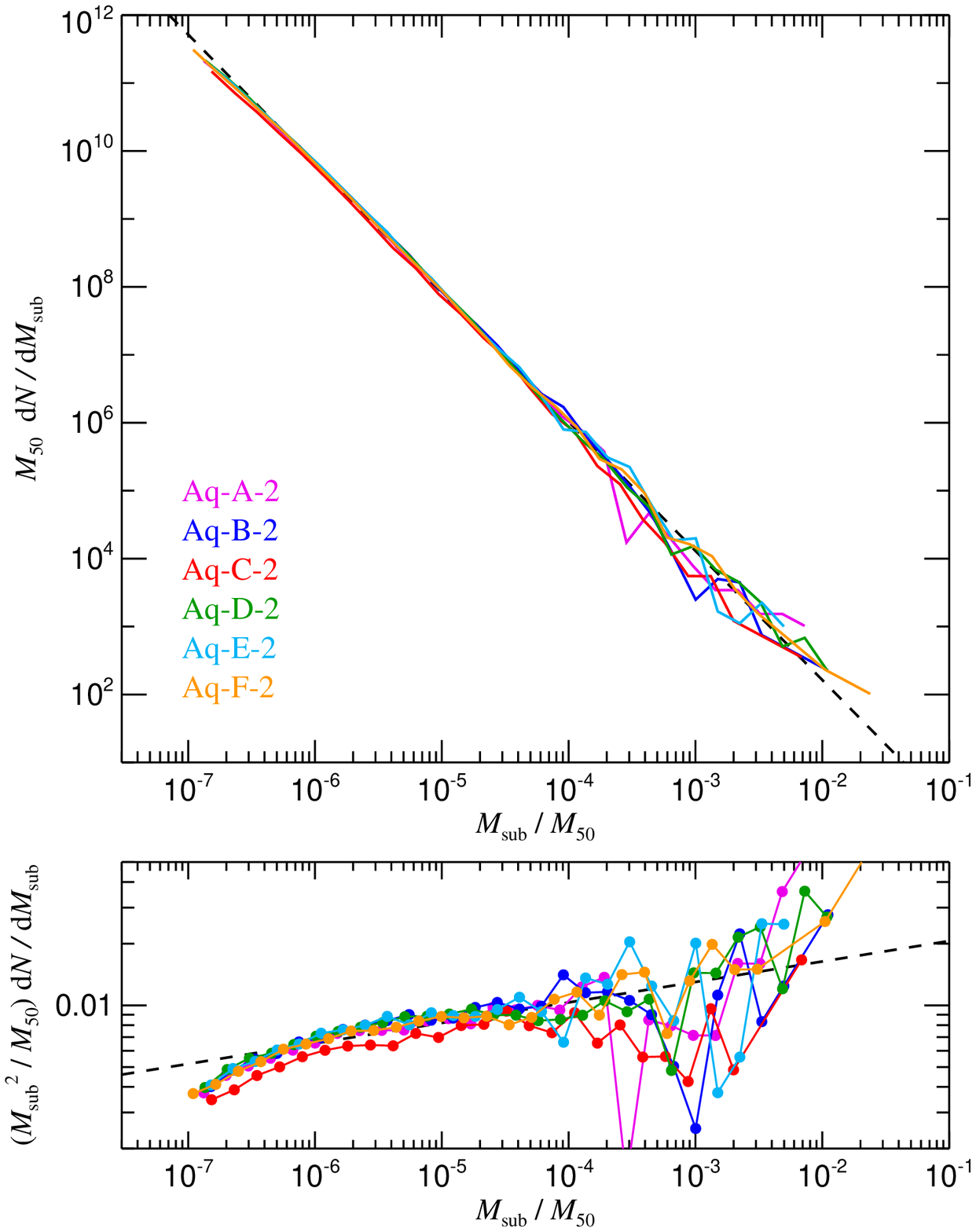}}
\caption{Differential subhalo abundance by mass for our six halos simulated at
  resolution level 2. We here count all subhalos inside $r_{50}$ and plot
  their abundance as a function of $M_{\rm sub}/M_{50}$. We see that at low
  mass the abundance of subhalos is universal to good accuracy when their mass
  is expressed in units of the mass of their parent
  halo. The bottom panel shows the same data but multiplied by a factor
$M^2_{\rm sub}$ to compress the vertical dynamic range. The dashed lines in both
panels show the power-law ${\rm d}N/{\rm d}M\propto M^{-1.9}$. 
\label{FigSubMassSpectrumAllHalos}}
\end{figure}

In Figure~\ref{FigSubMassSpectrumAllHalos}, we compare the differential
subhalo mass functions of these six halos, counting the numbers of subhalos
as a function of their mass normalized to the $M_{50}$ of their parent
halo. Interestingly, this shows that at small subhalo masses the subhalo
abundance per unit halo mass shows very little halo-to-halo scatter. In fact,
the mean differential abundance is well fit by equation (\ref{eqnfitA}) with
the parameters given above, and the {\it rms} halo-to-halo scatter in the
normalization is only $\sim 8\%$.

\begin{figure}
\resizebox{8.5cm}{!}{\includegraphics{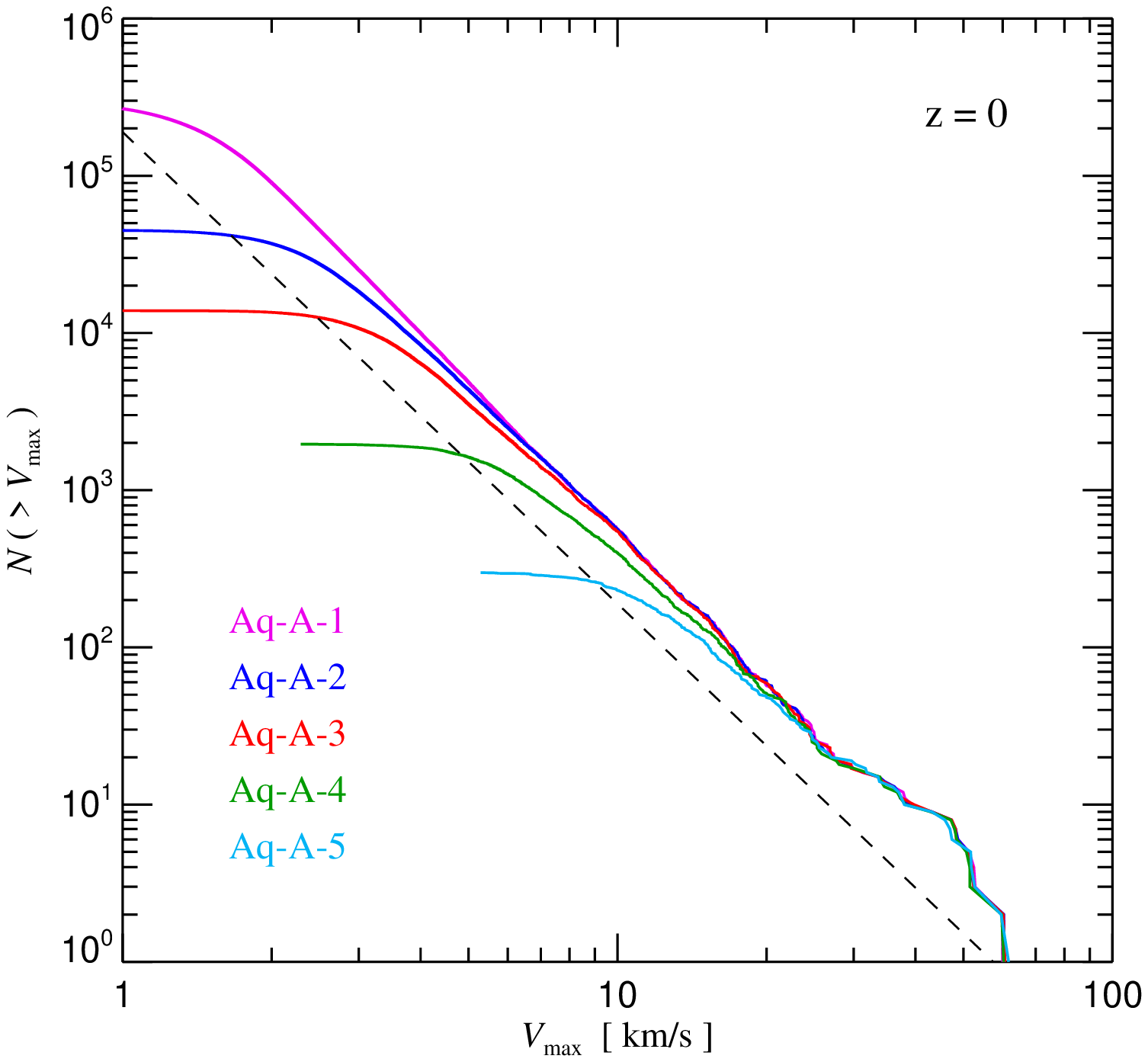}}
\resizebox{8.5cm}{!}{\includegraphics{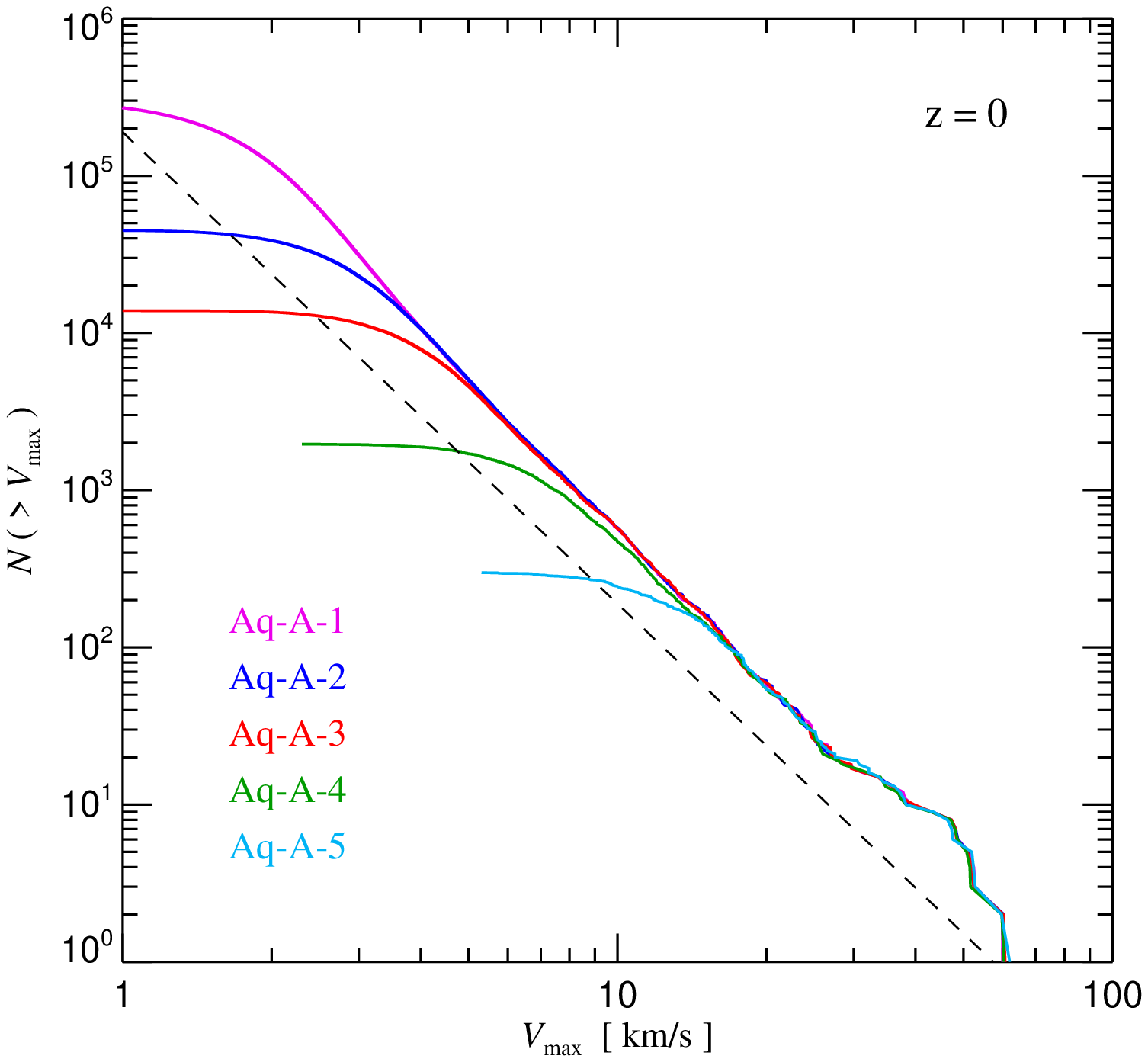}}
\caption{Cumulative subhalo abundance as a function of maximum subhalo
  circular velocity. The top panel shows the raw measurements from the
  simulations, while in the bottom panel, we have applied the correction of
  equation (\ref{EqVmaxCorrection}) to compensate approximately
  for the impact of
  the gravitational softening on $V_{\rm max}$. We show results for 5
  simulations of the Aq-A halo carried out with differing mass resolution. The
  dashed line is the fitting function given for their own simulations by
  \citet{Reed2005}, which also accurately matches the result for the `Via
  Lactea I' simulation \citep{Diemand2007}. This is clearly inconsistent with
  our own data.
\label{FigSubVelFuncC02}
}
\end{figure}

\begin{figure}
\resizebox{8.5cm}{!}{\includegraphics{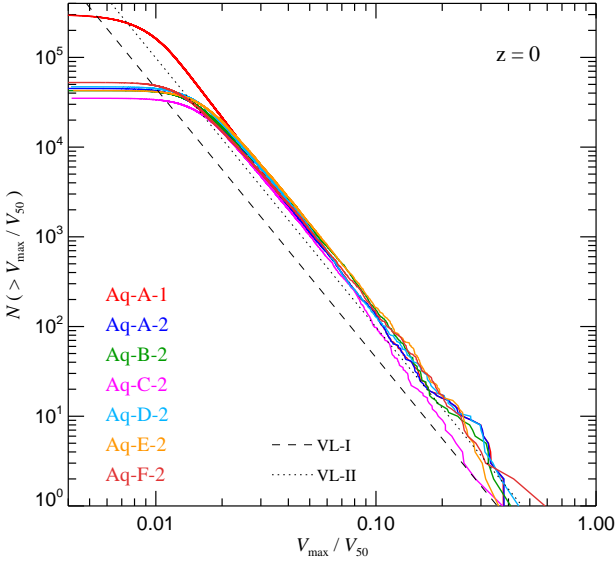}}
\caption{Cumulative subhalo abundance as a function of maximum subhalo
  circular velocity in units of the circular velocity of the main halo at
  $r_{50}$. We show results for all 6 of our halos at resolution level 2, and
  in addition we include our highest resolution result for the Aq-A-1 run. For
  comparison, we overplot fitting functions for the Via Lactea I and Via
  Lactea II simulations \citep{Diemand2007,Diemand2008}, appropriately
  rescaled from a normalization to $V_{\rm max,host}$ to one by $V_{\rm
    50,host}$.
\label{FigSubVelFuncAllSims}
}
\end{figure}

\begin{figure}
\resizebox{8.5cm}{!}{\includegraphics{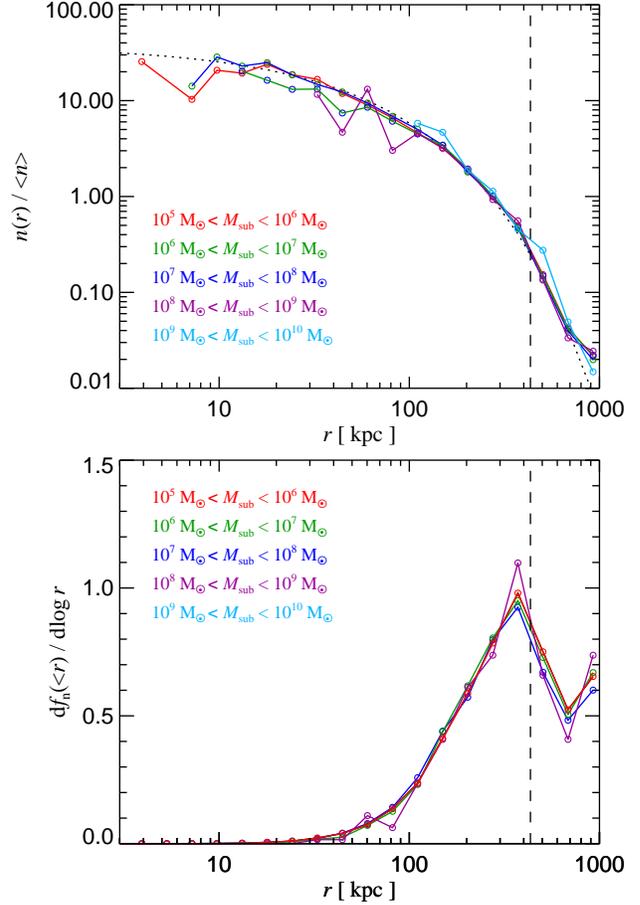}}
\caption{Subhalo number density profiles for different subhalo mass ranges in
  the Aq-A-1 simulation. In the top panel, the number density profiles for 5
  logarithmic mass bins are shown, normalized to the mean number density
  within $r_{50}$ (vertical dashed line). The profile shape appears
  independent of subhalo mass, and is well fit by an Einasto profile with
  $\alpha=0.678$ and $r_{-2}=199\,{\rm kpc}=0.81\,r_{200}$. The bottom panel shows the number fraction of
  subhalos per logarithmic interval in radius, on a linear-log plot. The area
  under the curves is proportional to subhalo number, showing that most
  subhalos are found in the outermost parts of the halo.
  \label{FigNumberDensProfile}}
\end{figure}

\begin{figure}
\resizebox{8.5cm}{!}{\includegraphics{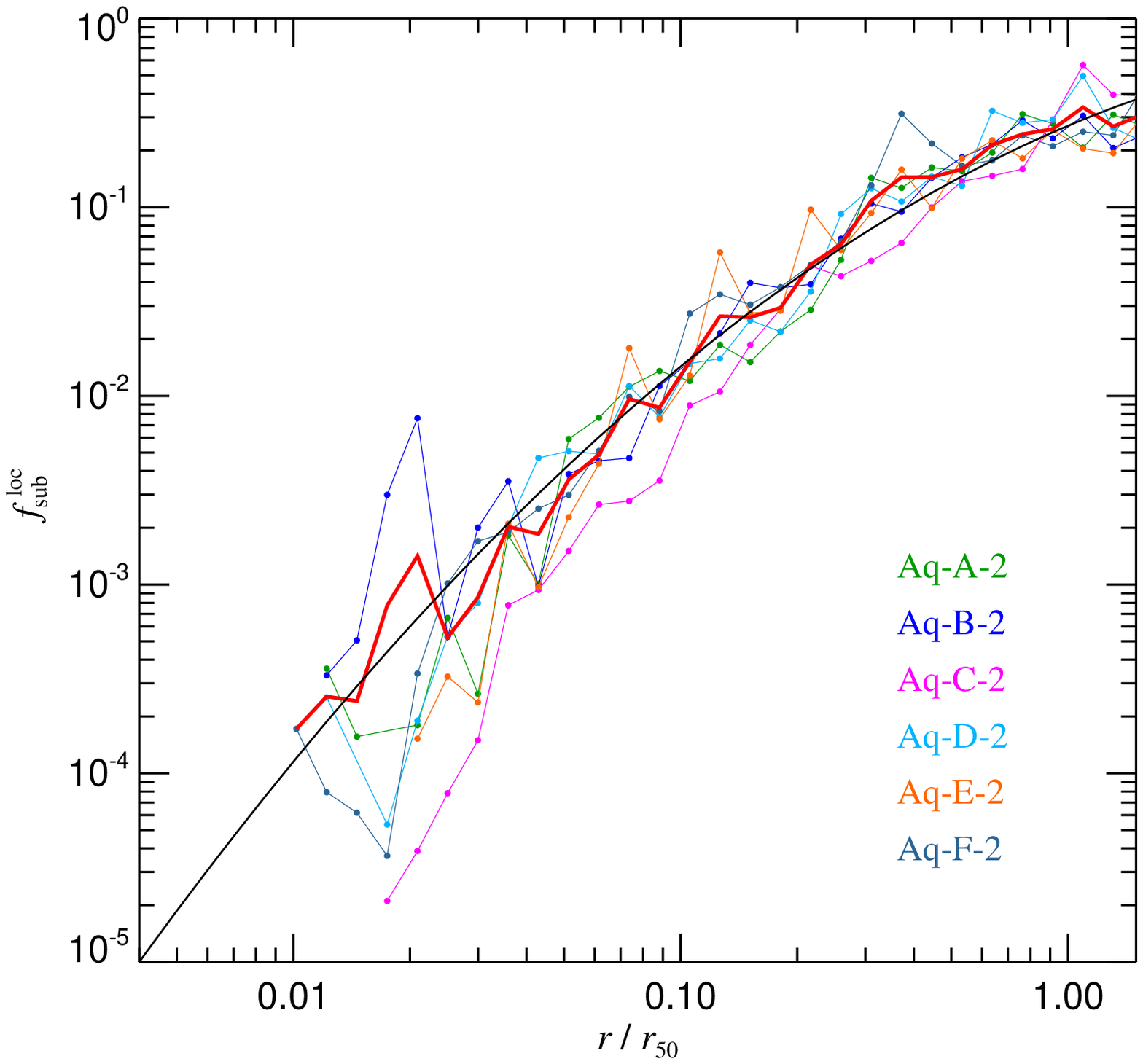}}\\
\resizebox{8.5cm}{!}{\includegraphics{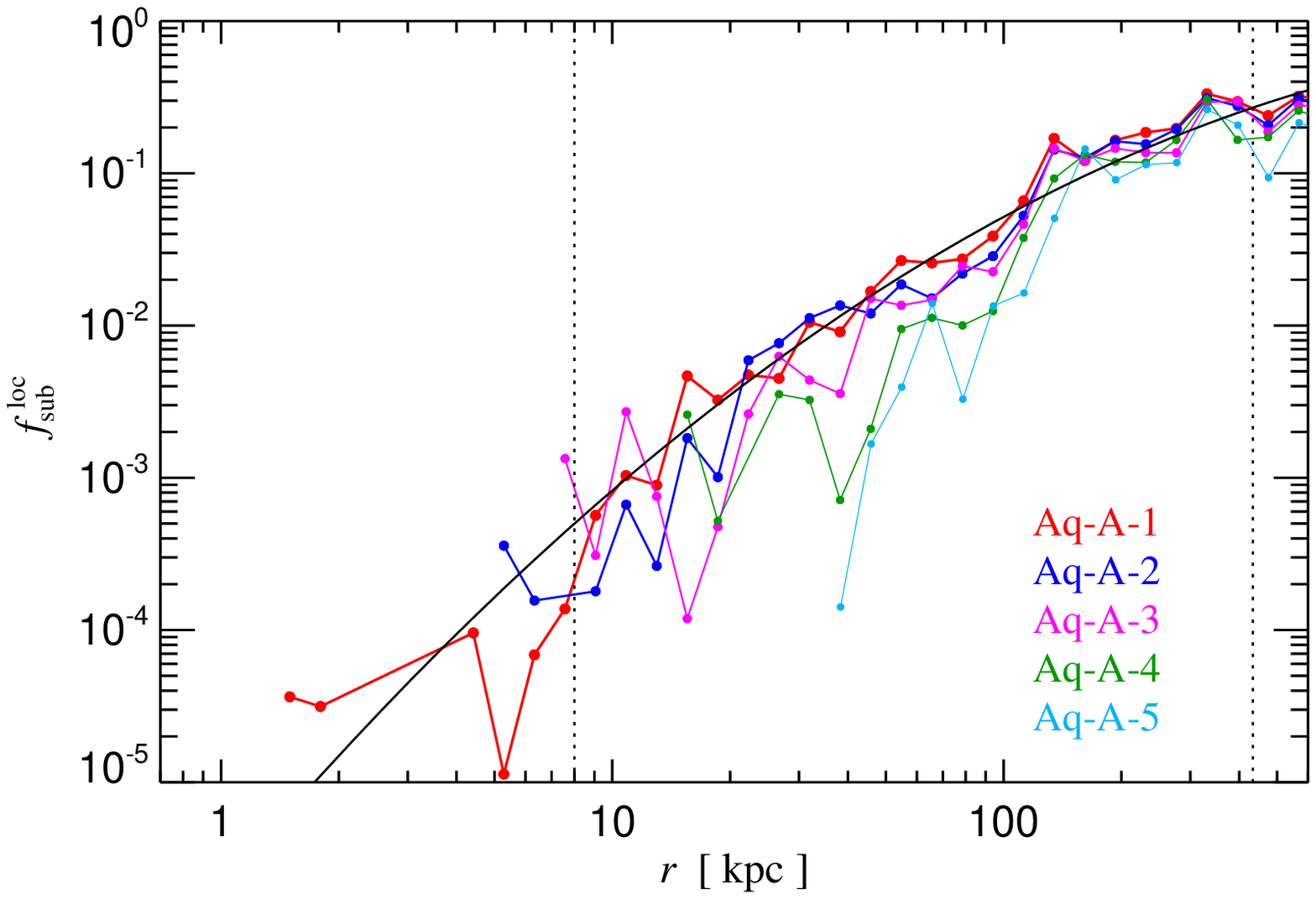}}\\
\resizebox{8.5cm}{!}{\includegraphics{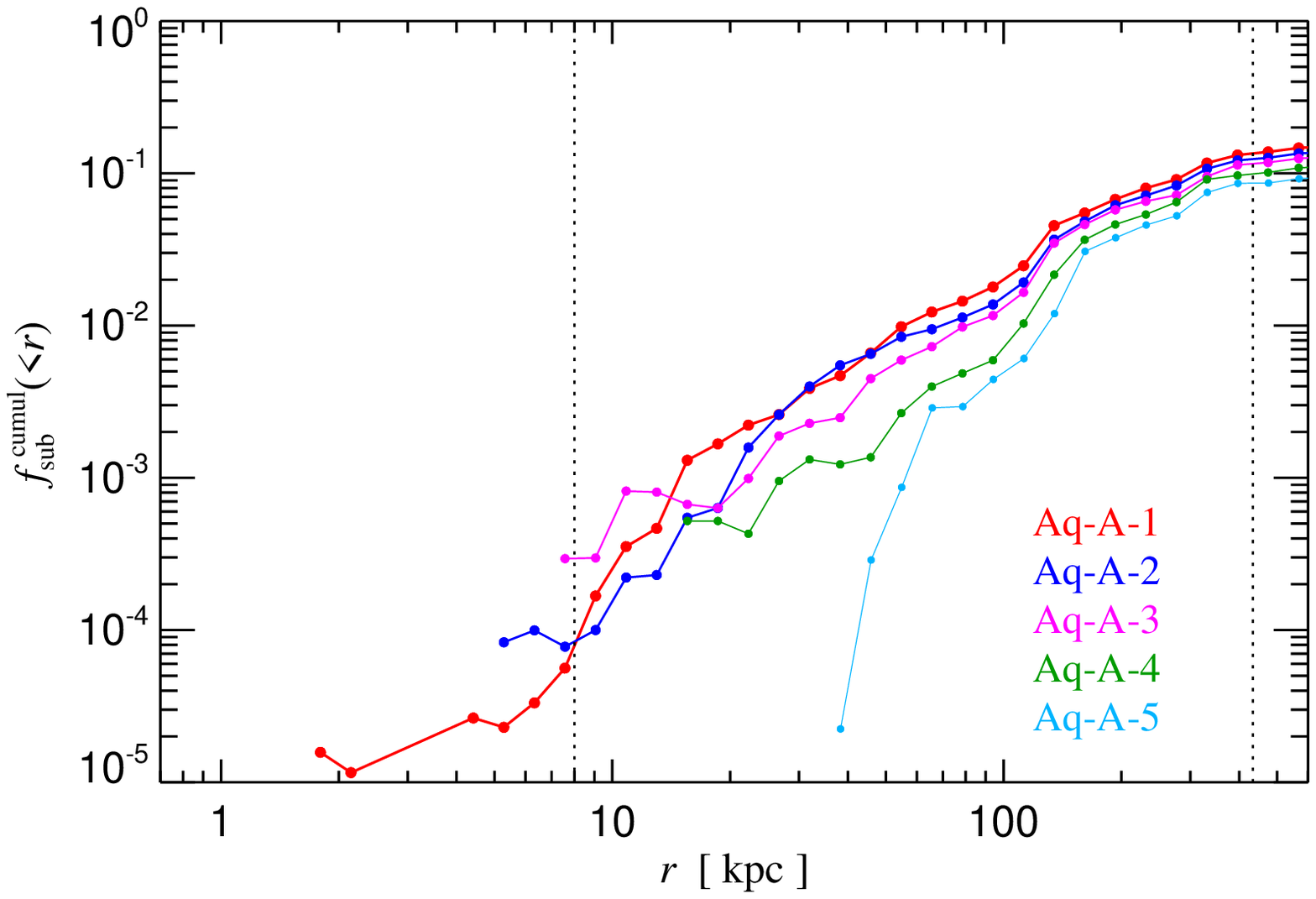}}
\caption{The mass fraction in subhalos as a function of radius.  In the top
  panel, we show results for the {\em local} mass fraction in substructures
  for our six different halos, as a function of radius normalized by $r_{50}$.
  The thick solid line shows the average of all the runs. In the middle panel,
  we consider the same quantity for the different resolution simulations of
  the Aq-A halo, while in the bottom panel we show the corresponding {\em
    cumulative} substructure fractions in the Aq-A halo. The solid line in the
  two upper panels is an empirical fit with a slowly running power law index.
  The vertical dotted lines at $8\,{\rm kpc}$ in the middle and bottom panels
  mark the position of the Solar circle; here the expected local mass
  fraction in subhalos has dropped well below $10^{-3}$. The outer vertical
  dotted lines mark $r_{50}$ for the Aq-A halo.
\label{FigRadSubMassFraction}
}
\end{figure}

In Table~\ref{TabProperties}, we also list a few other basic structural
properties of our halos, namely their maximum circular velocity $V_{\rm max}$,
the radius $r_{\rm max}$ at which this velocity is attained, a simple measure
for halo concentration, and the redshift at which the halo formed.  One
way to characterize the concentration of a halo is to express the mean
overdensity within $r_{\rm max}$ in units of the critical density.
This corresponds to the definition
\begin{equation}
\delta_V = \frac{\overline{\rho}(r_{\rm max})}{\rho_{\rm crit}} = 2
\left(\frac{V_{\rm max}}{H_0\,r_{\rm max}}\right)^2. \label{EqnConc}
\end{equation}
We can relate this quantity to a more familiar concentration measure
based on the NFW density profile:
\begin{equation}
\rho = \frac{\rho_s}{(r/r_s)(1 + r/r_s)^2},
\end{equation}
where $\rho_s$ and $r_s$ are a characteristic density and radius
respectively. Assuming this profile shape, the pair of values $V_{\rm max}$ and
$r_{\rm max}$ are sufficient to determine the density profile uniquely.  The
characteristic NFW overdensity $\delta_c$ is then
\begin{equation}
\delta_c = \frac{\rho_s}{\rho_{\rm crit}} =  7.213\, \delta_V ,
\end{equation}
which can be converted to the NFW concentration $c$ through the equation
\begin{equation}
\delta_c = \frac{200}{3} \frac{c^3}{\ln(1+c) - c/(1+c)}.
\end{equation}
We list the ``NFW'' concentrations calculated in this way as $c^*_{\rm NFW}$
in Table~\ref{TabProperties}. We note however that fits to the full density
profile may yield slightly different results, since in this case the circular
velocity curve of the fit will not necessarily peak exactly at $r=r_{\rm
  max}$.

Defining a precise value for the total mass of a subhalo requires an
operational definition of its `outer edge'. Different substructure detection
algorithms define different effective boundaries and so produce systematically
different total mass estimates. It may therefore be more robust to count
subhalos as a function of their peak circular velocity, which typically lies
well within the object and so is insensitive to definitions of its edge. Note,
however, that in small systems maximum circular velocity estimates can be more
sensitive to numerical resolution effects than total mass estimates. 

In Figure~\ref{FigSubVelFuncC02}, we show the cumulative abundance of
subhalos as a function of maximum circular velocity for our different
resolution simulations of Aq-A. Again, there is good
convergence. Indeed, at the massive end, the curves lie essentially on
top of each other, showing that we are really seeing the {\em same}
subhalos, and that they are reproduced with the {\em same} maximum
circular velocity in all the simulations. This suggests that we are
also achieving good convergence for the internal structure of
individual subhalos, an issue that we will investigate further below.

However, it is noticeable that the individual measurements for the velocity
functions peel away from their higher resolution counterparts comparatively
early at low velocities, which suggests  worse convergence than found for the
subhalo mass functions at the low mass end. This behaviour can be understood
as an effect of the gravitational softening length $\epsilon$, which lowers
the maximum circular velocities of subhalos for which $r_{\rm max}$ is not
much larger than $\epsilon$. To estimate the strength of this effect, we can
imagine that the gravitational softening for an existing subhalo is
adiabatically lowered from $\epsilon$ to zero. The angular momentum of
individual particle orbits is then an adiabatic invariant. Assuming for
simplicity that all particles are on circular orbits, and that the
gravitational softening can be approximated as a Plummer force with softening
length $\epsilon$, the expected change of the maximum circular velocity is
then
\begin{equation}
V_{\rm max}' = V_{\rm max} \, [ 1 + (\epsilon / r_{\rm max})^2]^{1/2} .
\label{EqVmaxCorrection}
\end{equation}
In the lower panel of Figure~\ref{FigSubVelFuncC02}, we plot the cumulative
velocity functions for these corrected maximum circular velocities. Clearly,
the measurements line up more tightly down to lower $V_{\rm max}$,
demonstrating explicitely that the convergence in the number of objects
counted as a function of (corrected) circular velocity is in principle as good
as that counted as a function of mass. Note that a similar correction can also be
applied to the measured $r_{\max}$ values. However, for the remainder of this
paper, we focus on the raw measurements from the simulations without applying
a gravitational softening correction.

The dashed line in Figure~\ref{FigSubVelFuncC02} shows the fit which
\citet{Reed2005} quote for the subhalo abundance as a function of maximum
circular velocity in their own simulations, $N(>V_{\rm max}) = (1/48)(V_{\rm
  max,sub}/V_{\rm max,host})^{-3}$.  \citet{Diemand2007} found this formula to
fit the results from their own Via Lactea I simulation very
well. Figure~\ref{FigSubVelFuncC02} thus confirms the indication from subhalo
mass fractions that our simulations show substantially more substructure than
reported for Via Lactea I. This is particularly evident at lower subhalo
masses which are unaffected by the small number effects which cause scatter in
the abundance of massive subhalos. With the help of J.  Diemand and his
collaborators, we have checked that this abundance difference is not a result
of the different subhalo detection algorithms used in our two projects.

We do not think that this discrepancy can be explained by halo-to-halo
scatter since it is much larger than the variation in substructure
abundance among our own sample of halos. This is demonstrated in
Figure~\ref{FigSubVelFuncAllSims}, which shows the cumulative subhalo
abundance distributions within $r_{\rm 50}$ as a function of maximum
subhalo circular velocity for all our resolution level 2 halos. We
plot subhalo count against subhalo maximum circular velocity
normalized to $V_{50}$, the circular velocity of the main halo at
$r_{\rm 50}$. Because the slope of the abundance curve is very close
to $-3$, this is equivalent to plotting subhalo count normalized by
the total parent halo mass within $r_{\rm 50}$ (which is proportional
to $V_{50}^3$) against subhalo maximum circular velocity.  There is
remarkably little scatter between our simulations when normalized in
this way; the {\it rms} scatter in amplitude in the power-law regime
is around 10\%.  The figure also shows the substructure abundance
reported for the Via Lactea I (dashed) and II (dotted) simulations
\citep{Diemand2007,Diemand2008}, after rescaling to the normalization
we prefer here.\footnote{Note that $V_{\rm 50,host}$ unambiguously
  characterizes the enclosed mass within $r_{50}$, the region in which
  subhalos are counted. This is not the case for $V_{\rm max,host}$,
  the velocity scale chosen by \citet{Diemand2008}, because it is
  additionally affected by halo concentration.}. There is a difference
of a factor of 3.1 between the mean abundance of small subhalos in our
simulations and in `Via Lactea I'. The \citet{Diemand2008} abundance
for Via Lactea II differs substantially from that for Via Lactea I and
is much closer to our results. Nevertheless, the abundance of small
subhalos in Via Lactea II is still 31\% lower than the mean for our
set of 6 halos, which is more than three times the {\it rms} scatter
in abundance between our halos.

These results lead us to disagree with the assertion by \citet{Madau2008} and
\citet{Diemand2008} that differences of this magnitude lie within the
halo-to-halo scatter.  Instead, the substantial difference between `Via Lactea
I' and Via Lactea II' must have a systematic origin.  We also think it
unlikely that the higher abundance in our simulations reflects the small
differences in the background cosmology assumed in the two projects, as
suggested by \citet{Madau2008}, even though this is a possibility we
  cannot exclude. For example, the Via Lactea simulations assumed a lower
value for $\sigma_8$ than we used, and we believe that lowering $\sigma_8$
should result in slightly {\it more} substructure in objects {\em of given
  mass}, simply because these halos then tend to form more recently which
increases the number of surviving subhalos within them
\citep[e.g.][]{DeLucia2004}. We have explicitly confirmed this effect by
comparing the substructure abundances in the Millennium simulation (with
$\sigma_8=0.9$) with those in the simulations of \citet{Wang2008}, which used
the same cosmology except for taking $\sigma_8=0.722$. On the other hand,
  the different tilt assumed for the primoridial power spectrum of the Via
  Lactea II simulation may have reduced the subhalo abundance and could
  perhaps be responsible for the difference \citep{Zentner2003}.

We note that the small halo-to-halo scatter in substructure abundance which we
find also contradicts the recent suggestion by \citet{Ishiyama2007} that the
halo-to-halo variation in subhalo abundance could be very large, and that the
apparent paucity of dwarfs surrounding the Milky Way might simply reflect the
fact that our Galaxy happens to live in a low density environment.

\begin{figure*}
\resizebox{15cm}{!}{\includegraphics{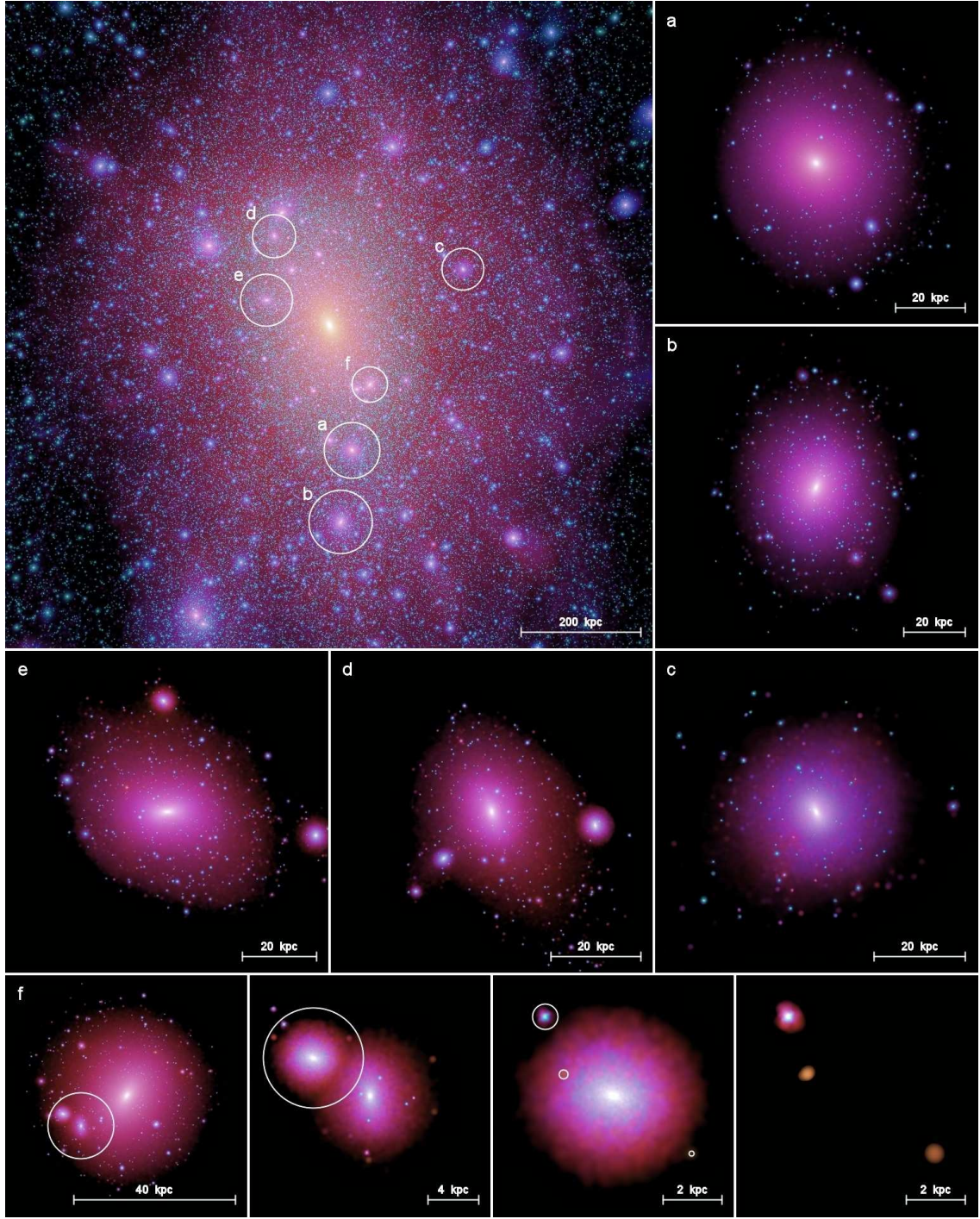}}
\caption{Images of substructure within substructure. The top left panel shows
  the dark matter distribution in a cubic region of side $2.5\times r_{\rm
    50}$ centred on the main halo in the Aq-A-1 simulation. The circles mark
  six subhalos that are shown enlarged in the surrounding panels, and in the
  bottom left panel, as indicated by the labels.  All these first generation
  subhalos contain other, smaller subhalos which are clearly visible in the
  images. {\small SUBFIND} finds these second generation subhalos and
  identifies them as daughter subhalos of the larger subhalos. If these
  (sub-)subhalos are large enough, they may contain a third generation of
  (sub-)subhalos, and sometimes even a fourth generation. The bottom panels
  show an example of such a situation. The subhalo shown on the bottom left
  contains another subhalo (circled) which is really made up of two main
  components and several smaller ones (bottom, second from left). The smaller
  of the two components is a third generation substructure (bottom, third from
  left) which itself contains three subhalos which are thus fourth generation
  objects (bottom right).\label{FigSubSubExamples}}
\end{figure*}

\subsection{The spatial distribution of subhalos}

In Figure~\ref{FigNumberDensProfile}, we show the radial distribution of
subhalos of different mass within our Aq-A-1 simulation. In the top panel, we
plot the number density profile for different subhalo mass ranges, each
normalized to the mean number density of subhalos of this mass within
$r_{50}$.  The number density of subhalos increases towards halo centre, but
much more slowly than the dark matter density, consistent with previous work
\citep[e.g.][]{Ghigna1998,Gao2004,Nagai2005,Diemand2004b,Diemand2007}. As a
result, most subhalos of a given mass are found in the outer parts of a halo,
even though the number density of subhalos is highest in the central
regions. Another view of this behaviour is given in the bottom panel of
Figure~\ref{FigNumberDensProfile}, which histograms the abundance of subhalos
as a function of log radius so that the area under the curves is proportional
to the total number of subhalos. Clearly, the vast majority of subhalos are
found between $\sim 100\,{\rm kpc}$ and the outer radius of the halo.

Perhaps the most remarkable aspect of Figure~\ref{FigNumberDensProfile} is
that there appears to be no trend in the shape of the number density profiles
with subhalo mass. Previous work has already hinted at this behaviour
\citep{Diemand2004b,Ludlow2008}, which is here confirmed with much better
statistics and over a much larger dynamic range in mass. We note that this
disagrees with a tentative finding by \citet{DeLucia2004}, who suggested that
more massive substructures have a radial profile that is more strongly
antibiased with respect to the mass than that of low mass subhalos.  Like
\citet{Ludlow2008} we find that the number density profile is well described
by an Einasto profile (a fit to our measurements yields a shape parameter
$\alpha=0.678$ and scale radius $r_{-2}=199\,{\rm kpc} = 0.81\,r_{200}$).  It
is thus tempting to conjecture that this behaviour continues to (arbitrarily)
small subhalo masses. If true, an interesting corollary is that there must be
a smooth dark matter component which dominates the inner regions of
halos. Only the outer parts may have a substantial mass fraction in lumps (see
also Figure~\ref{FigExtrapolatedMassFraction}). This contrasts with previous
speculations \citep{Calcaneo2000,Moore2001} that all the mass of a halo may be
bound in subhalos.

Further light on this question is shed by Figure~\ref{FigRadSubMassFraction},
where we show the local mass fraction in subhalos as a function of radius. In
the top panel, we compare results for our six different halos, with the
radial coordinate normalized by $r_{50}$. While there is some scatter between
the different halos, the general behaviour is rather similar and shows a rapid
decline of the local mass fraction in substructures towards the inner parts of
each halo. The mean of the six simulations (thick red line) is well fit by a
gently curving power-law. It can be parameterized by
\begin{equation}
f_{\rm sub}= \exp\left[\gamma + \beta \ln(r/r_{50}) + 0.5\,\alpha  \ln^2(r/ r_{50})\right],
\end{equation}
with parameters $\alpha=-0.36$, $\beta=0.87$, and $\gamma=-1.31$. This fit is
shown in the upper two panels of Figure~\ref{FigRadSubMassFraction} as a thin
black line.  The middle panel is the same measurement, but for all the
different resolution simulations of the Aq-A halo, while the bottom panel is
the corresponding cumulative plot.  These two panels give an impression of how
well numerical convergence is achieved for this quantity.

An interesting implication from Figure~\ref{FigRadSubMassFraction} is an
estimate of the fraction of the mass in substructures near the Solar Circle
(marked by a vertical dashed line).  At $r=8\,{\rm kpc}$, the expected local
mass fraction in substructure has dropped well below $10^{-3}$. This
measurement appears converged, and accounting for unresolved substructure does
not raise the fraction above $10^{-3}$ (compare
Figure~\ref{FigExtrapolatedMassFraction}). The dark matter distribution
through which the Earth moves should therefore be mostly smooth, with only a
very small contribution from gravitationally bound subhalos.

\section{Subhalos inside subhalos}  \label{SecSubsInSubs}

In our simulations, we find several levels of substructure within
substructure. Figure~\ref{FigSubSubExamples} illustrates this by
showing individually 6 of the largest Aq-A-1 subhalos in enlarged
frames. Clearly, all of these subhalos have embedded
substructures. Sometimes these second-generation subhalos contain a
further (third) level of substructure and, in a few cases, we even
find a fourth generation of subhalos embedded within these. An example is
given in the bottom row of Figure~\ref{FigSubSubExamples}, which zooms
recursively on regions of the subhalo labeled ``f'' in the top-left
panel.  As shown in the bottom-left panel, subhalo ``f'' has several
components, each of which has identifiable sub-components; we are able
to identify up to four levels of this hierarchy of substructure in
this system. We note that the hierarchy of nested structures is
established directly by the recursive nature of the {\small SUBFIND}
algorithm; at each level, a given substructure and its parent
structure are surrounded by a common outer density contour that
separates them from the next level in the hierarchy.

It is important to quantify in detail the hierarchical nature of substructure,
since this may have a number of consequences regarding indirect and direct
dark matter search strategies.  Recently, \citet{Shaw2007} suggested that the
\mbox{(sub-)substructure} distribution in subhalos might be a scaled version
of the substructure distribution in main halos. This claim has been echoed by
\citet{Diemand2008}, who report roughly equal numbers of substructures inside
radii enclosing a mean overdensity of 1000 times the cosmic average value
($r_{250}$ in our notation) and centred at either subhalos or the main
halo. This result has been interpreted by \citet{Kuhlen2008} to imply that the
(sub-)subhalo abundance per unit mass of a subhalo should be roughly constant
and equal to that of the main halo. This, however, seems unlikely because, as
we have seen, local substructure abundance is a strong function of radius in
main halos, with most of the substructure found in the outer regions.

In this section, we present the first convergence studies ever
attempted for (sub-)substructure inside subhalos in order to assess the
alleged self-similarity of the substructure hierarchy. We begin by
discussing a suitable definition for the outer edge of a subhalo,
which allows us to measure the (sub-)substructure mass fractions of
subhalos in a consistent manner. We then study the number and mass of
sub-subhalos within that radius and compare them with the expectation
from self-similarity. In order to compare with recent work by
\citet{Diemand2008}, we also carry out, for a few subhalos, the same
analysis within a radius of fixed overdensity, $r_{250}$.

\begin{figure}
\resizebox{8cm}{!}{\includegraphics{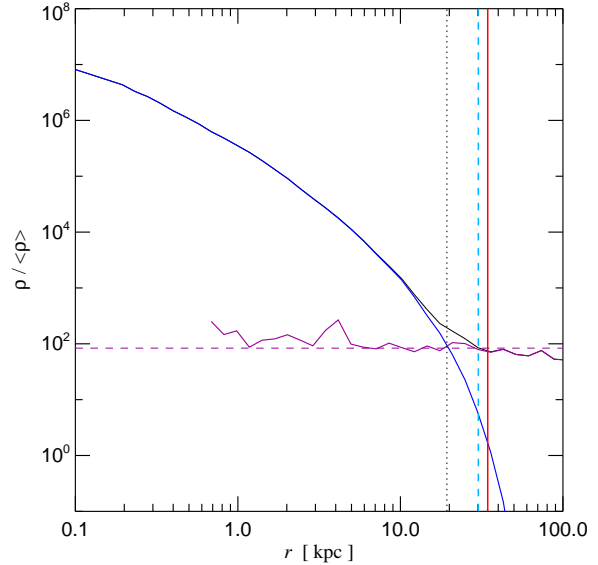}}%
\caption{An example of the determination of the subhalo radius $r_{\rm sub}$.
  The thick blue line shows the spherically averaged density profile of
  particles identified by {\small SUBFIND} as gravitationally bound to the
  subhalo, while the thin black line is the density profile of all particles.
  The purple line is the density profile of the difference, i.e.~of all
  particles that are not bound to the subhalo. The dashed horizontal line is
  the mean local background density $\rho_{\rm loc}$ at the subhalo's
  location, estimated within the radius marked by the vertical dotted line,
  which encloses $M_{\rm sub}$ in the profile of the total mass. The red
  vertical line marks our estimated radius $r_{\rm sub}$, at which point the
  bound density profile has dropped to $0.02\times \rho_{\rm loc}$. The dashed
  blue vertical line shows the estimated tidal radius of the subhalo based on
  $M_{\rm sub}$.
  \label{FigRsubExample}}
\end{figure}

\begin{figure}
\resizebox{8cm}{!}{\includegraphics{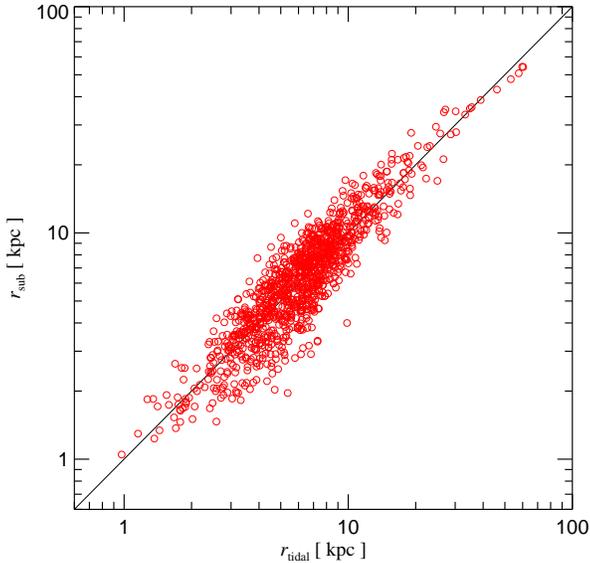}}%
\caption{Comparison between estimated subhalo radii and tidal radii for the
  Aq-A-1 simulation. Only subhalos with at least $10^4$ particles have been
  included in this plot.
\label{FigTidalRadii}}
\end{figure}

\subsection{The radius of a subhalo}

{\small SUBFIND} identifies substructures as locally overdense regions
relative to the average background density. It is thus  able to find
substructures and assign masses to them without large biases
throughout most of the halo. The procedure, however, may break down
near the centre, where the average density of the main halo may
overwhelm even the highest density peak of embedded substructures,
leading to potential biases in the masses assigned to subhalos
by {\small SUBFIND}. It is therefore desirable to find a definition for
the radius of a subhalo that is simple and physically meaningful and, at the same
time, relatively insensitive to the accuracy with which {\small
SUBFIND} assigns subhalo masses.

We have settled on the following operational procedure for determining
a subhalo radius, $r_{\rm sub}$. Starting from the centre of a
particular subhalo, we first measure spherically averaged density
profiles for all the mass and for the gravitationally bound mass (as
determined by {\small SUBFIND}). We then obtain a measure for the
local density of the main halo at the position of the subhalo location
by: (a) determining the radius $r'$ enclosing total mass equal to
$M_{\rm sub}$; (b) determining
the amount of unbound mass $M'$ inside this radius (this is simply the
difference between $M_{\rm sub}$ and the bound mass within $r'$); (c)
defining $\rho_{\rm loc} = M'/(4\pi r'^3/3)$. With this in hand, we
operationally define the bounding radius $r_{\rm sub}$ of the subhalo
to be the radius at which the spherically averaged density profile of
the bound mass has dropped below $0.02\times
\rho_{\rm loc} $. 

The bound density profile drops very steeply near the edge
of the subhalo, so changing the prefactor 0.02 has only a minor
influence on the radius determined in this way. We found that, with
this choice, the edge of the subhalo is robustly determined right at
the transition between the falling density profile of the subhalo, and
the approximately constant local background density of the halo. Also,
this radius tends to agree well with the minimum in the circular
velocity curve constructed using {\em all} the mass around the
subhalos's centre. In Figure~\ref{FigRsubExample}, we show a typical
example to illustrate this procedure. Note that the determination of
$\rho_{\rm loc}$, and hence the value of $r_{\rm sub}$ obtained
through the above procedure, is relatively insensitive to the precise
value of $M_{\rm sub}$.

In Figure~\ref{FigTidalRadii}, we show how the subhalo radii
determined in this way compare with tidal radii derived from the
distances of subhalos to halo centre, and their {\small SUBFIND}
masses $M_{\rm sub}$.  We calculate the tidal radius
\citep{BinneyTremaine87,Tormen1998} of a subhalo of mass $M_{\rm sub}$
and distance $r$ from the centre of the main halo as
\begin{equation}
r_{\rm t}= \left(\frac{ M_{\rm sub}}{[2-{\rm d}\ln M/{\rm d}\ln
r]\,M(<r)}\right)^{1/3}r
\end{equation}
where $M(< r)$ is the main halo mass within a sphere of radius
$r$. Reassuringly, there is good agreement between $r_{\rm t}$ and $r_{\rm sub}$ in the mean, with some
scatter. This gives us further confidence that our subhalo radii are
physically meaningful, and that {\small SUBFIND} correctly identifies
the self-bound regions of subhalos.

\subsection{Matching individual subhalos}

In order to study convergence not only of the main halo but also of
individual embedded subhalos, we need an appropriate method to
identify the {\em same} subhalo in simulations with different mass
resolution. This is less straightforward than it may seem at first, since one
cannot expect the subhalo to be at exactly the same {\em position} in
different simulations. When the numerical resolution is changed, small
phase offsets build up in the orbits of individual subhalos, causing
them to be at slightly different positions in different simulations,
even though their histories and their internal structure may agree in
detail \citep[see, e.g.,][]{Frenk1999}.

One solution to this problem is to match subhalos at very early times,
by tracking the particles of a particular subhalo back to the
(unperturbed) initial conditions. For each particle, we can then find
the closest particle in another realization of the initial conditions
created for the {\em same halo} but at different numerical
resolution. These matched particles can then be tracked forward in
time in the second simulation to see where they end up. This yields a
set of possible subhalo matches in the second simulation, among which
the one containing the largest number of matched particles is selected
as the partner to the original subhalo. To increase robustness, one may
require that the same match be obtained when the procedure is carried
out in reverse, i.e.~starting with the subhalo in the second
simulation.

We have found this procedure to work quite robustly for our simulation set. To
speed up the matching procedure in the unperturbed initial conditions, we have
successfully applied the following trick. Our IDs are constructed as 63-bit
Peano-Hilbert keys \citep{Springel2005b}, i.e.~they correspond to positions
along a space-filling fractal that tessellates our simulation volume with a
fiducial grid of $2^{21}$ cells per dimension. This corresponds to a comoving
spatial resolution of around $65\, {\rm pc}$, which is still considerably
smaller than the mean particle spacing in the high resolution region, even for
the Aq-A-1 simulation. Exploiting the fact that positions that are close on
the Peano-Hilbert curve are always close in 3D space (the reverse is, however, 
not always true), we can accelerate the matching by finding the particle with
the nearest Peano-Hilbert key in the second simulation. This always finds a
particle that is very close, although it does not guarantee that it is the
closest. This procedure turns out to be quite sufficient for the task at hand
here.

\begin{figure}
\resizebox{8cm}{!}{\includegraphics{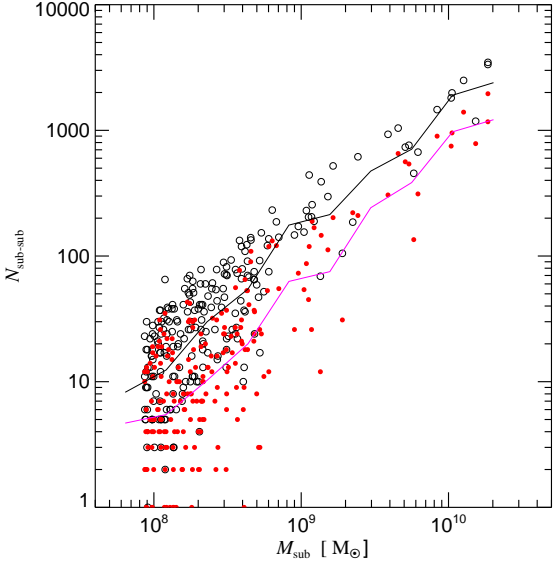}}%
\caption{(Sub-)substructure count within subhalos as a function of their
  mass. The plot includes all subhalos with more than 50,000 particles in the
  Aq-A-1 simulation. The actual \mbox{(sub-)subhalo} counts for these subhalos are
  shown as red dots and their average in logarithmically spaced mass bins is
  indicated by the purple line. Open circles show the (sub-)subhalo count
  which these same subhalos would have if they were scaled-down copies of the
  main halo, with the corresponding average indicated as a function of mass by
  the solid black line. When estimating these predictions, we correct for the
  lower effective resolution within subhalos in comparison to the main halo
  by only counting main halo subhalos above the appropriately scaled mass limit.
  \label{FigSubSubNumber}}
\end{figure}

\begin{figure}
\resizebox{8cm}{!}{\includegraphics{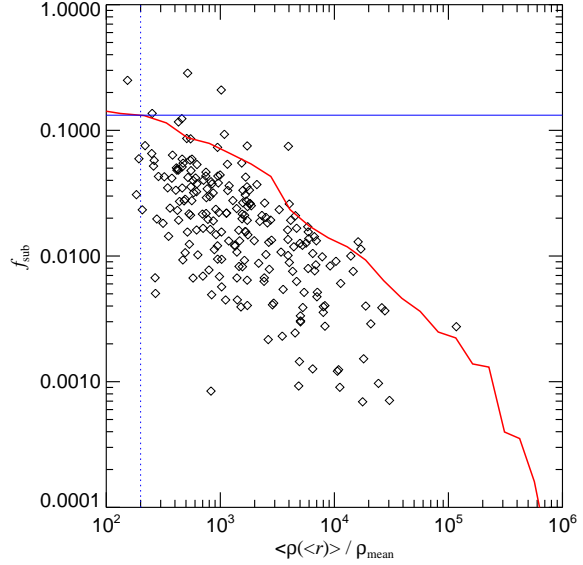}}%
\caption{(Sub-)substructure mass fraction in subhalos as a function of their
  radius within the main halo, expressed as the mean enclosed density in units
  of the cosmic mean density. The innermost subhalos are thus on the
  right. The plot includes all subhalos with more than 50,000 particles in
  Aq-A-1 (shown as symbols). All (sub-)subhalos within the estimated subhalo
  radius $r_{\rm sub}$ are counted, and their total mass is normalized by the
  total mass within $r_{\rm sub}$. To be able to compare all measured values
  of $f_{\rm sub}$ on an equal footing, we add a correction to account for
  \mbox{(sub-)subhalos} below our resolution limit but above the mass which
  corresponds to the resolution limit in the main halo after scaling down to
  each subhalo. The solid red line gives the substructure mass fraction which
  the subhalos would have if they were scaled down versions of the main halo,
  but with all material outside the tidal radius removed.  The horizontal blue
  line is the substructure mass fraction for the main halo within $r_{50}$.
  \label{FigSubSubMassFrac}}
\end{figure}

\begin{figure*}
\resizebox{8.5cm}{!}{\includegraphics{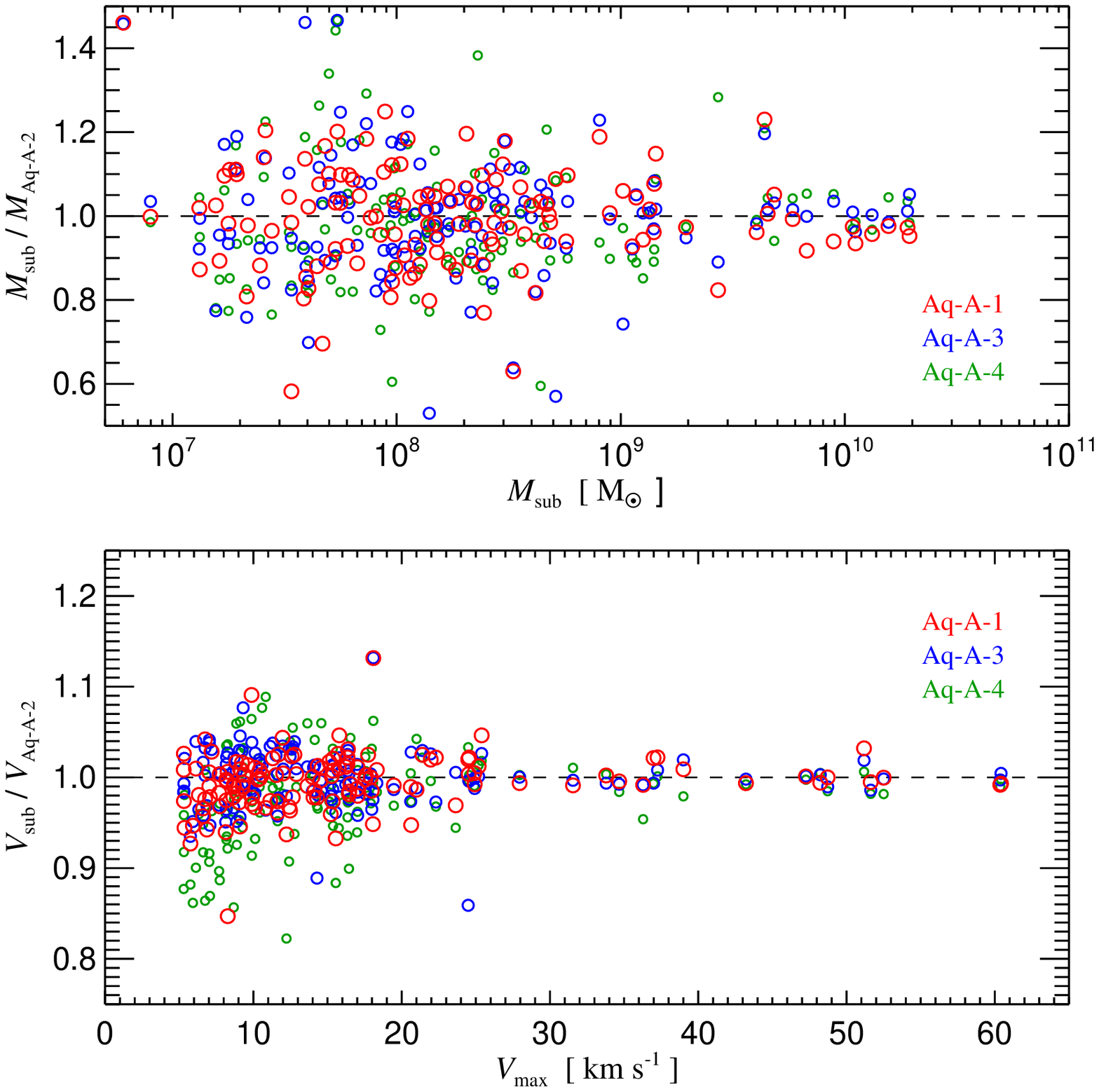}}\ %
\resizebox{8.5cm}{!}{\includegraphics{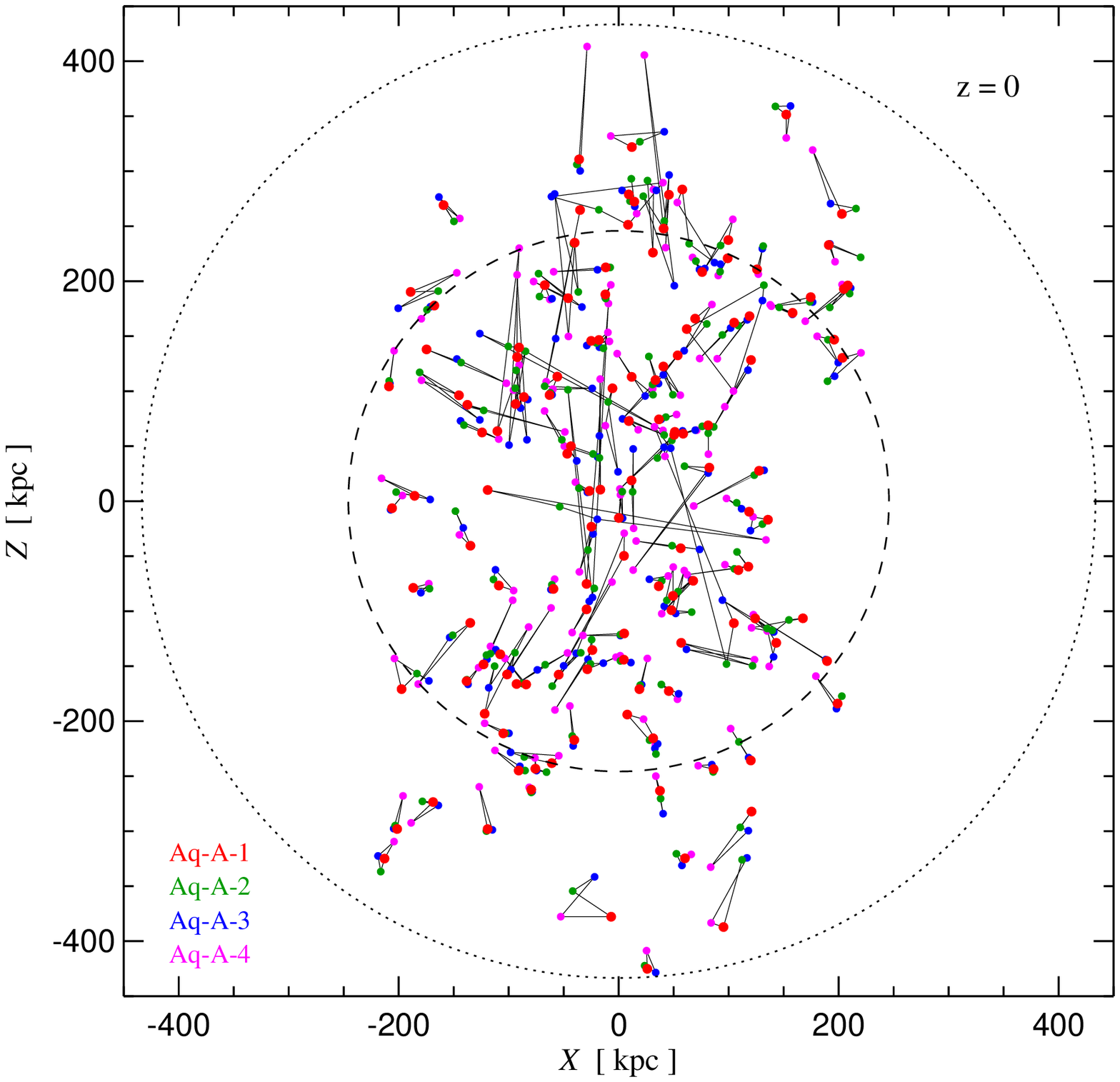}}
\caption{The two panels on the left compare the $M_{\rm sub}$ and $V_{\rm
    max}$ values of individual matched subhalos in Aq-A-1, Aq-A-2, Aq-A-3, and
  Aq-A-4. In both panels, the measured values are ratioed against the value
  found in Aq-A-2, our second highest resolution simulation of the `A' halo.
  In the panel on the right, we use small circles to indicate the projected
  positions of the matched subhalos on the $xz$-plane. The positions of each
  subhalo in the 4 different simulations are joined by thin lines.  The mean
  positional off-set between matched subhalos in Aq-A-4 and Aq-A-1 is $\sim
  54\,{\rm kpc}$. This shrinks to 28 and 26 kpc for matches of Aq-A-3 and
  Aq-A-2 to Aq-A-1, respectively. The inner and outer circles mark $r_{200}$
  and $r_{50}$.\label{FigMatchedSubhalos} }
\end{figure*}

In Figure~\ref{FigMatchedSubhalos} we compare properties of subhalos matched
in this way in the Aq-A series of simulations. The two panels on the left show
the masses and maximum circular velocities of subhalos in the Aq-A-1, Aq-A-3, and
Aq-A-4 simulations in units of the values measured for their counterparts in the
Aq-A-2 simulation. Remarkably, the agreement is excellent with a surprisingly
small scatter, and there is no obvious systematic offset between the different
resolutions. Further analysis of these data is given in \citet{Springel2008},
where we show that the maximum circular velocity of subhalos can be trusted in
the mean with an accuracy of 10\% down to $V_{\rm max} \sim 1.5\,{\rm km\,
  s^{-1}}$. Convergence for $r_{\rm max}$ is more difficult to achieve, but is
still good in the mean down to $r_{\rm max}\sim 165\,{\rm pc}$. In the right
panel of Fig.~\ref{FigMatchedSubhalos} we compare the projected spatial
positions of matching subhalos in the Aq-A-1, Aq-A-2, Aq-A-3 and Aq-A-4
simulations. While the agreement for the absolute coordinates is not perfect,
especially for subhalo quartets close to the main halo's centre, matching
subhalos are generally found quite close together. The mean spatial offset is
of order $\sim 30\,{\rm kpc}$, which is much better than we have typically
found in our older simulation work \citep{Stoehr2003}. This is a tribute to
the improved integration accuracy in {\small GADGET-3}, and to the high
quality of our initial conditions.

\begin{figure*}
\resizebox{15.0cm}{!}{\includegraphics{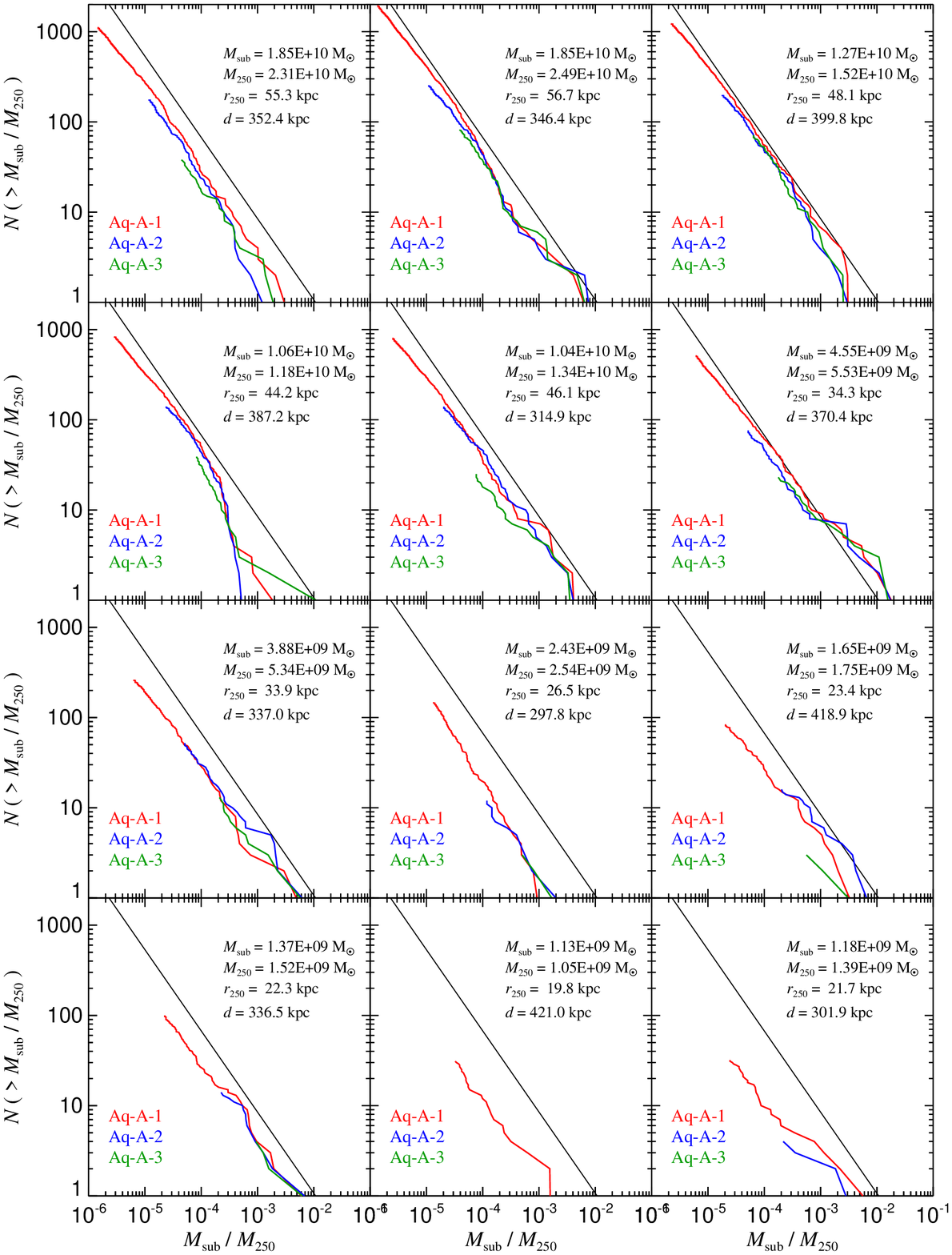}}
\caption{Cumulative mass function of (sub-)subhalos within subhalos. In each
  panel, we show results for one subhalo, and we compare results for different
  numerical resolutions, corresponding to simulations Aq-A-1, Aq-A-2 and
  Aq-A-3. Only substructures within a distance $r_{250}$ of the subhalo centre are
  counted, and the substructure mass is normalized to the mass within this
  radius $M_{250}$. The thin black power-law shows the subhalo abundance of
  main halos within a radius enclosing the same mean overdensity. The labels
  in each panel give the subhalo mass $M_{\rm sub}$ as determined by {\small
    SUBFIND}, the values of $M_{250}$ and $r_{250}$, and the distance $d$ of
  the subhalo from the centre of the main halo.
  \label{FigConvSubSubCumMassFunc}}
\end{figure*}

\subsection{The abundance of substructures within subhalos}

Using our matched sets of subhalos we are now ready to consider the
convergence of the properties of substructure inside subhalos. We
begin by considering (sub-)substructures within the subhalo radius,
$r_{\rm sub}$, and extend the analysis later to regions of fixed
overdensity, in order to compare with the results reported by
\citet{Diemand2008}.

We start by simply counting all subhalos within $r_{\rm sub}$ of a
given subhalo's centre and comparing the count with the number
expected from the assumption that subhalos are simply scaled-down
copies of the main halo. This expected number must be adjusted to take
into account that, unlike the main halo, subhalos have a different
``edge'' ($r_{\rm sub}$), as well as comparatively poorer mass
resolution. In practice, the self-similar expected number is computed
by considering in the main halo only subhalos with masses exceeding
$20\,m_{\rm p}\, M_{\rm main}/M_{r_{\rm sub}}$, where $20\,m_{\rm p}$
is our absolute {\small SUBFIND} mass limit for subhalo detection,
$M_{\rm main}$ is the main halo's mass within a radius that encloses
the same overdensity as that of the subhalo within $r_{\rm sub}$, and
$M_{r_{\rm sub}}$ is the {\em total} mass within $r_{\rm sub}$ of the
subhalo's centre.

In Figure~\ref{FigSubSubNumber} we compare the actual (sub-)subhalo
counts with the self-similar expectation, as a function of subhalo
mass.  Interestingly, we always find fewer (sub-)subhalos than
expected from the self-similar hypothesis. The suppression in
substructure abundance is not uniform; a few subhalos have almost the
full expected abundance, while others lie well below it. This is not
entirely unexpected, and it is probably related to the time since each
subhalo was accreted into the main halo and the extent to which it has
lost mass to tides. Typically the number of (sub-)subhalos is a factor
of 2 or 3 below that predicted by self-similarity.

The mass fraction of subhalos in the form of embedded substructure provides
another way of quantifying \mbox{(sub-)substructure} abundance. To estimate
this we simply measure the total mass of all (sub-)subhalos within $r_{\rm
  sub}$ and divide it by the {\em total} mass $M(r_{\rm sub})$ contained
within this same radius. Because the minimum subhalo size identified by {\small
  SUBFIND} is always 20 particles regardless of the mass of the parent object
(main halo or subhalo), we correct the measured cumulative substructure mass by
extrapolating the mass function from $20\,m_{\rm p}$ down to $20\,m_{\rm
  p}\,M(r_{\rm sub})/M_{\rm main}$ using a slope $N(>m)\propto m^{-0.9}$. Note
that $M_{\rm main}$ is here  the mass of the main halo measured within the
radius that encloses an equal mean overdensity as the subhalo within $r_{\rm
  sub}$.

In Figure~\ref{FigSubSubMassFrac}, we compare the substructure mass
fraction of subhalos to those of the main halo. The latter is computed
within the radius where the main halo density equals that of the
subhalo ($\overline{\rho}(r)=\rho_{\rm sub}$).  As shown in
Figure~\ref{FigSubSubMassFrac}, computed in this way the main halo
prediction is a monotonically decreasing function of $\rho_{\rm sub}$.
Interestingly, this line forms an accurate upper bound to the measured
substructure mass fractions of subhalos. Close to the line, the
self-similar expectation is fulfilled, but the fact that most subhalos
lie considerably below the line confirms that substructure is on
average significantly under-abundant in subhalos when compared to the
main halo.

For completeness, and in order to ease comparison with previous work, we show
in Figure~\ref{FigConvSubSubCumMassFunc} the cumulative subhalo abundance for
12 of our more massive subhalos in the Aq-A halo. For the first time, we
consider the convergence of the (sub-)substructure mass function for
individual subhalos, based on the Aq-A-1, Aq-A-2, and Aq-A-3 simulations and
our matched set of subhalos.  Here we count all substructures within a sphere
(of radius $r_{250}$) centred on the subhalo, and of mean overdensity 250
times the critical density \citep[and so 1000 times the cosmic mean density,
as chosen by][]{Diemand2008}.  A further criterion for selecting the subhalos
shown in Figure~\ref{FigConvSubSubCumMassFunc} was that their tidal radius
$r_t$ should exceed $r_{250}$, thus ensuring that the region where
(sub-)subhalos are counted really lies {\em inside} the subhalo.  Note that
this criterion is actually rather restrictive, as it precludes subhalos from
study that are at radii $r < r_{250}= 258\,{\rm kpc}$.

\begin{figure}
\resizebox{8.5cm}{!}{\includegraphics{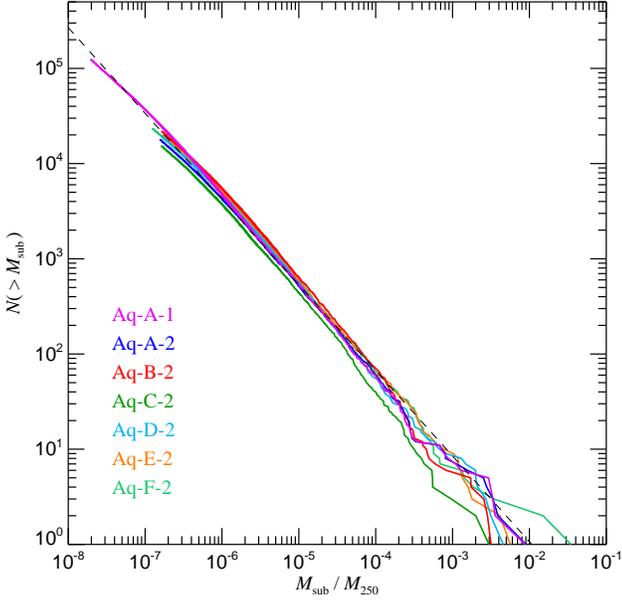}}
\caption{Cumulative subhalo mass function within $r_{250}$ for our 6 different
  halos at level 2 resolution, as well as for our highest resolution run
  Aq-A-1. Subhalo masses are measured in units of $M_{250}$ in each case. The
  dashed line is a fit to the mean mass function.}
\label{FigCumSubMassFuncInsideR1000}
\end{figure}

The (sub-)subhalos shown in Figure~\ref{FigConvSubSubCumMassFunc} are
counted as a function of their mass normalized by $M_{250}$, the total
mass of each subhalo within its own $r_{250}$. When the substructure
in main halos is counted in this way, a near-universal cumulative
subhalo mass function is found, as we show explicitly in
Figure~\ref{FigCumSubMassFuncInsideR1000} for our sample of 6 halos
simulated at resolution level 2. There is a well-defined mean
relationship with little scatter. For reference, we include a
power-law fit to this relation in the various panels of
Figure~\ref{FigConvSubSubCumMassFunc}. Clearly, also in this case most
subhalos show a cumulative substructure mass function with similar
slope but with a normalization that is typically substantially
lower. Only a few subhalos have a substructure abundance which
approaches that predicted by scaling our results the main halo.

\begin{figure}
\resizebox{7.5cm}{!}{\includegraphics{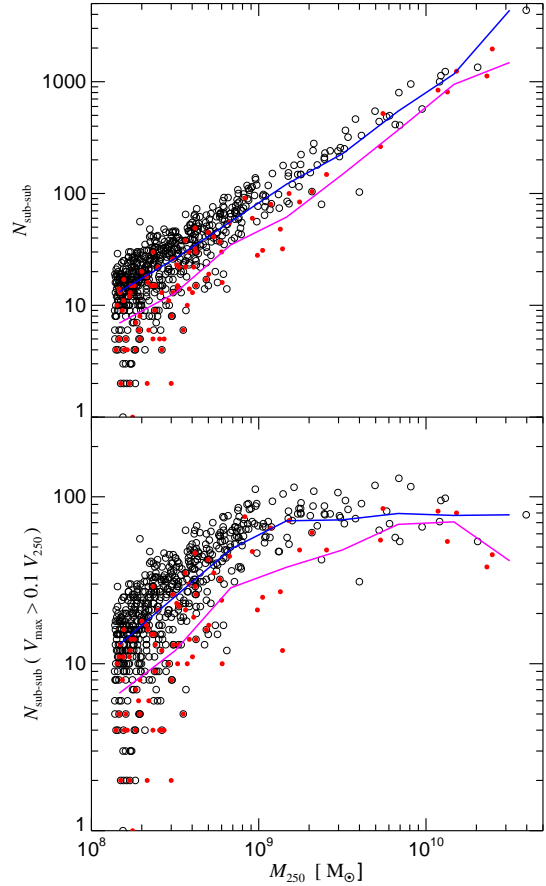}}%
\caption{Substructure count in subhalos of Aq-A-1 within $r_{250}$ (red
    filled circles) compared with field halos (hollow black circles) inside
    $r_{250}$, as a function of $M_{250}$. The top panel shows the
    substructure count down to our resolution limit, while the bottom panel
    gives the count above a limiting maximum circular velocity equal to
    $0.1\,V_{250}$, which effectively measures the amplitude of the subhalo
    velocity function. The solid lines give averages for independent
    logarithmic mass bins. We see that there is a systematic offset in the
    substructure abundance of field halos and genuine subhalos.  The downturn
    of the count above $0.1\,V_{250}$ for masses below $\sim 10^9\,{\rm
      M}_\odot$ is due to resolution limitations.
    \label{FigSubSubFieldComparison}}
\end{figure}

We may also compare the substructure abundance of subhalos inside $r_{250}$
with that of {\em field halos} of equal $M_{250}$, such that both are equally
well resolved and biases due to different numerical resolutions are
excluded. Figure~\ref{FigSubSubFieldComparison} shows our results for this
comparison, both in terms of the count of all substructures down to the
resolution limit, and by just counting subhalos with a maximum circular
velocity larger then $0.1\, V_{250}$, which effectively measures the amplitude
of the (sub-)subhalo velocity function. We here used uncontaminated field
halos found in the high-resolution region around the main halo in Aq-A-1, and
compared them to subhalos in the outer parts of the main halo (with $r >
258\,{\rm kpc}$, such that their tidal radius is larger than $r_{250}$).
Again we find an offset of about a factor of 2 in the mean substructure
abundance between field halos and genuine subhalos. 

We conclude that the (sub-)substructure abundance in subhalos is {\em
not}, in general, a scaled version of that in main halos. Rather, the
self-similar expectation provides an upper limit on the abundance of
these second generation substructures; less (sub-)substructure is
typically found. This reflects the fact that the substructure
abundance of a subhalo is not only diminished by tidal truncation once
it falls into a larger structure; in addition, its retained
substructures continue to lose mass to the main subhalo through tidal
effects and, in strong contrast to the situation for main halos, they
are not continually replenished by infall of new subhalos from the
field. The substructure deficit in subhalos when compared to main
halos {\em at the same mean overdensity} is expected to grow with time
as they orbit within their main halos. It will therefore be more
marked in subhalos in the inner halo, which are typically ``older''.

Our results thus caution strongly against the assumption that subhalos
typically have mass fractions in substructure similar to the main halo
\citep[as suggested by, e.g.,][]{Shaw2007,Kuhlen2008}.

\section{Internal structure of subhalos}  \label{SecSubInternalStructure}

In this section we study the structural properties of subhalos, and we compare
them to the properties of similar mass isolated halos. We analyze how the
density profiles of individual subhalos converge as numerical resolution is
increased, and we measure the concentration of subhalos as a function of mass,
circular velocity and radial distance. We then compare with the corresponding
relations for field halos.

\subsection{Density profiles for subhalos} \label{SecProfiles}

The internal density structure of nonlinear dark matter halos is one of the
most important predictions obtained from numerical simulations of the CDM
paradigm. The density profile directly affects the rotation curves of
galaxies, the gravitational lensing properties of dark matter halos, and the
X-ray luminosity and SZ-signal of galaxy groups and clusters. The density
profile of subhalos also determines the kinematics of the stars in satellite
galaxies, which are observationally accessible for the dwarf spheroidals
around the Milky Way \citep[e.g.][]{Stoehr2002,Strigari2007}. Furthermore, the
inner density profiles of halos and subhalos are critical for estimating the
luminosity in dark matter annihilation radiation, in case such a decay channel
exists.

About a decade ago, \citet{Navarro1996,Navarro1997} were able to show that the
spherically averaged density profiles of dark matter halos have an
approximately universal shape that is well described by a simple fitting formula
\begin{equation}
\rho(r) = \frac{\delta_c \,\rho_{\rm crit}}{({r}/{r_s})({r}/{r_s}+1)^2}
\end{equation}
that has become known as the NFW-profile. In this double power law, the local
logarithmic slope gradually changes from a value of $-3$ in the outer parts to
an asymptotic slope of $-1$ in the inner parts. The spatial scale $r_s$ of
this transition is treated as a fitting parameter and is often parameterized
in terms of the concentration $c=r_{200}/r_s$ of the halo, which is, in fact,
simply a reparameterization of $\delta_c$, the characteristic overdensity
relative to the critical density: $\delta_c = (200/3)\,c^3 / [{\rm
    ln}(1+c)-c/(1+c)]$. NFW showed the concentration to depend
systematically on halo mass, a finding that can be interpreted as reflecting
the density of the universe at the time of halo formation. A number of
analytic fitting functions for this dependence have been proposed in the
literature \citep{Navarro1997,Bullock2001,Eke2001}, but only recent
large-volume simulations have been able to calibrate it reliably for rare
objects like massive galaxy clusters \citep{Neto2007,Gao2007}.

Ever since the discovery of the NFW profile, the structure of the inner cusp
has been the subject of much discussion and controversy.  As computing power
has increased, many groups have reexamined the value of the inner slope using
ever bigger and better resolved simulations, but no consensus has yet emerged
\citep{Fukushige1997,Moore1998,Moore1999,Ghigna2000,Jing2000,Fukushige2001,
  Klypin2001,Jing2002,Fukushige2003,Power2003,Navarro2004,Fukushige2004,
  Diemand2005,Stoehr2006,Knollmann2008}.

\begin{figure*}
\resizebox{15cm}{!}{\includegraphics{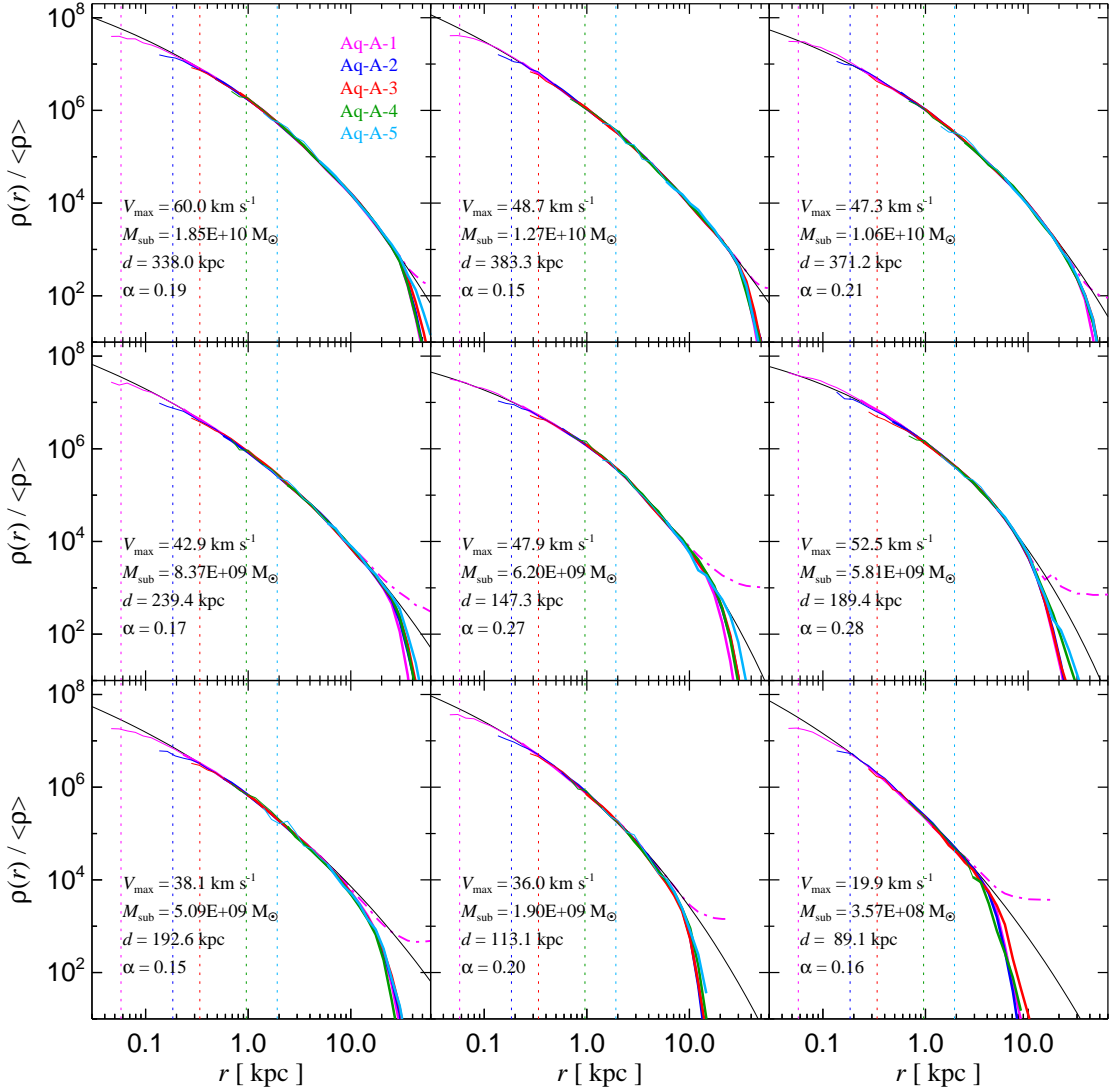}}
\caption{Subhalo density profiles for nine different subhalos in the Aq-A
  halo, simulated with varying resolution. The profiles show the bound mass
  only and are drawn with thick lines for the radial range where convergence is
  expected, based on the criterion of \citet{Power2003}. They are continued
  with thin lines down to the scale $2\,\epsilon$. Vertical dashed lines mark
  the radii where the force law becomes Newtonian ($2.8\,\epsilon$). The
  dot-dashed purple line in each panel is the density profile of all the mass
  around the subhalo's centre (i.e. including unbound mass). The thin black
  line shows a fit with the Einasto profile. The labels in each panel give the
  maximum circular velocity, mass, and distance $d$ to halo centre for each
  subhalo. $\alpha$ is the shape parameter of the Einasto profile, which we
  here allowed to vary freely in our fits.
\label{FigSubhaloDensProfiles}
}
\end{figure*}

\begin{figure*}
\resizebox{15cm}{!}{\includegraphics{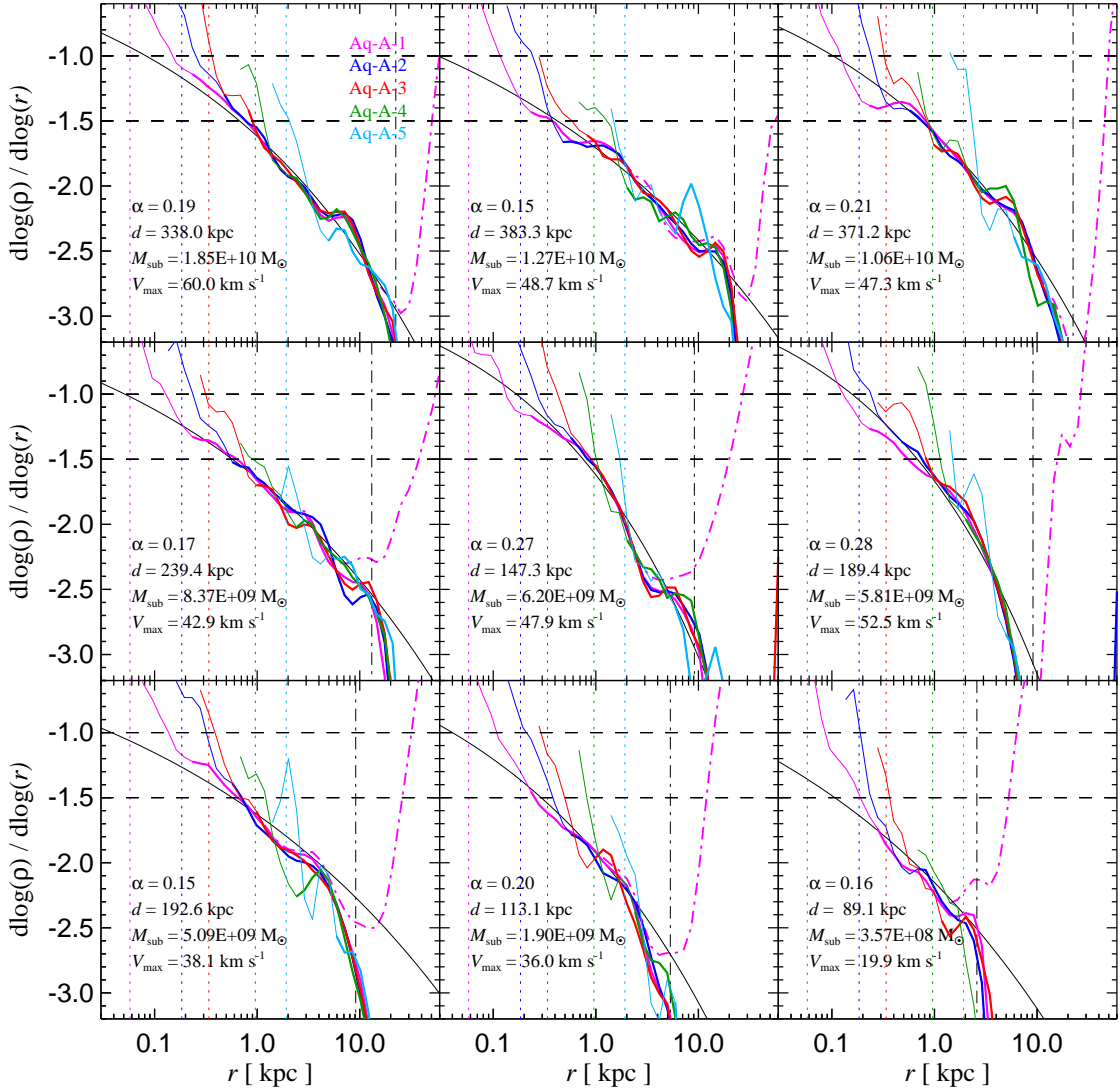}}
\caption{Local logarithmic slope (obtained by numerical differentiation) of
  the measured density profile for 9 different subhalos, at varying numerical
  resolution.  As in Figure~\ref{FigSubhaloDensProfiles}, the thick line style
  is used in regions where numerical convergence is expected. The thin solid
  line shows a fit with the Einasto profile. Slopes of $-1.5$ (corresponding
  to the Moore profile) and of $-1.0$ (the NFW profile) are marked with
  horizontal dashed lines.
\label{FigSubhaloDiffSlopes}}
\end{figure*}

\begin{figure*}
\resizebox{15cm}{!}{\includegraphics{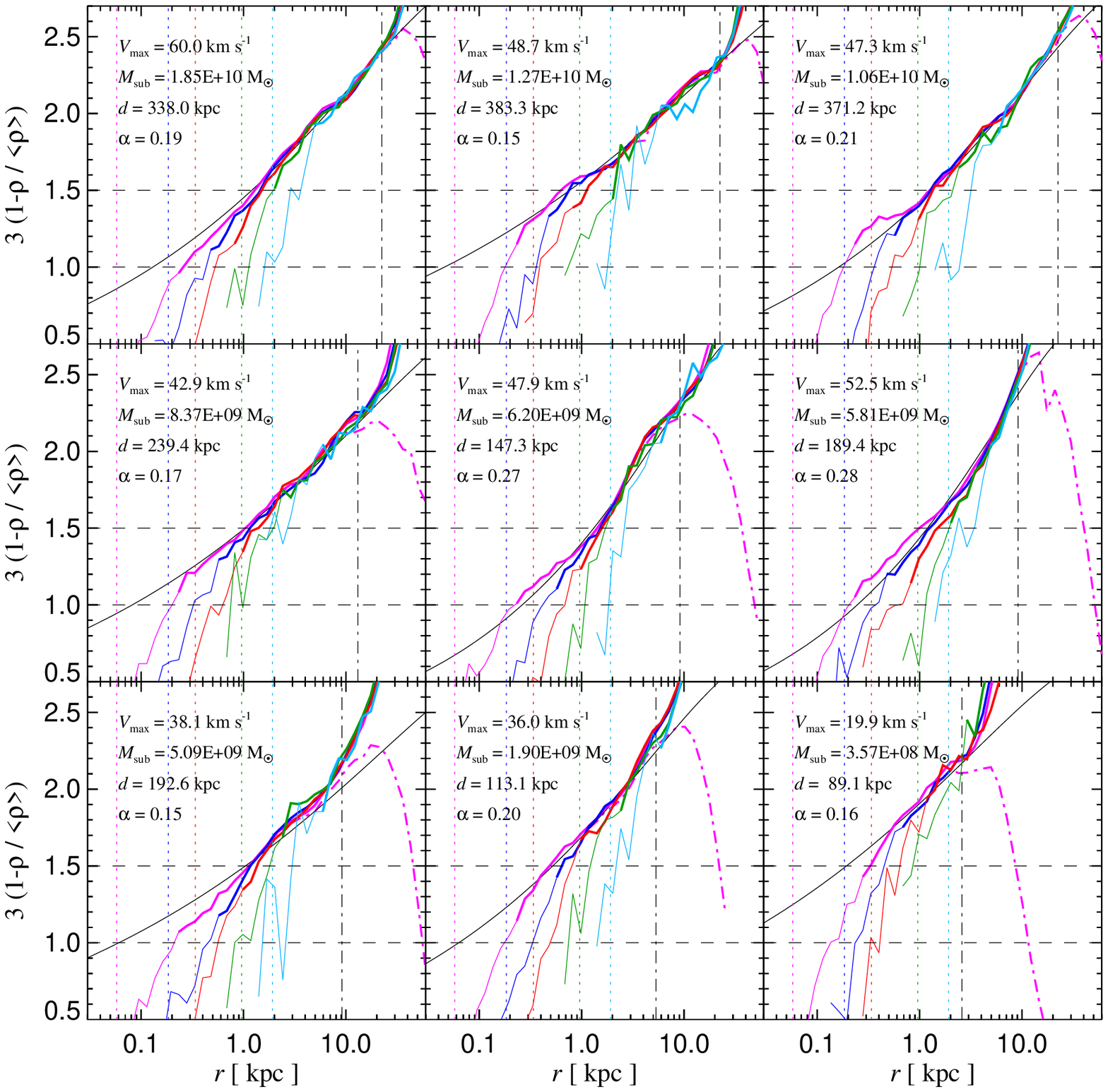}}
\caption{The maximum possible inner asymptotic slope of the density profiles
  of 9 subhalos in the Aq-A-1 simulation. The plot uses the same linestyles as
  in Fig.~\ref{FigSubhaloDiffSlopes}. Rather than the local slope of the
  density profile, this plot shows the most negative slope that is consistent
  both with the locally enclosed mass, and the local spherically averaged
  density. \label{FigSubhaloSlopeLimit} }
\end{figure*}

It has often been claimed that the inner cusps of halos and subhalos
may have slopes less than $-1$, with some studies even proposing
an asymptotic slope of $-1.5$ \citep{Moore1999,Fukushige2001}. For
main halos this proposition has been ruled out in recent years by
newer generations of simulations. Nevertheless, the idea that the
asymptotic slope is typically steeper than $-1$ (e.g.~$\sim -1.2$) is
still widespread and has been reiterated in recent papers, even though
this is clearly inconsistent with, for example,
Fig.~\ref{DensProfilesC02} or the numerical data in
\citet{Navarro2004}. 

With respect to the density profiles of subhalos, the situation is
even more unclear. So far few studies have examined this question
directly.  \citet{Stoehr2006} found that the circular velocity curves
of subhalos are best fit by a parabolic function relating $\log V$ to
$\log r$, implying that the density profiles become shallower in the
centre than NFW. On the other hand, \citet{Diemand2008} recently
argued that subhalos have steep cusps with a mean asymptotic slope of
$-1.2$.

We want to emphasize from the outset that the nature of halo and subhalo
density profiles, becoming gradually and monotonically shallower towards the
centre, makes it easy to arrive at the wrong conclusion for the structure of
the inner cusp. Almost all numerical simulations to date have been able to
produce demonstrably converged results for the density profile only in regions
where the local slope is significantly steeper than $-1$. They have also all
shown that the slope at the innermost measured point is significantly
shallower than at radii a factor of a few further out.  Thus, although no
slope as shallow as $-1$ has been found, there is also no convincing evidence
that the values measured are close to the asymptotic value, if one exists.
Most claims of steep inner cusp slopes are simply based on the assertion
that the slope measured at the innermost resolved point continues all the
way to the centre.

\citet{Navarro2004} argued that the local logarithmic slope of halo profiles
changes smoothly with radius and is poorly fit by models like those of NFW or
Moore that tend to an asymptotic value on small scales. They showed that in
their simulation data the radial change of the local logarithmic slope can be
well described by a power-law in radius, of the form
\begin{equation}
\frac{{\rm d}\log \rho}{{\rm d}\log r} = -2\left(\frac{r}{r_{-2}}\right)^{\alpha},
\end{equation}
which corresponds to a density profile
\begin{equation}
\rho(r) = \rho_{-2}\, \exp \left(- \frac{2}{\alpha}\left[
    \left(\frac{r}{r_{-2}}\right)^\alpha-1\right] \right).
\end{equation}
Here $\rho_{-2}$ and $r_{-2}$ are the density and radius at the point where
the local slope is $-2$.  This profile was first used by \citet{Einasto1965}
to describe the stellar halo of the Milky Way, so we refer to it as the
Einasto profile. The introduction of a shape parameter, $\alpha$ may be
expected, of course, to provide improved fits, but we note that fixing
$\alpha\sim 0.16$ gives a two-parameter function which still fits mean halo
profiles much better than the NFW form over a wide range of halo masses
\citep[i.e. with maximum residuals of a few percent rather than 10
  percent,][]{Gao2007}. Further evidence for a profile where local slope
changes gradually has been presented by \citet{Stoehr2003, Stoehr2006,
  Graham2006}. For reference, we note that the enclosed mass for the Einasto
profile is
\begin{equation}
M(r) = 
\frac{4\pi r_{-2}^3 \rho_{-2}}{\alpha}\,\exp\left(\frac{3\ln \alpha +2 - \ln 8}{\alpha}\right)\,
\gamma\left[ 
\frac{3}{\alpha}, \frac{2}{\alpha}\left(\frac{r}{r_{-2}}\right)^{\alpha}
\right]
\end{equation}
where $\gamma(a,x)$ is the lower incomplete gamma function.  For a value of
$\alpha=0.18$ the radius where the maximum circular velocity peaks is given by
$r_{\rm max} = 2.189\, r_{-2}$, and the maximum circular velocity is related to
the parameters of the profile by $V_{\rm max}^2 = 11.19\, G
r_{-2}^2\,\rho_{-2}$.

No published simulation to date has had enough dynamic range to measure the
logarithmic slope of the density profile in the region where the Einasto model
would predict it to be shallower than $-1$, so only indirect arguments could
be advanced for this behaviour \citep{Navarro2004}. This situation has changed
with the Aquarius Project, as can be seen from Fig.~\ref{DensProfilesC02}, and
in \citet{Navarro2008} we provide a detailed analysis of this question.  In
the following, we focus on the density profiles of dark matter {\it subhalos},
where the available particle number is, of course, much smaller. Our best
resolved subhalos in the Aq-A-1 simulation contain more than 10 million
particles, allowing a relatively precise characterization of their density
profiles.  Until recently, such particle numbers represented the
state-of-the-art for simulations of {\em main} halos.

In Figure~\ref{FigSubhaloDensProfiles}, we show spherically averaged density
profiles for 9 subhalos within the Aq-A halo. For each we compare up to 5
different resolutions, covering a factor of $\sim 1835$ in particle mass.  The
density profiles line up quite well outside their individual resolution
limits, as predicted by the convergence criterion of \cite{Power2003} in the
form given in equation~(\ref{EqnPower}). Individual profiles in the panels are
plotted as thick solid lines at radii where convergence is expected according
to this criterion, but they are extended inwards as thin lines to twice the
gravitational softening length (the gravitational force is exactly Newtonian
outside the radii marked by vertical dashed lines). These density profiles are
based on particles that are gravitationally bound to the subhalos, but for
comparison we also show a profile for each subhalo that includes all the mass
(i.e.  including unbound particles; thick dashed lines). It is clear that the
background density dominates beyond the `edge' of each subhalo. It is
therefore important that this region is excluded when fitting analytic model
density profiles to the subhalos.

\begin{figure}
\resizebox{8.5cm}{!}{\includegraphics{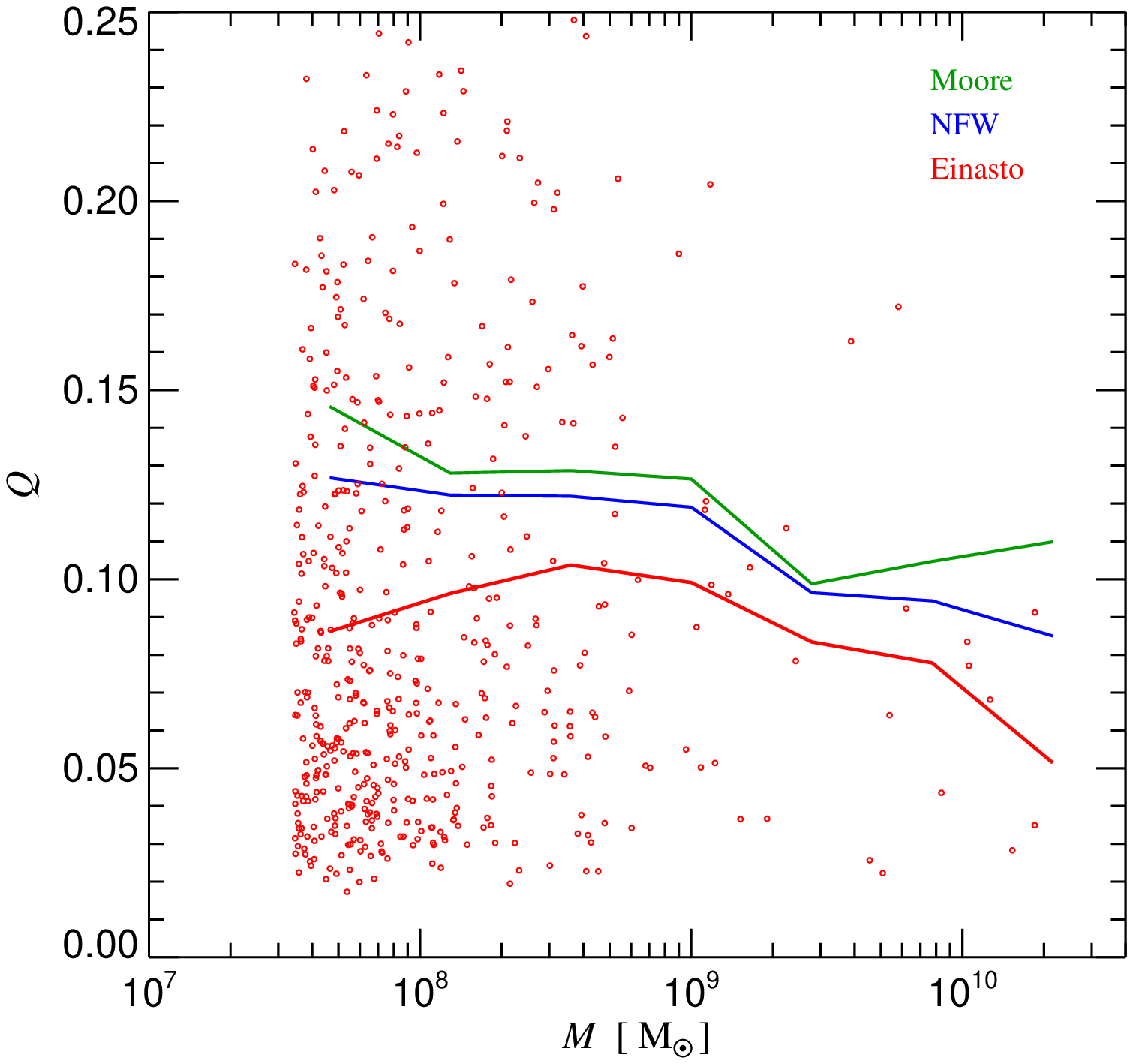}}
\caption{Quality of fits to subhalo density profiles, based on three different
  two-parameter models, an NFW profile, a Moore profile, and an Einasto
  profile with $\alpha=0.18$. The circles show a measure for the mean
  deviation per bin, $Q$, for 526 subhalos in the main halo of the Aq-A-1
  simulation. The subhalos considered contain between 20,000 and nearly $\sim
  10$ million particles. The lines in different colours show averages in
  logarithmic mass bins for each of the three
  profiles.\label{FigSubhaloFitQuality} }
\end{figure}

\begin{figure}
  \resizebox{8cm}{!}{\includegraphics{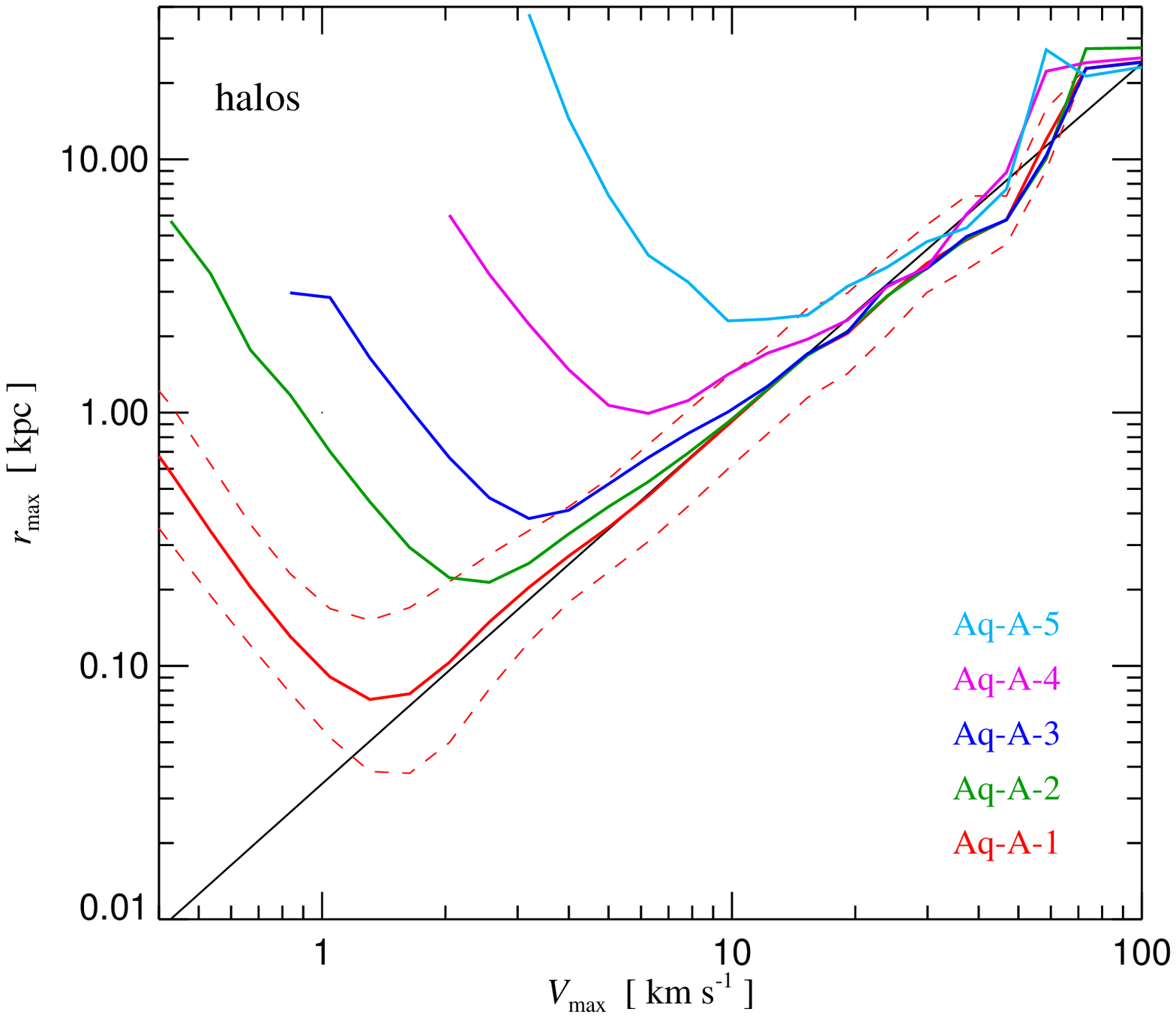}}\\
  \resizebox{8cm}{!}{\includegraphics{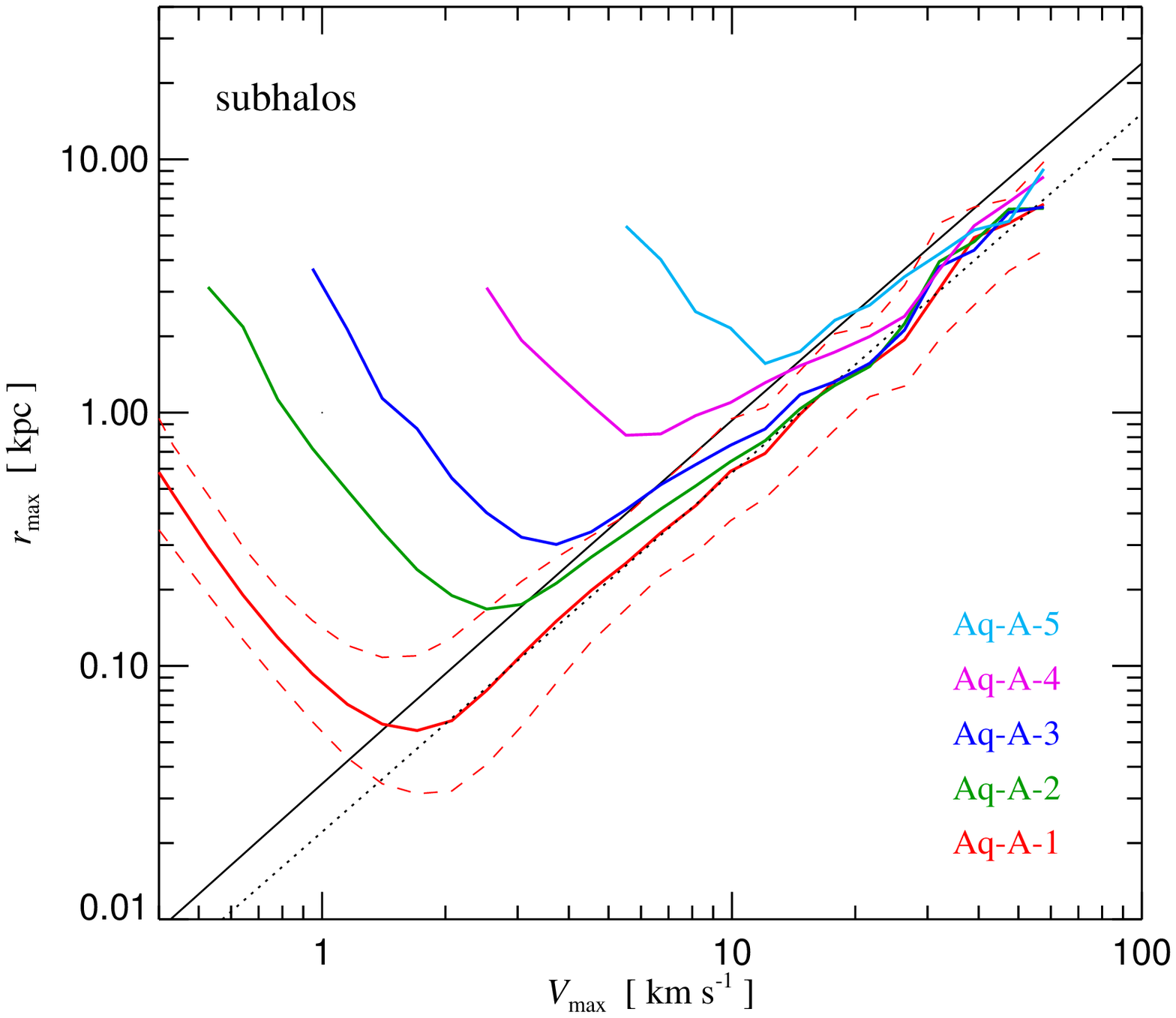}}
\caption{Relation between $r_{\rm max}$ and $V_{\rm max}$ for main halos (top)
  and subhalos (bottom) in the Aq-A series of simulations. We compare results
  for simulations of different resolution for this halo, and we use solid
  lines to mark the mean of $\log r_{\rm max}$ in each bin. The dashed red
  lines enclose 68\% of the distribution for the Aq-A-1 simulation. The solid
  line is an extrapolation to smaller mass of the result of \cite{Neto2007}
  for the halos of the Millennium simulation, while the dotted power law in the
  lower panel is a fit to our results for subhalos, lying a factor 0.62 lower.
 \label{FigRmaxVsVmax}
}
\end{figure}

In making such fits, we restrict ourselves to the radial range between the
convergence radius (equation \ref{EqnPower}) and the largest radius where
the density of bound mass exceeds 80\% of the total mass density. The
density profiles themselves are measured in a set of radial shells spaced
equally in $\log r$.  To define the best fit, we minimize the sum of the
squared differences in the log between measurement and model, i.e.~we
characterize the goodness of fit by a quantity
\begin{equation}
 Q^2 = \frac{1}{N_{\rm bins}}\sum_i [\ln \rho_i - \ln \rho^{\rm
   model}(r_i)]^2 \label{eqnQ} ,
\end{equation}
where the sum extends over all bins $i$. We then minimize $Q$ with respect to
the parameters of the model profile.  We have included such fits as thin solid
lines in Figure~\ref{FigSubhaloDensProfiles}, based on the Einasto profile,
allowing the third parameter $\alpha$ to vary as well. The resulting values of
$\alpha$ and the maximum circular velocities of the subhalos, as well as their
mass and distance to the main halo's centre are shown as labels in the
individual panels.

It is clear from Figure~\ref{FigSubhaloDensProfiles} that the Einasto profile
provides a good description of subhalo radial density profiles, but due to the
large dynamic range on the vertical axis combined with the narrow radial range
over which the density profile can be fit, it is not clear in this
representation whether the Einasto fit is significantly better than fits
with other analytic functions, like the NFW or Moore profiles.

Further insight can be obtained by studying the local logarithmic slopes
of the subhalo density profiles as a function of radius, which we show in
Figure~\ref{FigSubhaloDiffSlopes}, obtained by finite differencing of the
measured density profiles.  Again, we compare the differing resolutions
available for Aq-A, and plot the results as thick lines for radii where we
expect convergence according to \citet{Power2003}, continuing them with thin
lines towards smaller scales. The convergence criterion appears to work quite
well and in most cases accurately delineates a limit beyond which the profiles
suddenly start to become significantly flatter. At larger radii, the local
slopes change continuously and smoothly with radius. For several subhalos, we
have direct evidence that for the local slope is significantly shallower than
$-1.5$ in the innermost converged bin, thereby ruling out the Moore profile
for at least some dark matter subhalos. In one case, we find convergence to a
slope which is clearly shallower than $-1.2$. As for main halos, 
extrapolation of the shape of these curves to smaller radii suggests that
profiles that will become significantly shallower before reaching an
asymptotic inner slope, if one exists. From these results it seems very
unlikely that typical dark matter subhalos could have power law cusps with 
slopes as steep as $-1.2$, as recently suggested by \citet{Diemand2008}.

Another way to arrive at a similar conclusion is not to consider the
numerically differentiated density profile, but rather the maximum asymptotic
inner slope
\begin{equation}
\beta(r)= 3 [1-\rho(r)/\overline{\rho}(r)]
\end{equation} 
that can be supported by the enclosed mass at a certain radius. This quantity
was introduced by \citet{Navarro2004}. It requires converged values for both
the local density and the enclosed mass at each radius $r$.  This is a more
stringent convergence requirement than asking that the density alone be
converged. Nevertheless, it can provide a powerful lower limit on the profile
slope in the inner regions; there cannot possibly be a cusp steeper than
$\rho\propto r^{-\beta}$ since there is simply not enough mass enclosed to
support it. In Figure~\ref{FigSubhaloSlopeLimit}, we show $\beta(r)$ as a
function of radius for the same subhalos as before, using the same approach to
mark the \citet{Power2003} convergence radius. We see that this convergence
criterion is not conservative enough in some of cases, where the enclosed mass
is not fully converged for the last bin. The \citet{Power2003} criterion was
actually designed for this quantity, but it has only been tested for main
halos, and it is not surprising that we find subhalos to be somewhat more
demanding.  Nevertheless, this figure reinforces our earlier conclusion. For
most of the subhalos, a central dark matter cusp as steep as the Moore profile
can be safely excluded, and in a few cases, the limit is shallower than $\sim
-1.3$. Again, the shape of $\beta(r)$ suggests that limits are likely to
tighten considerably once still smaller scales can be resolved.

Finally, we would like to answer objectively the question whether the Einasto
model fits subhalo profiles better than the NFW or Moore models; in other
words, whether it produces smaller residuals overall. To test this question,
we fix $\alpha$ for the Einasto profile at $\alpha=0.18$ so that there are
only two free parameters left, as in the NFW and Moore profiles. (These are a
characteristic overdensity and a radial scale.) Our results are insensitive to
varying $\alpha$ in the range $\sim 0.16 - 0.20$. We estimate best fits for
526 subhalo profiles (considering all subhalos in Aq-A-1 with more than 20,000
particles) by minimizing the quantity $Q$ defined by equation (\ref{eqnQ})
over the radial range between the Power convergence radius and an outer radius
defined as above. In Figure~\ref{FigSubhaloFitQuality} we show the results. We
plot the mean residual per bin with symbols giving results for the Einasto
profile to illustrate the typical scatter. The solid coloured lines are means
for the three different profile shapes, calculated for logarithmic bins of
subhalo mass. We see that the Einasto profile consistently produces the lowest
residuals, followed by the NFW profile, while the Moore profile is
consistently the worst. The relatively small difference in the quality of the
fit between the NFW and Moore profiles is due to the fact that the resolution
limitations for the subhalos restrict the fits to comparatively large radii
where the two still have quite similar shape.  There appears to be no
systematic trend with subhalo mass. 

We conclude that the density profiles of subhalos show similar
behaviour to those of main halos; the local logarithmic slope becomes
gradually shallower with decreasing radius. There is no evidence that
a fixed asymptotic power law has been reached at the innermost
converged points. Inner cusps as steep as the Moore profile are
excluded for most objects, and for some objects we can already exclude
logarithmic slopes as steep as $-1.3$.

\subsection{The concentration of subhalos}

Because the density profiles of dark matter halos are not pure power
laws it is possible to assign them a characteristic density or
``concentration''. Perhaps the simplest such measure is the
overdensity (relative to critical) within the radius where the
circular velocity curve peaks (see equation \ref{EqnConc}). In many
studies it has been found that halos of a given mass exhibit a
well-defined characteristic concentration
\citep{Navarro1997,Eke2001,Bullock2001,Neto2007,Gao2007}, or in other
words, that the radius $r_{\rm max}$ at which the circular velocity
peaks is tightly correlated with the maximum circular velocity $V_{\rm
  max}$. Recently, \citet{Neto2007} have given an accurate fit to this
relation for halos with masses between about $10^{12}$ and
$10^{15}{\rm M}_\odot$, based on the good statistics provided by the
Millennium Simulation.

\begin{figure}
\resizebox{8cm}{!}{\includegraphics{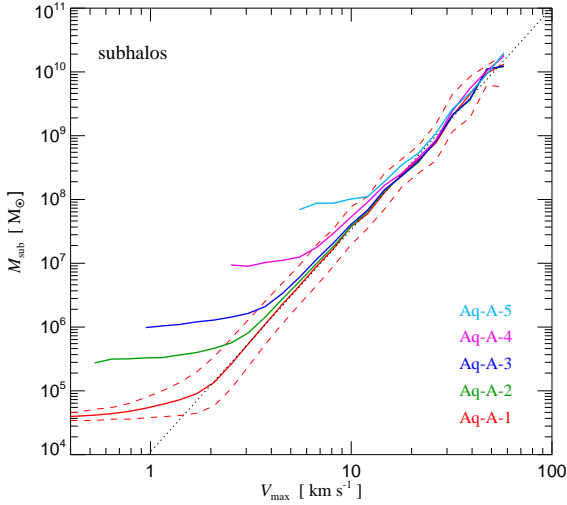}}
\caption{Mean relation between subhalo mass as assigned by {\small SUBFIND} and
  maximum halo circular velocity. The dashed red lines enclose 68\% of the
  distribution around the mean (calculated as an average of $\log M_{\rm
    sub}$) for the Aq-A-1 simulation. The dotted line is a power law fit, 
$M_{\rm sub} \simeq  3.37\times 10^7\,(V_{\rm max}/10\,{\rm km\,s^{-1}})^{3.49}$,
to the results of Aq-A-1.
\label{FigMsubVsVmax} }
\end{figure}

\begin{figure*}
\resizebox{8.0cm}{!}{\includegraphics{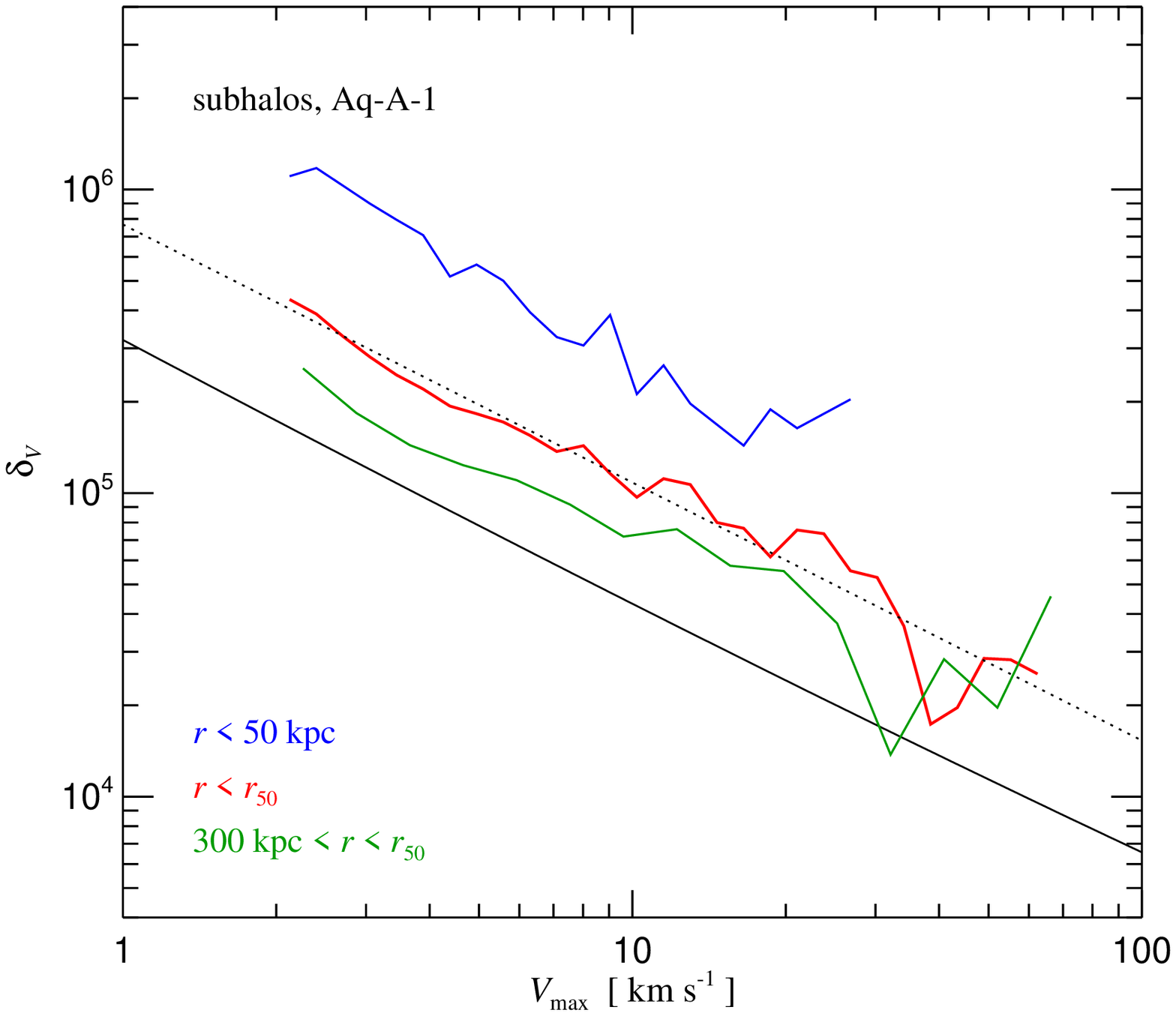}}%
\resizebox{8.0cm}{!}{\includegraphics{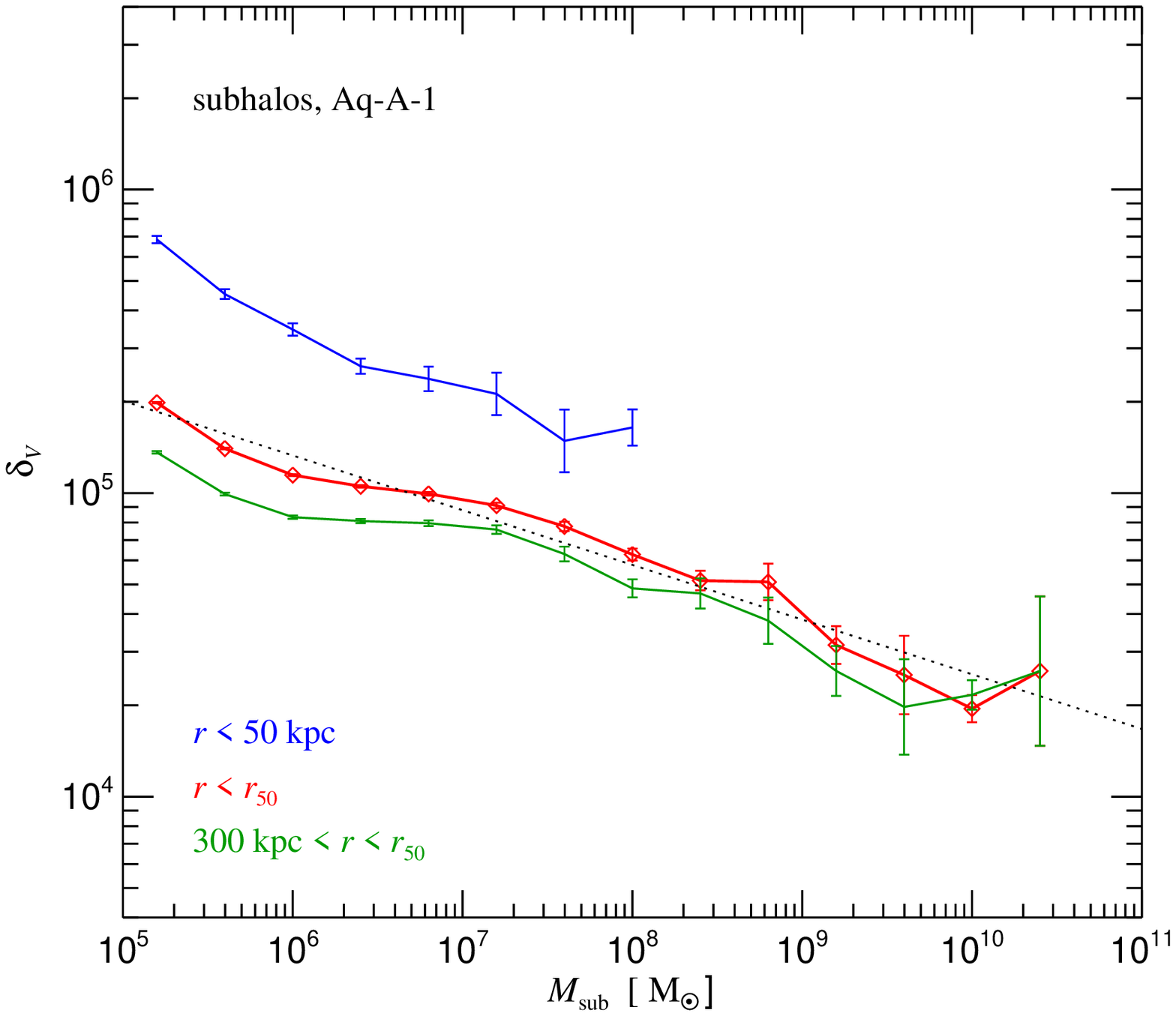}}
\caption{Subhalo characteristic density (which is a measure of concentration) 
  as a function of maximum circular velocity
  (left panel) and subhalo mass (right panel). We show results for all halos
  in the $r_{50}$ radius, as well as separately for the inner halo within 50
  kpc, and for an outer shell between 300 kpc and $r_{50}$. The solid line in
  the left panel gives an extrapolation of the result which \citet{Neto2007}
  quote for halos in the Millennium simulation, while the dotted power law 
lies a factor 2.6 higher and fits our results for the subhalos within $r_{50}$.
The dotted line in the right panel, $\delta_V \simeq 5.80\times 10^8\,(M_{\rm
  sub}/10^8\,{\rm M}_\odot)^{-0.18}$, is a fit to our results for  all 
 subhalos within $r_{50}$.
 \label{FigConcVsVmaxMsub}}
\end{figure*}

In the top panel of Figure~\ref{FigRmaxVsVmax}, we show the relationship
between $r_{\rm max}$ and $V_{\rm max}$ as measured for main halos in our Aq-A
simulation. These are halos that are outside of the main Milky-Way sized halo,
but are still contained in the high-resolution region. We make sure to include
only halos that are free of any contamination by boundary particles.
Comparing the various resolutions available for the `A' halo, it can be seen
that the correlation can be trusted down to about $V_{\rm max} \sim 1.5\,{\rm
  km\,s^{-1}}$ for our highest resolution calculation, the Aq-A-1
run. Remarkably, we find that the power-law relation of \citet{Neto2007}
describes our measurements very accurately, despite the fact that this is an
extrapolation by several orders of magnitude into a regime which was
previously unconstrained by numerical data.

The bottom panel of Figure~\ref{FigRmaxVsVmax} shows the equivalent
measurements for subhalos that are contained within $r_{50}$ of the main
halo. Clearly, these subhalos are typically more concentrated than halos of
the same circular velocity in the field, as first found by \citet{Ghigna1998}.
At equal $V_{\rm max}$, the $r_{\rm max}$ values of subhalos are on average
62\% of those of field halos, corresponding to a 2.6 times higher
characteristic density.  This can be understood as a result of tidal mass
loss. As \cite{Penarrubia2008} show, stripping reduces both $V_{\rm max}$ and
$r_{\rm max}$, but the reduction in $r_{\rm max}$ is larger, so that the
concentration increases \citep[see
also][]{Hayashi2003,Kazantzidis2004,Bullock2005}.  We note that this effect
also increases the characteristic density and the dark matter annihilation
luminosity of subhalos relative to halos in the field when they are compared
at {\em equal mass}, contrary to the arguments of \citet{Strigari2007b}.

The subhalo masses $M_{\rm sub}$ are tightly correlated with $V_{\rm
  max}$ as well, as shown in Figure~\ref{FigMsubVsVmax}. However, the
slope of this relation, $M_{\rm sub}\propto V_{\rm max}^{3.5}$, is
somewhat steeper than expected for a self-similar scaling of subhalo
structure with size. This is again a consequence of tidal mass loss,
which affects the mass of a subhalo more than its maximum circular
velocity.

Another interesting quantity to consider is the mean characteristic density
contrast $\delta_V = V_{\rm max}^2 / (H_0 r_{\rm max})^2$ of subhalos. In
Figure~\ref{FigConcVsVmaxMsub}, we show the dependence of this measure of
concentration on circular velocity and subhalo mass. Clearly, the
concentration increases strongly with decreasing subhalo mass. Interestingly,
this trend is equally strong for subhalo samples at different radii, but the
absolute values of the concentrations are larger at smaller radii. This is
illustrated in Fig.~\ref{FigConcVsVmaxMsub} which compares results for the
inner halo ($r<50\,{\rm kpc}$) and for a shell at large radii ($r>300\,{\rm
  kpc}$) with results for the halo as a whole.

\begin{figure}
\resizebox{8.0cm}{!}{\includegraphics{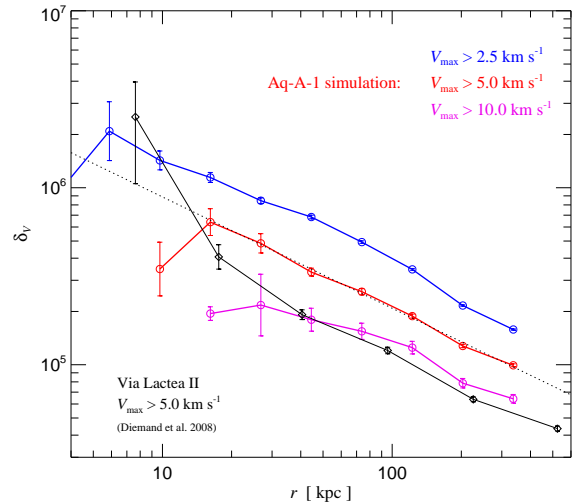}}
\caption{Subhalo concentration as a function of radius, for subhalos with
  maximum circular velocity larger than 2.5, 5, or $10\,{\rm km\, s^{-1}}$.
  The dotted line is a fit to our result for the $5\,{\rm km\, s^{-1}}$
  sample, which yields $\delta_V \simeq 3.77\times 10^6\, (r/{\rm
    kpc})^{-0.63}$. For comparison, we have also included the result quoted
  recently by  \citet{Diemand2008} for the Via Lactea II simulation, which
  selected subhalos with $V_{\rm max} > 5\,{\rm km\, s^{-1}}$. Clearly, our
  subhalos are more concentrated than theirs at the same circular velocity.
  \label{FigConcVsRadius}}
\end{figure}

This radial trend is more directly displayed in Figure~\ref{FigConcVsRadius},
where we show the mean characteristic density contrast as a function of radius
for samples selected above different lower cut-offs in circular velocity.  In
general, subhalo concentrations rise towards halo centers, as found by
\citet{Diemand2007b,Diemand2008}.  For comparison, we also show results for Via
Lactea II, as recently published by \citet{Diemand2008} where a cut-off of
$5\,{\rm km\, s^{-1}}$ was used. Interestingly, our subhalos are substantially
more concentrated than those in Via Lactea II for the same lower cut-off. The
Via Lactea II subhalos are actually slightly less concentrated than our
subhalos selected above $10\,{\rm km\, s^{-1}}$. The origin of this difference
is unclear, but it may be related to the discrepancy in the abundance of
subhalos that we discussed earlier in Section~\ref{SecSubAbundance}.

\section{Summary}  \label{SecSummary}

In this paper, we have presented first results from the Aquarius
Project, a Virgo Consortium\footnote{The Virgo Consortium is an
  international collaboration of astronomers working on supercomputer
  simulations of cosmic structure formation, see
  http://www.virgo.dur.ac.uk} programme to carry out high-resolution
dark matter simulations of Milky-Way-sized halos in the $\Lambda$CDM
cosmology.  This project seeks clues to the formation of galaxies and
to the nature of the dark matter by designing strategies for exploring
the formation of our Galaxy and its luminous and dark satellites, for
searching for signals from dark matter annihilation, and for designing
experiments for the direct detection of dark matter.

In our approach, we pay great attention to validating our numerical results to
careful convergence studies. In addition, we explore possible uncertainties in
predictions for the Milky Way resulting from the scatter in properties between
otherwise similar halos. Thus, we simulate not just one realization at ultra
high resolution, but rather a sample of (currently) 6 different halos.  Our
ambition is to redefine the state-of-the-art in this field with respect to the
accuracy of the cosmological N-body simulations, and the rigour with which
quantitative statements about halo structure can be made.

Our new simulation code {\small GADGET-3}, developed specifically for the
Aquarius Project, is a highly efficient, massively parallel N-body code. It
offers much better scalability to large numbers of compute cores and a higher
basic speed than its parent code {\small GADGET-2} \citep{Springel2005b}.  It
is able to cope efficiently with the tight coupling of around 1.5 billion
particles in a single nonlinear object, split up across 1024 CPUs. Some of our
simulations at resolution level 2 were run on an even larger number of compute
cores, using up to 4096 cores of a Bluegene/P computer. Here we used a novel
feature in {\small GADGET-3} that can exploit additional compute cores in
shared-memory nodes by means of threads (based on the POSIX pthreads library)
yielding a mix of distributed and shared memory parallelism.  The ability to
simulate this high degree of clustering and nonlinearity on massively parallel
architectures is a prerequisite for exploiting the power of upcoming petaflop
computers for the next generation of high-precision simulations of
cosmological structure formation.

The results presented above demonstrate that we have created a remarkably
accurate set of simulations, reaching very good convergence for the dark
matter density profile and the substructure mass function over the maximum
range that could be expected. Even the location, mass and internal structure
of individual large dark matter subhalos reproduce well between simulations of
differing resolution, a level of convergence which exceeds anything previously
reported in the literature.

The abundance of dark matter subhalos is remarkably uniform across our
halo sample when normalized to parent halo mass, and when considering
subhalos sufficiently small that fluctuations due to counting
statistics are unimportant. The differential subhalo mass function is
tilted to a slope slightly shallower than the critical value $-2$, so
that, even when extrapolated to arbitrarily small masses, the total
mass fraction in substructures remains small, less than 3\% within 100
kpc of halo centre, and less than 20\% within $r_{50}\sim 400$~kpc.
Adopting the logarithmically divergent slope $-2$ (which our results
appear to exclude) does not increase these mass fractions by more than
a factor of 2 or 3 for lower mass limits in the range $10^{-6}$ to
$10^{-12},{\rm M}_\odot$, which plausibly correspond to the thermal
free-streaming limit if the dark matter is the lightest sypersymmetric
particle. The inner halo is dominated by a smoothly distributed dark
matter component, not by substructure.

Independent of their present mass, substructures have a {\em strong}
preference to be found in the outer regions of halos. For example, we estimate
that at most a fraction of $10^{-3}$ of the dark matter at the Solar circle is
in bound subclumps. The rest is smoothly distributed.  Note, however, that
this smooth component is expected to have a rich structure in velocity
space, being composed of a large number (perhaps $10^5$ or more) of
cold streams \citep{Helmi2003,Vogelsberger2008}.

Contrary to previous claims, we find that substructure in subhalos is
not a scaled-down version of substructure in main halos. Subhalos typically
have less substructure than main halos. This is due to two causes.  Tidal
stripping removes the outer substructure-rich parts of halos when they fall
into a larger system and become subhalos. In addition, as the retained
substructure ages it decreases in mass and number and is not replaced by the
infall of new objects. As a result, the substructure mass fraction in subhalos
is often much smaller than in main halos, particularly for subhalos in the
inner regions which are the most heavily stripped and also, typically, the
``oldest''.

We have presented the first detailed convergence study of the shape of
subhalo density profiles to be based on simulation sets where the {\em
  same} subhalo can be identified in simulations of differing mass
resolution. We find that the inner regions of subhalos, well inside
their tidal truncation radii, can be well fit by NFW or Einasto
profiles. Einasto fits are typically preferred, even when the shape
parameter $\alpha$ is fixed to a standard value, e.g.,
$\alpha=0.18$. We have also studied how the local logarithmic slope of
the density profile varies with radius, finding profiles to become
gradually shallower towards the centre with no sign of approaching an
asymptotic power-law behaviour. This is very similar to the behaviour
of the central cusp in isolated dark matter halos \citep{Navarro2004}.
\citep[We will address main halo cusps using the Aquarius simulations
  in a forthcoming paper,][]{Navarro2008}.  We find many subhalos for
which the slope at the innermost converged point is substantially
shallower than $-1.5$, and a few where it is shallower than
$-1.2$. The Moore profile appears firmly excluded as a description of
the inner regions of subhalos. It should {\rm not} be used when
modelling the cold dark matter annihilation signal, as in a number of
recent papers \citep[e.g.][]{Baltz2008}.

The concentration of subhalos is higher than that of halos of the same
circular velocity or of the same mass in the field. This can be
understood as a consequence of tidal truncation and mass loss
\citep{Kazantzidis2004,Bullock2005,Penarrubia2008} which lead to a
larger reduction of $V_{\rm max}$ than of $r_{\rm
  max}$. Interestingly, we find that the relationship between $r_{\rm
  max}$ and $V_{\rm max}$ for {\em field} halos is very well fit by
the fitting function given by \citet{Neto2007} for the Millennium
simulation, even though this involves an extrapolation over many
orders of magnitude towards lower mass. At the same maximum circular
velocity, we find that the $r_{\rm max}$ values of subhalos are, on
average, only 62\% of those of field halos.

We note that our results disagree with those of the recent Via Lactea I and II
simulations \citep{Diemand2007,Madau2008,Diemand2008,Kuhlen2008} on several
important points.  We find substantially more substructure than reported for
the Via Lactea simulations, and the discrepancy with Via Lactea I is larger
than the expected halo-to-halo scatter, based on our own simulation set.  Our
subhalos are also more concentrated than those found in the Via Lactea II
simulation. We also differ in our conclusions about the amount of
(sub-)substructure in subhalos, which we demonstrate to be less than predicted
by the hypothesis that subhalos are tidally truncated, but otherwise
scaled-down versions of field halos. Finally, we disagree with the claim of
\citet{Diemand2008} that subhalos have central power-law cusps with a mean
slope of $-1.2$.

In future work, we will analyze the detailed formation history of the
`Aquarius' halos and the evolution of their substructure. We will also build a
new generation of semi-analytic models to follow the evolution of the baryonic
component, and we will compare these with full hydrodynamical simulations of
these same halos that we have already begun to carry out.  This should bring
new insights into galaxy formation, and directly address possible small-scale
challenges to the $\Lambda$CDM theory. The verdict about whether CDM works on
such scales is still pending.

\section*{Acknowledgements}
The simulations for the Aquarius Project were carried out at the Leibniz
Computing Center, Garching, Germany, at the Computing Centre of the
Max-Planck-Society in Garching, at the Institute for Computational Cosmology
in Durham, and on the `STELLA' supercomputer of the LOFAR experiment at the
University of Groningen. SW acknowledges the Aspen Center for Physics for
providing the perfect atmosphere for final editing of this paper.

\bibliographystyle{mnras}
\bibliography{paper}

\end{document}